# BioInformatics

**SABU M. THAMPI**
**Assistant Professor**
**Dept. of CSE**
**LBS College of Engineering**
**Kasaragod, Kerala-671542**
**smtlbs@yahoo.co.in**

# Introduction

Bioinformatics is a new discipline that addresses the need to manage and interpret the data that in the past decade was massively generated by genomic research. This discipline represents the convergence of genomics, biotechnology and information technology, and encompasses analysis and interpretation of data, modeling of biological phenomena, and development of algorithms and statistics. Bioinformatics is by nature a cross-disciplinary field that began in the 1960s with the efforts of Margaret O. Dayhoff, Walter M. Fitch, Russell F. Doolittle and others and has matured into a fully developed discipline. However, bioinformatics is wide-encompassing and is therefore difficult to define. For many, including myself, it is still a nebulous term that encompasses molecular evolution, biological modeling, biophysics, and systems biology. For others, it is plainly computational science applied to a biological system. Bioinformatics is also a thriving field that is currently in the forefront of science and technology. Our society is investing heavily in the acquisition, transfer and exploitation of data and bioinformatics is at the center stage of activities that focus on the living world. It is currently a hot commodity, and students in bioinformatics will benefit from employment demand in government, the private sector, and academia.

With the advent of computers, humans have become 'data gatherers', measuring every aspect of our life with inferences derived from these activities. In this new culture, everything can and will become data (from internet traffic and consumer taste to the mapping of galaxies or human behavior). Everything can be measured (in pixels, Hertz, nucleotide bases, etc), turned into collections of numbers that can be stored (generally in bytes of information), archived in databases, disseminated (through cable or wireless conduits), and analyzed. We are expecting giant pay-offs from our data: proactive control of our world (from earthquakes and disease to finance and social stability), and clear understanding of chemical, biological and cosmological processes. Ultimately, we expect a better life. Unfortunately, data brings clutter and noise and its interpretation cannot keep pace with its accumulation. One problem with data is its multi-dimensionality and how to uncover underlying signal (patterns) in the most parsimonious way (generally using nonlinear approaches. Another problem relates to what we do with the data. Scientific discovery is driven by falsifiability and imagination and not by purely logical processes that turn observations into understanding. Data will not generate knowledge if we use inductive principles.

The gathering, archival, dissemination, modeling, and analysis of biological data falls within a relatively young field of scientific inquiry, currently known as 'bioinformatics', 'Bioinformatics was spurred by wide accessibility of computers with increased compute power and by the advent of *genomic*s. Genomics made it possible to acquire nucleic acid sequence and structural information from a wide range of genomes at an unprecedented pace and made this information accessible to further analysis and experimentation. For example, sequences were matched to those coding for globular proteins of known structure (defined by crystallography) and were used in high-throughput combinatorial approaches (such as DNA microarrays) to study patterns of gene expression. Inferences from sequences and biochemical data were used to construct metabolic networks. These activities have generated terabytes of data that are now being analyzed with computer, statistical, and machine learning techniques. The sheer number of sequences and information derived from these endeavors has given the false impression that imagination and hypothesis do not play a role in acquisition of biological knowledge. However, bioinformatics becomes only a science when fueled by hypothesis-driven research and within the context of the complex and ever-changing living world.

The science that relates to bioinformatics has many components. It usually relates to biological molecules and therefore requires knowledge in the fields of biochemistry, molecular biology, molecular evolution, thermodynamics, biophysics, molecular engineering, and statistical mechanics, to name a few. It requires the use of computer science, mathematical, and statistical principles. Bioinformatics is in the cross roads of experimental and theoretical science. Bioinformatics is not only about modeling or data ‘mining’, it is about understanding the molecular world that fuels life from evolutionary and mechanistic perspectives. It is truly inter-disciplinary and is changing. Much like biotechnology and genomics, bioinformatics is moving from applied to basic science, from developing tools to developing hypotheses.

**The Genome**
The hereditary information that an organism passes to its offspring is represented in each of its cells. The representation is in the form of DNA molecules. Each DNA molecule is a long chain of chemical structures called *nucleotides* of four different types, which can be viewed abstractly as characters from the alphabet fA; C; T; Gg. The totality of this information is called the *genome* of the organism. In humans the genome consists of nucleotides. A major task of molecular biology is to _ Extract the information contained in the genomes of different organisms; _ Elucidate the structure of the genome; _ Apply this knowledge to the diagnosis and ultimately, treatment, of genetic diseases (about 4000 such diseases in humans have been identified); _ By comparing the genomes of different species, explain the process and mechanisms of evolution. These tasks require the invention of new algorithms.

**Genetics**
Genetics is the study of heredity. It began with the observations of Mendel on the inheritance of simple traits such as the color of peas. Mendel worked out the fundamental mathematical rules for the inheritance of such traits. The central abstraction in genetics is the concept of a *gen*e. In classical genetics an organism's genes were abstract attributes for which no biochemical basis was known. Over the past 40 years it has become possible to understand the mechanisms of heredity at the molecular level. Classical genetics was based on breeding experiments on plants and animals. Modern genetics is based on the molecular study of fungi, bacteria and viruses.

**Cells**
In the early 19th century development of the microscope enabled cells to be observed. A cell is an assortment of chemicals inside a sac bounded by a fatty layer called the *plasma membran*e. Bacteria and protozoa consist of single cells. Plants and animals are complex collections of cells of different types, with similar cells joined together to form tissues. A human has trillions of cells. In a *eukaryotic cell* the genetic material is inside a *nucleus* separated from the rest of the cell by a membrane, and the cell contains a number of discrete functional units called *organelle*s. The part of the cell outside the nucleus is called the *cytoplas*m. In a *prokaryotic cell* there is no nucleus and the structure is more homogeneous. Bacteria are prokaryotes, plants, animals and protozoa are eukaryotes. A eukaryotic cell has diameter 10-100 microns, a bacterial cell 1-10 microns.

The genetic material in cells is contained in structures called *chromosome*s. In most prokaryotes, each cell has a single chromosome. In eukaryotes, all the cells of any individual contain the same number of chromosomes, 8 in fruities, 46 in humans and bats, 84 in the rhinoceros. In mammals and many other eukaryotic organisms the chromosomes occur in *homologous* (similar in shape and structure) pairs, except for the *sex chromosome*s. The male has an X chromosome and a Y chromosome, and the female has two X chromosomes. Cells with homologous pairs of chromosomes are called *diploi*d. Cells with unpaired chromosomes are called *haploi*d.I*npolyploid* organisms the chromosomes come in homologous triplets, quadruplets etc. *Mitosis* is the process of cell division. In mitosis each chromosome gets duplicated, the chromosomes line up in the central part of the cell, then separate into two groups which move to the ends of the cell, and a cell membrane forms separating the two sister cells. During mitosis chromosomes can be observed well under the microscope. They can be distinguished from one another by their banding structure, and certain abnormalities observed. For example, *Down Syndrome* occurs when there is an extra copy of chromosome 21.

The *sperm* and *egg* cells *(gamete*s) in sexually reproducing species are haploid. The process of *meiosis* produces four sperm/egg cells from a single diploid cell. We describe the main steps in meiosis. The original diploid cell has two homologous copies of each chromosome (with the exception of the X and Y chromosomes). One of these is called the *paternal copy* (because it was inherited from the father) and the other is called the *maternal cop*y. Each of the two copies duplicates itself. The four resulting copies are then brought together, and material may be exchanged between the paternal and maternal copies. This process is called *recombination* or *crossove*r. Then a cell division occurs, with the two paternal copies (possibly containing some material from the maternal copies by virtue of recombination) going to one of the daughter cells, and the two maternal copies to the other daughter cell. Each daughter cell then divides, with one of the two copies going to each of its daughter cells. The result is that four haploid sperm or egg cells are produced. In *fertilization* a sperm cell and an egg cell combine to form a diploid *zygot*e, from which a new individual develops.

**Inheritance of Simple Traits**

Many traits have two different versions: for example, pea seeds may be either green or yellow, and either smooth or wrinkled. In the late 19th century Mendel postulated that there is an associated gene that can exist in two versions, called *allele*s. An individual has two copies of the gene for the trait. If the two copies are the same allele the individual is *homozygous* for the trait, otherwise *heterozygou*s. On the basis of breeding experiments with peas Mendel postulated that each parent contributes a random one of its two alleles to the child. He also postulated incorrectly that the alleles for different traits are chosen independently.

**Proteins**

A *protein* is a very large biological molecule composed of a chain of smaller molecules called *amino acid*s. Thousands of different proteins are present in a cell, the synthesis of each type of protein being directed by a different gene. Proteins make up much of the cellular structure (our hair, skin, and fingernails consist largely of protein). Some proteins are *enzymes* which catalyze (make possible or greatly increase the rate of) chemical reactions within the cell. As we shall discuss later, some proteins are *transcription factors* which regulate the manner in which genes direct the production of other proteins. Proteins on the surfaces of cells act as receptors for hormones and other signaling molecules. The genes control the ability of cells to make enzymes. Thus genes control the functioning of cells.

**DNA**

DNA was discovered in 1869. Most of the DNA in cells is contained in the chromosomes. DNA is chemically very different from protein. DNA is structured as a double helix consisting of two long strands that wind around a common axis. Each strand is a very long chain of *nucleotides* of four types, A,C, T and G. There are four different types of nucleotides, distinguished by rings called *base*s, together with a common portion consisting of a sugar called *deoxyribose* and a *phosphate grou*p. On the sugar there are two sites, called the 3' and 5' sites. Each phosphate is bonded to two successive sugars, at the 3' site of one and the 5' site of the other. These phosphate-sugar links form the backbones of the two chains. The bases of the two chains are weakly bonded together in *complementary pairs* each of the form CG or AT. Thus the chains have directionality (from 3' to 5'), and the sequence of bases on one chain determines the sequence on the other, by the rule of complementarity.

**The linear ordering of the nucleotides determines the genetic informatio**n. Because of complementarity the two strands contain the same information, and this redundancy is the basis of DNA repair mechanisms. A gene is a part of the sequence of nucleotides which codes for a protein in a manner that will be detailed below. Variations in the sequence of the gene create different alleles, and *mutations* (changes in the sequence, due, for example, to disturbance by ultraviolet light) create new alleles. The *genome* is the totality of DNA stored in chromosomes typical of each species. The genome contains most of the information needed to specify an organism's properties. DNA undergoes replication, repair, rearrangement and recombination. These reactions are catalyzed by enzymes. We describe DNA replication. During replication, the two strands unwind at a particular point along the double helix. In the presence of an enzyme called *DNA polymerase* the unwound chain serves as a template for the formation of a complementary sequence of nucleotides, which are adjoined to the complementary strand one-by-one. Many short segments are formed, and these are bonded together in a reaction catalyzed by an enzyme called *DNA ligas*e. There are mechanisms for repairing errors that occur in this replication process, which occurs during mitosis.

# Definition of Bioinformatics

Roughly, bioinformatics describes any use of computers to handle biological information.

In practice the definition used by most people is narrower; bioinformatics to them is a synonym for "computational molecular biology"- the use of computers to characterize the molecular components of living things.

*"Classical" bioinformatics:*

"The mathematical, statistical and computing methods that aim to solve biological problems using DNA and amino acid sequences and related information."

**"The Loose" definition:**
There are other fields-for example medical imaging/image analysis which might be considered part of bioinformatics. There is also a whole other discipline of biologically-inspired computation; genetic algorithms, AI, neural networks. Often these areas interact in strange ways. Neural networks, inspired by crude models of the functioning of nerve cells in the brain, are used in a program called PHD to predict, surprisingly accurately, the secondary structures of proteins from their primary sequences. What almost all bioinformatics has in common is the processing of large amounts of biologically-derived information, whether DNA sequences or breast X-rays.

Even though the three terms: *bioinformatics*, *computational biology* and *bioinformation infrastructure* are often times used interchangeably, broadly, the three may be defined as follows:

1. *bioinformatics* refers to database-like activities, involving persistent sets of data that are maintained in a consistent state over essentially indefinite periods of time;
2. *computational biology* encompasses the use of algorithmic tools to facilitate biological analyses; while
3. *bioinformation infrastructure* comprises the entire collective of information management systems, analysis tools and communication networks supporting biology. Thus, the latter may be viewed as a computational scaffold of the former two.

*Bioinformatics* is currently defined as the study of information content and information flow in biological systems and processes. It has evolved to serve as the bridge between observations (data) in diverse biologically-related disciplines and the derivations of understanding (information) about how the systems or processes function, and subsequently the application (knowledge). A more pragmatic definition in the case of diseases is the understanding of dysfunction (diagnostics) and the subsequent applications of the knowledge for therapeutics and prognosis.

**The National Center for Biotechnology Information (NCBI 2001) defines bioinformatics as:**

"Bioinformatics is the field of science in which biology, computer science, and information technology merge into a single discipline. There are three important sub-disciplines within bioinformatics: the development of new algorithms and statistics with which to assess relationships among members of large data sets; the analysis and interpretation of various types of data including nucleotide and amino acid sequences, protein domains, and protein structures; and the development and implementation of tools that enable efficient access and management of different types of information."

**A Bioinformaticist versus a Bioinformatician (1999):**
Bioinformatics has become a mainstay of genomics, proteomics, and all other *.omics (such as phenomics) that many information technology companies have entered the business or are considering entering the business, creating an IT (information technology) and BT (biotechnology) convergence.

**A bioinformaticist** is an expert who not only knows how to use bioinformatics tools, but also knows how to write interfaces for effective use of the tools.

**A bioinformatician**, on the other hand, is a trained individual who only knows to use bioinformatics tools without a deeper understanding.

Thus, a bioinformaticist is to *.omics as a mechanical engineer is to an automobile. A bioinformatician is to *.omics as a technician is to an automobile.

## Definitions of Fields Related to Bioinformatics:

**Computational Biology:**

Computational biologists interest themselves more with evolutionary, population and theoretical biology rather than cell and molecular biomedicine. It is inevitable that molecular biology is profoundly important in computational biology, but it is certainly not what computational biology is all about. In these areas of computational biology it seems that computational biologists have tended to prefer statistical models for biological phenomena over physico-chemical ones.

One computational biologist (Paul J Schulte) did object to the above and makes the entirely valid point that this definition derives from a popular use of the term, rather than a correct one. Paul works on water flow in plant cells. He points out that biological fluid dynamics is a field of computational biology in itself. He argues that this, and any application of computing to biology, can be described as "computational biology". Where we disagree, perhaps, is in the conclusion he draws from this-which I reproduce in full: "Computational biology is not a "field", but an "approach" involving the use of computers to study biological processes and hence it is an area as diverse as biology itself.

**Genomics:**
Genomics is a field which existed before the completion of the sequences of genomes, but in the crudest of forms, for example the referenced estimate of 100000 genes in the human genome derived from an famous piece of "back of an envelope" genomics, guessing the weight of chromosomes and the density of the genes they bear. Genomics is any attempt to analyze or compare the entire genetic complement of a species or species (plural). It is, of course possible to compare genomes by comparing more-or-less representative subsets of genes within genomes.

**Proteomics:**
Michael J.Dunn, the Editor-in-Chief of Proteomics defines the "proteome" as: "the Protein complement of the genome" and proteomics to be concerned with: "Qualitative and quantitative studies of gene expression at the level of the functional proteins themselves" that is: "an interface between protein biochemistry and molecular biology" Characterizing the many tens of thousands of proteins expressed in a given cell type at a given time---whether measuring their molecular weights or isoelectric points, identifying their ligands or determining their structures---involves the storage and comparison of vast numbers of data. Inevitably this requires bioinformatics.

**Pharmacogenomics**:
Pharmacogenomics is the application of genomic approaches and technologies to the identification of drug targets. Examples include trawling entire genomes for potential receptors by bioinformatics means, or by investigating patterns of gene expression in both pathogens and hosts during infection, or by examining the characteristic expression patterns found in tumours or patients samples for diagnostic purposes (or in the pursuit of potential cancer therapy targets).

**Pharmacogenetics:**
All individuals respond differently to drug treatments; some positively, others with little obvious change in their conditions and yet others with side effects or allergic reactions. Much of this variation is known to have a genetic basis. Pharmacogenetics is a subset of pharmacogenomics which uses genomic/bioinformatic methods to identify genomic correlates, for example SNPs (Single Nucleotide Polymorphisms), characteristic of particular patient response profiles and use those markers to inform the administration and development of therapies. Strikingly such approaches have been used to "resurrect" drugs thought previously to be ineffective, but subsequently found to work with in subset of patients or in optimizing the doses of chemotherapy for particular patients.

**Cheminformatics:**
The Web advertisement for Cambridge Healthtech Institute's Sixth Annual Cheminformatics conference describes the field thus: "the combination of chemical synthesis, biological screening, and data-mining approaches used to guide drug discovery and development" but this, again, sounds more like a field being identified by some of its most popular (and lucrative) activities, rather than by including all the diverse studies that come under its general heading.

The story of one of the most successful drugs of all time, penicillin, seems bizarre, but the way we discover and develop drugs even now has similarities, being the result of chance, observation and a lot of slow, intensive chemistry. Until recently, drug design always seemed doomed to continue to be a labour-intensive, trial-and-error process. The possibility of using information technology, to plan intelligently and to automate processes related to the chemical synthesis of possible therapeutic compounds is very exciting for chemists and biochemists. The rewards for bringing a drug to market more rapidly are huge, so naturally this is what a lot of cheminformatics works is about. The span of academic cheminformatics is wide and is exemplified by the interests of the cheminformatics groups at the Centre for Molecular and Biomolecular Informatics at the University of Nijmegen in the Netherlands. These interests include: · Synthesis Planning · Reaction and Structure Retrieval · 3-D Structure Retrieval · Modelling · Computational Chemistry · Visualisation Tools and Utilities

Trinity University's Cheminformatics Web page, for another example, concerns itself with cheminformatics as the use of the Internet in chemistry.

**Medical Informatics:**

"Biomedical Informatics is an emerging discipline that has been defined as the study, invention, and implementation of structures and algorithms to improve communication, understanding and management of medical information." Medical informatics is more concerned with structures and algorithms for the manipulation of medical data, rather than with the data itself. This suggests that one difference between bioinformatics and medical informatics as disciplines lies with their approaches to the data; there are bioinformaticists interested in the theory behind the manipulation of that data and there are bioinformatics scientists concerned with the data itself and its biological implications. Medical informatics, for practical reasons, is more likely to deal with data obtained at "grosser" biological levels---that is information from super-cellular systems, right up to the population level-while most bioinformatics is concerned with information about cellular and biomolecular structures and systems.

# Origin & History of Bioinformatics

Over a century ago, bioinformatics history started with an Austrian monk named Gregor Mendel. He is known as the "Father of Genetics". He cross-fertilized different colors of the same species of flowers. He kept careful records of the colors of flowers that he cross-fertilized and the color(s) of flowers they produced. Mendel illustrated that the inheritance of traits could be more easily explained if it was controlled by factors passed down from generation to generation.

Since Mendel, bioinformatics and genetic record keeping have come a long way. The understanding of genetics has advanced remarkably in the last thirty years. In 1972, Paul berg made the first recombinant DNA molecule using ligase. In that same year, Stanley Cohen, Annie Chang and Herbert Boyer produced the first recombinant DNA organism. In 1973, two important things happened in the field of genomics:

1. Joseph Sambrook led a team that refined DNA electrophoresis using agarose gel, and
2. Herbert Boyer and Stanely Cohen invented DNA cloning. By 1977, a method for sequencing DNA was discovered and the first genetic engineering company, Genetech was founded.

By 1981, 579 human genes had been mapped and mapping by insitu hybridization had become a standard method. Marvin Carruthers and Leory Hood made a huge leap in bioinformatics when they invented a mehtod for automated DNA sequencing. In 1988, the Human Genome organization (HUGO) was founded. This is an international organization of scientists involved in Human Genome Project. In 1989, the first complete genome map was published of the bacteria Haemophilus influenza.

The following year, the Human Genome Project was started. By 1991, a total of 1879 human genes had been mapped. In 1993, Genethon, a human genome research center in France Produced a physical map of the human genome. Three years later, Genethon published the final version of the Human Genetic Map. This concluded the end of the first phase of the Human Genome Project.

In the mid-1970s, it would take a laboratory at least two months to sequence 150 nucleotides. Ten years ago, the only way to track genes was to scour large, well documented family trees of relatively inbred populations, such as the Ashkenzai Jews from Europez. These types of genealogical searches 11 million nucleotides a day for its corporate clients and company research.

Bioinformatics was fuelled by the need to create huge databases, such as GenBank and EMBL and DNA Database of Japan to store and compare the DNA sequence data erupting from the human genome and other genome sequencing projects. Today, bioinformatics embraces protein structure analysis, gene and protein functional information, data from patients, pre-clinical and clinical trials, and the metabolic pathways of numerous species.

**Origin of internet:**
The management and, more importantly, accessibility of this data is directly attributable to the development of the Internet, particularly the World Wide Web (WWW). Originally developed for military purposes in the 60's and expanded by the National Science Foundation in the 80's, scientific use of the Internet grew dramatically following the release of the WWW by CERN in 1992.

**HTML:**
The WWW is a graphical interface based on hypertext by which text and graphics can be displayed and highlighted. Each highlighted element is a pointer to another document or an element in another document which can reside on any internet host computer. Page display, hypertext links and other features are coded using a simple, cross-platform HyperText Markup Language (HTML) and viewed on UNIX workstations, PCs and Apple Macs as WWW pages using a browser.

**Java:**
The first graphical WWW browser - Mosaic for X and the first molecular biology WWW server - ExPASy were made available in 1993. In 1995, Sun Microsystems released Java, an object-oriented, portable programming language based on C++. In addition to being a standalone programming language in the classic sense, Java provides a highly interactive, dynamic content to the Internet and offers a uniform operational level for all types of computers, provided they implement the 'Java Virtual Machine' (JVM). Thus, programs can be written, transmitted over the internet and executed on any other type of remote machine running a JVM. Java is also integrated into Netscape and Microsoft browsers, providing both the common interface and programming capability which are vital in sorting through and interpreting the gigabytes of bioinformatics data now available and increasing at an exponential rate.

**XML:**
The new XML standard 8 is a project of the World-Wide Web Consortium (W3C) which extends the power of the WWW to deliver not only HTML documents but an unlimited range of document types using customised markup. This will enable the bioinformatics community to exchange data objects such as sequence alignments, chemical structures, spectra etc., together with appropriate tools to display them, just as easily as they exchange HTML documents today. Both Microsoft and Netscape support this new technology in their latest browsers.

**CORBA:**
Another new technology, called CORBA, provides a way of bringing together many existing or 'legacy' tools and databases with a common interface that can be used to drive them and access data. CORBA frameworks for bioinformatics tools and databases have been developed by, for example, NetGenics and the European Bioinformatics Institute (EBI).

Representatives from industry and the public sector under the umbrella of the Object Management Group are working on open CORBA-based standards for biological information representation The Internet offers scientists a universal platform on which to share and search for data and the tools to ease data searching, processing, integration and interpretation. The same hardware and software tools are also used by companies and organisations in more private yet still global Intranet networks. One such company, Oxford GlycoSciences in the UK, has developed a bioinformatics system as a key part of its proteomics activity.

**ROSETTA:**

ROSETTA focuses on protein expression data and sets out to identify the specific proteins which are up- or down-regulated in a particular disease; characterise these proteins with respect to their primary structure, post-translational modifications and biological function; evaluate them as drug targets and markers of disease; and develop novel drug candidates OGS uses a technique called fluorescent IPG-PAGE to separate and measure different protein types in a biological sample such as a body fluid or purified cell extract. After separation, each protein is collected and then broken up into many different fragments using controlled techniques. The mass and sequence of these fragments is determined with great accuracy using a technique called mass spectrometry. The sequence of the original protein can then be theoretically reconstructed by fitting these fragments back together in a kind of jigsaw. This reassembly of the protein sequence is a task well-suited to signal processing and statistical methods.

ROSETTA is built on an object-relational database system which stores demographic and clinical data on sample donors and tracks the processing of samples and analytical results. It also interprets protein sequence data and matches this data with that held in public, client and proprietary protein and gene databases. ROSETTA comprises a suite of linked HTML pages which allow data to be entered, modified and searched and allows the user easy access to other databases. A high level of intelligence is provided through a sophisticated suite of proprietary search, analytical and computational algorithms. These algorithms facilitate searching through the gigabytes of data generated by the Company's proteome projects, matching sequence data, carrying out de novo peptide sequencing and correlating results with clinical data. These processing tools are mostly written in C, C++ or Java to run on a variety of computer platforms and use the networking protocol of the internet, TCP/IP, to co-ordinate the activities of a wide range of laboratory instrument computers, reliably identifying samples and collecting data for analysis.

The need to analyse ever increasing numbers of biological samples using increasingly complex analytical techniques is insatiable. Searching for signals and trends in noisy data continues to be a challenging task, requiring great computing power. Fortunately this power is available with today's computers, but of key importance is the integration of analytical data, functional data and biostatistics. The protein expression data in ROSETTA forms only part of an elaborate network of the type of data which can now be brought to bear in Biology. The need to integrate different information systems into a collaborative network with a friendly face is bringing together an exciting mixture of talents in the software world and has brought the new science of bioinformatics to life.

## Origin of bioinformatic/biological databases

The first bioinformatic/biological databases were constructed a few years after the first protein sequences began to become available. The first protein sequence reported was that of bovine insulin in 1956, consisting of 51 residues. Nearly a decade later, the first nucleic acid sequence was reported, that of yeast alanine tRNA with 77 bases. Just a year later, Dayhoff gathered all the available sequence data to create the first bioinformatic database. The Protein Data Bank followed in 1972 with a collection of ten X-ray crystallographic protein structures, and the SWISSPROT protein sequence database began in 1987.A huge variety of divergent data resources of different type sand sizes are now available either in the public domain or more recently from commercial third parties. All of the original databases were organised in a very simple way with data entries being stored in flat files, either one perentry, or as a single large text file. Re-write - Later on lookup indexes were added to allow convenient keyword searching of header information.

**Origin of tools**

After the formation of the databases, tools became available to search sequence databases - at first in a very simple way, looking for keyword matches and short sequence words, and then more sophisticated pattern matching and alignment based methods. The rapid but less rigorous BLAST algorithm has been the mainstay of sequence database searching since its introduction a decade ago, complemented by the more rigorous and slower FASTA and Smith Waterman algorithms. Suites of analysis algorithms, written by leading academic researchers at Stanford, CA, Cambridge, UK and Madison, WI for their in-house projects, began to become more widely available for basic sequence analysis. These algorithms were typically single function black boxes that took input and produced output in the form of formatted files. UNIX style commands were used to operate the algorithms, with some suites having hundreds of possible commands, each taking different command options and input formats. Since these early efforts, significant advances have been made in automating the collection of sequence information.

Rapid innovation in biochemistry and instrumentation has brought us to the point where the entire genomic sequence of at least 20 organisms, mainly microbial pathogens, are known and projects to elucidate at least 100 more prokaryotic and eukaryotic genomes are currently under way. Groups are now even competing to finish the sequence of the entire human genome. With new technologies we can directly examine the changes in expression levels of both mRNA and proteins in living cells, both in a disease state or following an external challenge. We can go on to identify patterns of response in cells that lead us to an understanding of the mechanism of action of an agent on a tissue. The volume of data arising from projects of this nature is unprecedented in the pharma industry, and will have a profound effect on the ways in which data are used and experiments performed in drug discovery and development projects. This is true not least because, with much of the available interesting data being in the hands of commercial genomics companies, pharmcos are unable to get exclusive access to many gene sequences or their expression profiles.

The competition between co-licensees of a genomic database is effectively a race to establish a mechanistic role or other utility for a gene in a disease state in order to secure a patent position on that gene. Much of this work is carried out by informatics tools. Despite the huge progress in sequencing and expression analysis technologies, and the corresponding magnitude of more data that is held in the public, private and commercial databases, the tools used for storage, retrieval, analysis and dissemination of data in bioinformatics are still very similar to the original systems gathered together by researchers 15-20 years ago. Many are simple extensions of the original academic systems, which have served the needs of both academic and commercial users for many years. These systems are now beginning to fall behind as they struggle to keep up with the pace of change in the pharma industry. Databases are still gathered, organised, disseminated and searched using flat files. Relational databases are still few and far between, and object-relational or fully object oriented systems are rarer still in mainstream applications. Interfaces still rely on command lines, fat client interfaces, which must be installed on every desktop, or HTML/CGI forms. Whilst they were in the hands of bioinformatics specialists, pharmcos have been relatively undemanding of their tools. Now the problems have expanded to cover the mainstream discovery process, much more flexible and scalable solutions are needed to serve pharma R&D informatics requirements.

There are different views of origin of Bioinformatics- *From T K Attwood and D J Parry-Smith's "Introduction to Bioinformatics", Prentice-Hall 1999 [Longman Higher Education; ISBN 0582327881]:* "The term bioinformatics is used to encompass almost all computer applications in biological sciences, but was originally coined in the mid-1980s for the analysis of biological sequence data."

From *Mark S. Boguski's article in the "Trends Guide to Bioinformatics" Elsevier, Trends Supplement 1998 p1:* "The term "bioinformatics" is a relatively recent invention, not appearing in the literature until 1991 and then only in the context of the emergence of electronic publishing... The **National Center for Biotechnology Information (NCBI)**, is celebrating its 10th anniversary this year, having been written into existence by US Congressman Claude Pepper and President Ronald Reagan in 1988. So bioinformatics has, in fact, been in existence for more than 30 years and is now middle-aged."

## A Chronological History of Bioinformatics

- **1953** - Watson & Crick proposed the double helix model for DNA based x-ray data obtained by Franklin & Wilkins.
- **1954** - Perutz's group develop heavy atom methods to solve the phase problem in protein crystallography.
- **1955** - The sequence of the first protein to be analysed, bovine insulin, is announed by F.Sanger.
- **1969** - The ARPANET is created by linking computers at Standford and UCLA.
- **1970** - The details of the Needleman-Wunsch algorithm for sequence comparison are published.
- **1972** - The first recombinant DNA molecule is created by Paul Berg and his group.
- **1973** - The Brookhaven Protein DataBank is announeced (Acta.Cryst.B,1973,29:1764). Robert Metcalfe receives his Ph.D from Harvard University. His thesis describes Ethernet.
- **1974** - Vint Cerf and Robert Khan develop the concept of connecting networks of computers into an "internet" and develop the Transmission Control Protocol (TCP).
- **1975** - Microsoft Corporation is founded by Bill Gates and Paul Allen.
  Two-dimensional electrophoresis, where separation of proteins on SDS polyacrylamide gel is combined with separation according to isoelectric points, is announced by P.H.O'Farrel.
- **1988** - The National Centre for Biotechnology Information (NCBI) is established at the National Cancer Institute.
  The Human Genome Intiative is started (commission on Life Sciences, National Research council. Mapping and sequencing the Human Genome, National Academy Press: wahington, D.C.), 1988.
  The FASTA algorith for sequence comparison is published by Pearson and Lupman.
  A new program, an Internet computer virus desined by a student, infects 6,000 military computers in the US.
- **1989** - The genetics Computer Group (GCG) becomes a privatae company.
  Oxford Molceular Group,Ltd.(OMG) founded, UK by Anthony Marchigton, David Ricketts, James Hiddleston, Anthony Rees, and W.Graham Richards. Primary products: Anaconds, Asp, Cameleon and others (molecular modeling, drug design, protein design).
- **1990** - The BLAST program (Altschul,et.al.) is implemented.
  Molecular applications group is founded in California by Michael Levitt and Chris Lee. Their primary products are Look and SegMod which are used for molecular modeling and protein deisign.
  InforMax is founded in Bethesda, MD. The company's products address sequence analysis, database and data management, searching, publication graphics, clone construction, mapping and primer design.
- **1991** - The research institute in Geneva (CERN) announces the creation of the protocols which make -up the World Wide Web.
  The creation and use of expressed sequence tags (ESTs) is described.
  Incyte Pharmaceuticals, a genomics company headquartered in Palo Alto California, is formed.
  Myriad Genetics, Inc. is founded in Utah. The company's goal is to lead in the discovery of major common human disease genes and their related pathways. The company has discovered and sequenced, with its academic collaborators, the following major genes: BRCA1, BRACA1 , CHD1, MMAC1, MMSC1, MMSC2, CtIP, p16, p19 and MTS2.

- **1993** - CuraGen Corporation is formed in New Haven, CT.
  Affymetrix begins independent operations in Santa Clara, California.
- **1994** - Netscape Communications Corporation founded and releases Naviagator, the commerical version of NCSA's Mozilla.
  Gene Logic is formed in Maryland.
  The PRINTS database of protein motifs is published by Attwood and Beck.
  Oxford Molecular Group acquires IntelliGenetics.
- **1995** - The Haemophilus influenzea genome (1.8) is sequenced.
  The Mycoplasma genitalium genome is sequenced.
- **1996** - The genome for Saccharomyces cerevisiae (baker's yeadt, 12.1 Mb) is sequenced.
  The prosite database is reported by Bairoch, et.al.
  Affymetrix produces the first commerical DNA chips.
- **1997** - The genome for E.coli (4.7 Mbp) is published.Oxford Molecualr Group acquires the Genetics Computer Group. LION bioscience AG founded as an intergrated genomics company with strong focus on bioinformatics. The company is built from IP out of the European Molecualr Biology Laboratory (EMBL), the European Bioinformtics Institute (EBI), the GErman Cancer Research Center (DKFZ), and the University of Heidelberg.paradigm Genetics Inc., a company focussed on the application of genomic technologies to enhance worldwide food and fiber production, is founded in Research Triangle Park, NC. deCode genetics publishes a paper that described the location of the FET1 gene, which is responsible for familial essential tremor, on chromosome 13 (Nature Genetics).
- **1998** - The genomes for Caenorhabitis elegans and baker's yeast are published.The Swiss Institute of Bioinformatics is established as a non-profit foundation.Craig Venter forms Celera in Rockville, Maryland. PE Informatics was formed as a center of Excellence within PE Biosystems. This center brings together and leverges the complementary expertise of PE Nelson and Molecualr Informatics, to further complement the genetic instrumention expertise of Applied Biosystems.Inpharmatica, a new Genomics and Bioinformatics company, is established by University College London, the Wolfson Institute for Biomedical Research, five leading scientists from major British academic centres and Unibio Limited. GeneFormatics, a company dedicated to the analysis and predication of protein structure and function, is formed in San Diego.Molecualr Simulations Inc. is acquired by Pharmacopeia.
- **1999** - deCode genetics maps the gene linked to pre-eclampsia as a locus on chromosome 2p13.
- **2000** - The genome for Pseudomonas aeruginosa (6.3 Mbp) is published. The Athaliana genome (100 Mb) is secquenced.The D.melanogaster genome (180 Mb) is sequenced.Pharmacopeia acquires Oxoford Molecular Group.
- **2001** - The huam genome (3,000 Mbp) is published.

## Biological Database

A biological database is a large, organized body of persistent data, usually associated with computerized software designed to update, query, and retrieve components of the data stored within the system. A simple database might be a single file containing many records, each of which includes the same set of information. For example, a record associated with a nucleotide sequence database typically contains information such as contact name; the input sequence with a description of the type of molecule; the scientific name of the source organism from which it was isolated; and, often, literature citations associated with the sequence.

For researchers to benefit from the data stored in a database, two additional requirements must be met:

- Easy access to the information; and
- A method for extracting only that information needed to answer a specific biological question.

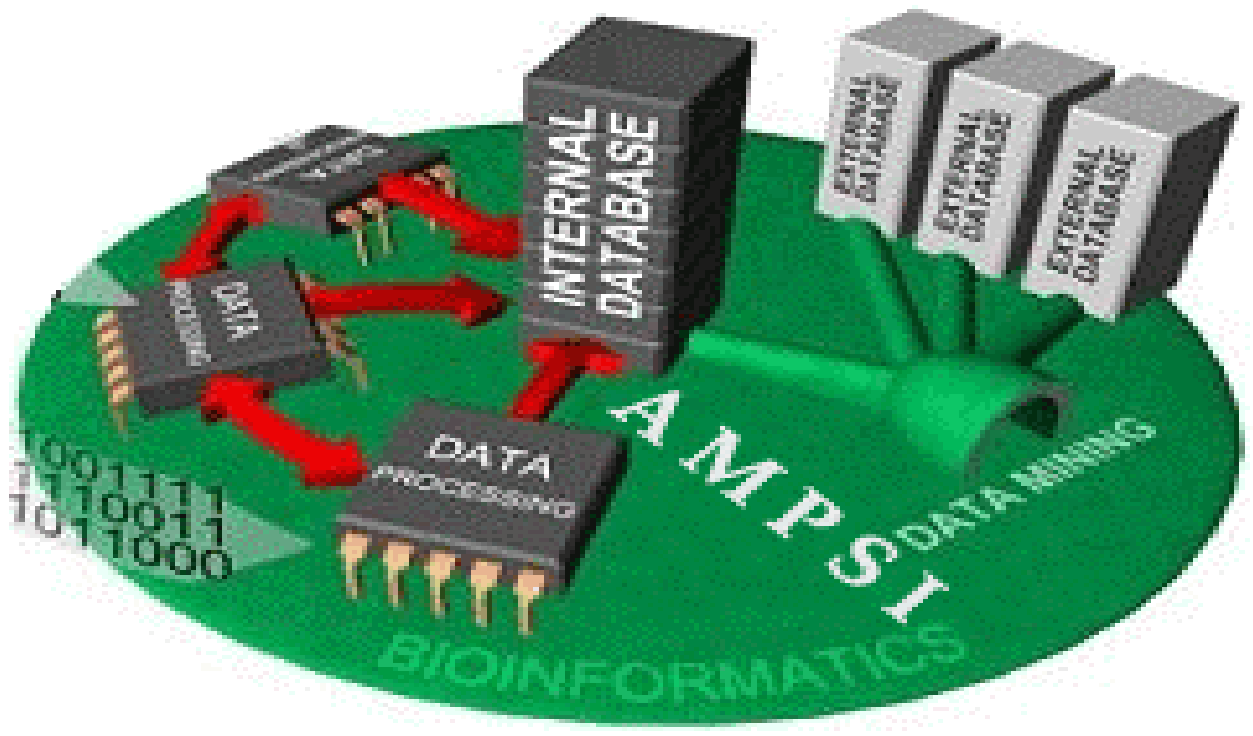

Currently, a lot of bioinformatics work is concerned with the technology of databases. These databases include both "public" repositories of gene data like GenBank or the Protein DataBank (the PDB), and private databases like those used by research groups involved in gene mapping projects or those held by biotech companies. Making such databases accessible via open standards like the Web is very important since consumers of bioinformatics data use a range of computer platforms: from the more powerful and forbidding UNIX boxes favoured by the developers and curators to the far friendlier Macs often found populating the labs of computer-wary biologists. RNA and DNA are the proteins that store the hereditary information about an organism. These macromolecules have a fixed structure, which can be analyzed by biologists with the help of bioinformatic tools and databases. A few popular databases are GenBank from NCBI (National Center for Biotechnology Information), SwissProt from the Swiss Institute of Bioinformatics and PIR from the Protein Information Resource.

**GenBank:**
GenBank (Genetic Sequence Databank) is one of the fastest growing repositories of known genetic sequences. It has a flat file structure, that is an ASCII text file, readable by both humans and computers. In addition to sequence data, GenBank files contain information like accession numbers and gene names, phylogenetic classification and references to published literature.There are approximately 191,400,000 bases and 183,000 sequences as of June 1994.

**EMBL:**
The EMBL Nucleotide Sequence Database is a comprehensive database of DNA and RNA sequences collected from the scientific literature and patent applications and directly submitted from researchers and sequencing groups. Data collection is done in collaboration with GenBank (USA) and the DNA Database of Japan (DDBJ). The database currently doubles in size every 18 months and currently (June 1994) contains nearly 2 million bases from 182,615 sequence entries.

**SwissProt:**
This is a protein sequence database that provides a high level of integration with other databases and also has a very low level of redundancy (means less identical sequences are present in the database).

**PROSITE:**
The PROSITE dictionary of sites and patterns in proteins prepared by Amos Bairoch at the University of Geneva.

**EC-ENZYME:**
The 'ENZYME' data bank contains the following data for each type of characterized enzyme for which an EC number has been provided: EC number, Recommended name, Alternative names, Catalytic activity, Cofactors, Pointers to the SWISS-PROT entrie(s) that correspond to the enzyme, Pointers to disease(s) associated with a deficiency of the enzyme.

**PDB:**
The X-ray crystallography Protein Data Bank (PDB), compiled at the Brookhaven National Laboratory.

**GDB:**
The GDB Human Genome Data Base supports biomedical research, clinical medicine, and professional and scientific education by providing for the storage and dissemination of data about genes and other DNA markers, map location, genetic disease and locus information, and bibliographic information.

**OMIM:**
The Mendelian Inheritance in Man data bank (MIM) is prepared by Victor Mc Kusick with the assistance of Claire A. Francomano and Stylianos E. Antonarakis at John Hopkins University.

**PIR-PSD:**
PIR (Protein Information Resource) produces and distributes the PIR-International Protein Sequence Database (PSD). It is the most comprehensive and expertly annotated protein sequence database.The PIR serves the scientific community through on-line access, distributing magnetic tapes, and performing off-line sequence identification services for researchers. Release 40.00: March 31, 1994 67,423 entries 19,747,297 residues.

Protein sequence databases are classified as primary, secondary and composite depending upon the content stored in them. PIR and SwissProt are primary databases that contain protein sequences as 'raw' data. Secondary databases (like Prosite) contain the information derived from protein sequences. Primary databases are combined and filtered to form non-redundant composite database

**Genethon Genome Databases**

PHYSICAL MAP: computation of the human genetic map using DNA fragments in the form of YAC contigs. GENETIC MAP: production of micro-satellite probes and the localization of chromosomes, to create a genetic map to aid in the study of hereditary diseases. GENEXPRESS (cDNA): catalogue the transcripts required for protein synthesis obtained from specific tissues, for example neuromuscular tissues.

**21 Bdb: LBL's Human Chr 21 database:**

This is a W3 interface to LBL's ACeDB-style database for Chromosome 21, 21Bdb, using the ACeDB gateway software developed and provided by Guy Decoux at INRA.

**MGD: The Mouse Genome Databases:**

MGD is a comprehensive database of genetic information on the laboratory mouse. This initial release contains the following kinds of information: Loci (over 15,000 current and withdrawn symbols), Homologies (1300 mouse loci, 3500 loci from 40 mammalian species), Probes and Clones (about 10,000), PCR primers (currently 500 primer pairs), Bibliography (over 18,000 references), Experimental data (from 2400 published articles).

**ACeDB (A Caenorhabditis elegans Database) :**

Containing data from the Caenorhabditis Genetics Center (funded by the NIH National Center for Research Resources), the C. elegans genome project (funded by the MRC and NIH), and the worm community. Contacts: Mary O'Callaghan (moc@mrc-lmb.cam.ac.uk) and Richard Durbin.

ACeDB is also the name of the generic genome database software in use by an increasing number of genome projects. The software, as well as the C. elegans data, can be obtained via ftp.

ACeDB databases are available for the following species: C. elegans, Human Chromosome 21, Human Chromosome X, Drosophila melanogaster, mycobacteria, Arabidopsis, soybeans, rice, maize, grains, forest trees, Solanaceae, Aspergillus nidulans, Bos taurus, Gossypium hirsutum, Neurospora crassa, Saccharomyces cerevisiae, Schizosaccharomyces pombe, and Sorghum bicolor.

**MEDLINE:**

MEDLINE is NLM's premier bibliographic database covering the fields of medicine, nursing, dentistry, veterinary medicine, and the preclinical sciences. Journal articles are indexed for MEDLINE, and their citations are searchable, using NLM's controlled vocabulary, MeSH (Medical Subject Headings). MEDLINE contains all citations published in Index Medicus, and corresponds in part to the International Nursing Index and the Index to Dental Literature. Citations include the English abstract when published with the article (approximately 70% of the current file).

## The Database Industry

Because of the high rate of data production and the need for researchers to have rapid access to new data, public databases have become the major medium through which genome sequence data are published. Public databases and the data services that support them are important resources in bioinformatics, and will soon be essential sources of information for all the molecular biosciences. However, successful public data services suffer from continually escalating demands from the biological community. It is probably important to realize from the very beginning that the databases will never completely satisfy a very large percentage of the user community. The range of interest within biology itself suggests the difficulty of constructing a database that will satisfy all the potential demands on it. There is virtually no end to the depth and breadth of desirable information of interest and use to the biological community.

EMBL and GenBank are the two major nucleotide databases. EMBL is the European version and GenBank is the American. EMBL and GenBank collaborate and synchronize their databases so that the databases will contain the same information. The rate of growth of DNA databases has been following an exponential trend, with a doubling time now estimated to be 9-12 months. In January 1998, EMBL contained more than a million entries, representing more than 15,500 species, although most data is from model organisms such as Saccharomyces cerevisiae, Homo sapiens, Caenorhabditis elegans, Mus musculus and Arabidopsis thaliana. These databases are updated on a daily basis, but still you may find that a sequence referred to in the latest issue of a journal is not accessible. This is most often due to the fact that the release-date of the entry did not correlate with the publication date, or that the authors forgot to tell the databases that the sequences have been published.

**The principal requirements on the public data services are:**

- **Data quality** - data quality has to be of the highest priority. However, because the data services in most cases lack access to supporting data, the quality of the data must remain the primary responsibility of the submitter.
- **Supporting data** - database users will need to examine the primary experimental data, either in the database itself, or by following cross-references back to network-accessible laboratory databases.
- **Deep annotation** - deep, consistent annotation comprising supporting and ancillary information should be attached to each basic data object in the database.
- **Timeliness** - the basic data should be available on an Internet-accessible server within days (or hours) of publication or submission.
- **Integration** - each data object in the database should be cross-referenced to representation of the same or related biological entities in other databases. Data services should provide capabilities for following these links from one database or data service to another.

## The Creation of Sequence Databases

Most biological databases consist of long strings of nucleotides (guanine, adenine, thymine, cytosine and uracil) and/or amino acids (threonine, serine, glycine, etc.). Each sequence of nucleotides or amino acids represents a particular gene or protein (or section thereof), respectively. Sequences are represented in shorthand, using single letter designations. This decreases the space necessary to store information and increases processing speed for analysis.

While most biological databases contain nucleotide and protein sequence information, there are also databases which include taxonomic information such as the structural and biochemical characteristics of organisms. The power and ease of using sequence information has however, made it the method of choice in modern analysis.
In the last three decades, contributions from the fields of biology and chemistry have facilitated an increase in the speed of sequencing genes and proteins. The advent of cloning technology allowed foreign DNA sequences to be easily introduced into bacteria.
In this way, rapid mass production of particular DNA sequences, a necessary prelude to sequence determination, became possible.

Oligonucleotide synthesis provided researchers with the ability to construct short fragments of DNA with sequences of their own choosing. These oligonucleotides could then be used in probing vast libraries of DNA to extract genes containing that sequence. Alternatively, these DNA fragments could also be used in polymerase chain reactions to amplify existing DNA sequences or to modify these sequences. With these techniques in place, progress in biological research increased exponentially.

For researchers to benefit from all this information, however, two additional things were required: 1) ready access to the collected pool of sequence information and 2) a way to extract from this pool only those sequences of interest to a given researcher. Simply collecting, by hand, all necessary sequence information of interest to a given project from published journal articles quickly became a formidable task. After collection, the organization and analysis of this data still remained. It could take weeks to months for a researcher to search sequences by hand in order to find related genes or proteins. Computer technology has provided the obvious solution to this problem. Not only can computers be used to store and organize sequence information into databases, but they can also be used to analyze sequence data rapidly. The evolution of computing power and storage capacity has, so far, been able to outpace the increase in sequence information being created.

Theoretical scientists have derived new and sophisticated algorithms which allow sequences to be readily compared using probability theories. These comparisons become the basis for determining gene function, developing phylogenetic relationships and simulating protein models. The physical linking of a vast array of computers in the 1970's provided a few biologists with ready access to the expanding pool of sequence information. This web of connections, now known as the Internet, has evolved and expanded so that nearly everyone has access to this information and the tools necessary to analyze it. Databases of existing sequencing data can be used to identify homologues of new molecules that have been amplified and sequenced in the lab. The property of sharing a common ancestor, homology, can be a very powerful indicator in bioinformatics.

## Acquisition of sequence data

Bioinformatics tools can be used to obtain sequences of genes or proteins of interest, either from material obtained, labelled, prepared and examined in electric fields by individual researchers/groups or from repositories of sequences from previously investigated material.

## Analysis of data

Both types of sequence can then be analysed in many ways with bioinformatics tools. They can be assembled. Note that this is one of the occasions when the meaning of a biological term differs markedly from a computational one. Computer scientists, banish from your mind any thought of assembly language. Sequencing can only be performed for relatively short stretches of a biomolecule and finished sequences are therefore prepared by arranging overlapping "reads" of monomers (single beads on a molecular chain) into a single continuous passage of "code". This is the bioinformatic sense of assembly.

They can be mapped (that is, their sequences can be parsed to find sites where so-called "restriction enzymes" will cut them). They can be compared, usually by aligning corresponding segments and looking for matching and mismatching letters in their sequences. Genes or proteins which are sufficiently similar are likely to be related and are therefore said to be "homologous" to each other---the whole truth is rather more complicated than this. Such cousins are called "homologues". If a homologue (a related molecule) exists then a newly discovered protein may be modelled---that is the three dimensional structure of the gene product can be predicted without doing laboratory experiments.

Bioinformatics is used in primer design. Primers are short sequences needed to make many copies of (amplify) a piece of DNA as used in PCR (the Polymerase Chain Reaction). Bioinformatics is used to attempt to predict the function of actual gene products. Information about the similarity, and, by implication, the relatedness of proteins is used to trace the "family trees" of different molecules through evolutionary time. There are various other applications of computer analysis to sequence data, but, with so much raw data being generated by the Human Genome Project and other initiatives in biology, computers are presently essential for many biologists just to manage their day-to-day results Molecular modelling/structural biology is a growing field which can be considered part of bioinformatics. There are, for example, tools which allow you (often via the Net) to make pretty good predictions of the secondary structure of proteins arising from a given amino acid sequence, often based on known "solved" structures and other sequenced molecules acquired by structural biologists. Structural biologists use "bioinformatics" to handle the vast and complex data from X-ray crystallography, nuclear magnetic resonance (NMR) and electron microscopy investigations and create the 3-D models of molecules that seem to be everywhere in the media.

**Note:** Unfortunately the word "map" is used in several different ways in biology/genetics/bioinformatics. The definition given above is the one most frequently used in this context, but a gene can be said to be "mapped" when its parent chromosome has been identified, when its physical or genetic distance from other genes is established and---less frequently---when the structure and locations of its various coding components (its "exons") are established.

# Bioinformatics-Programs & Tools

Bioinformatic tools are software programs that are designed for extracting the meaningful information from the mass of data & to carry out this analysis step.

**Factors that must be taken into consideration when designing these tools are**

- The end user (the biologist) may not be a frequent user of computer technology
- These software tools must be made available over the internet given the global distribution of the scientific research community

**Major categories of Bioinformatics Tools**

There are both standard and customized products to meet the requirements of particular projects. There are data-mining software that retrieves data from genomic sequence databases and also visualization tools to analyze and retrieve information from proteomic databases. These can be classified as homology and similarity tools, protein functional analysis tools, sequence analysis tools and miscellaneous tools. Here is a brief description of a few of these. Everyday bioinformatics is done with sequence search programs like BLAST, sequence analysis programs, like the EMBOSS and Staden packages, structure prediction programs like THREADER or PHD or molecular imaging/modelling programs like RasMol and WHATIF.

**Homology and Similarity Tools**

Homologous sequences are sequences that are related by divergence from a common ancestor. Thus the degree of similarity between two sequences can be measured while their homology is a case of being either true of false. This set of tools can be used to identify similarities between novel query sequences of unknown structure and function and database sequences whose structure and function have been elucidated.

**Protein Function Analysis**

These groups of programs allow you to compare your protein sequence to the secondary (or derived) protein databases that contain information on motifs, signatures and protein domains. Highly significant hits against these different pattern databases allow you to approximate the biochemical function of your query protein.

**Structural Analysis**

These sets of tools allow you to compare structures with the known structure databases. The function of a protein is more directly a consequence of its structure rather than its sequence with structural homologs tending to share functions. The determination of a protein's 2D/3D structure is crucial in the study of its function.

**Sequence Analysis**

This set of tools allows you to carry out further, more detailed analysis on your query sequence including evolutionary analysis, identification of mutations, hydropathy regions,

CpG islands and compositional biases. The identification of these and other biological properties are all clues that aid the search to elucidate the specific function of your sequence.

## Some examples of Bioinformatics Tools

**BLAST**

BLAST (**B**asic **L**ocal **A**lignment **S**earch **T**ool) comes under the category of homology and similarity tools. It is a set of search programs designed for the Windows platform and is used to perform fast similarity searches regardless of whether the query is for protein or DNA. Comparison of nucleotide sequences in a database can be performed. Also a protein database can be searched to find a match against the queried protein sequence. NCBI has also introduced the new queuing system to BLAST (Q BLAST) that allows users to retrieve results at their convenience and format their results multiple times with different formatting options.

Depending on the type of sequences to compare, there are different programs:

- blastp compares an amino acid query sequence against a protein sequence database
- blastn compares a nucleotide query sequence against a nucleotide sequence database
- blastx compares a nucleotide query sequence translated in all reading frames against a protein sequence database
- tblastn compares a protein query sequence against a nucleotide sequence database dynamically translated in all reading frames
- tblastx compares the six-frame translations of a nucleotide query sequence against the six-frame translations of a nucleotide sequence database.

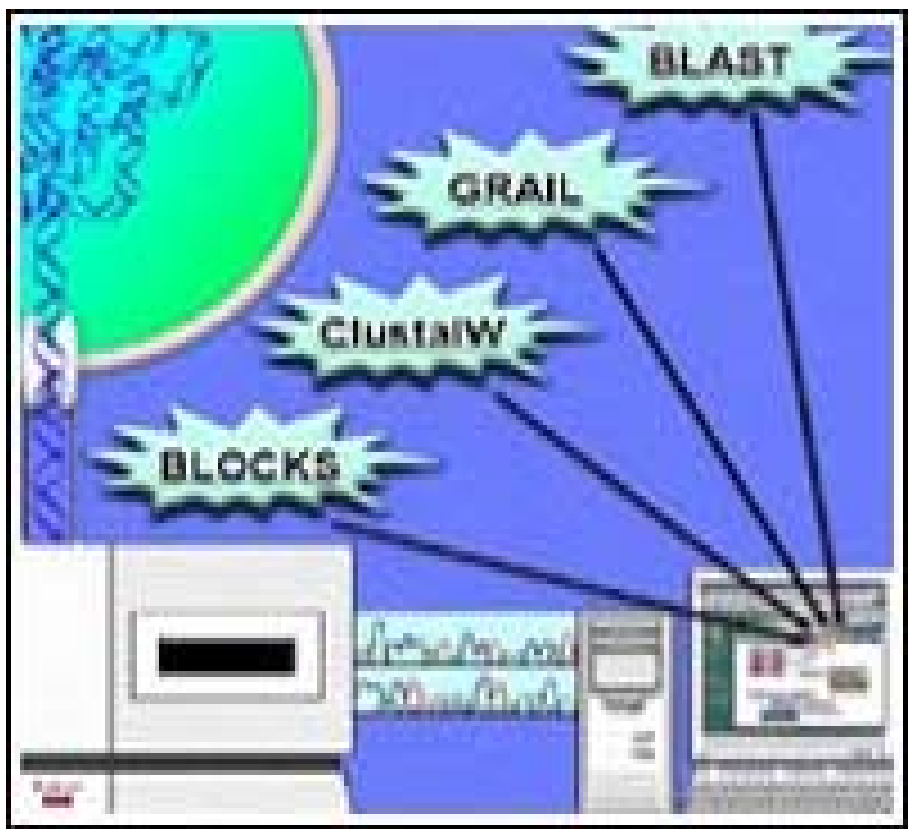

**FASTA**

**FAST** homology search **A**ll sequences .An alignment program for protein sequences created by Pearsin and Lipman in 1988. The program is one of the many heuristic algorithms proposed to speed up sequence comparison. The basic idea is to add a fast prescreen step to locate the highly matching segments between two sequences, and then extend these matching segments to local alignments using more rigorous algorithms such as Smith-Waterman.

**EMBOSS**

EMBOSS (European Molecular Biology Open Software Suite) is a software-analysis package. It can work with data in a range of formats and also retrieve sequence data transparently from the Web. Extensive libraries are also provided with this package, allowing other scientists to release their software as open source. It provides a set of sequence-analysis programs, and also supports all UNIX platforms.

**Clustalw**
It is a fully automated sequence alignment tool for DNA and protein sequences. It returns the best match over a total length of input sequences, be it a protein or a nucleic acid.

**RasMol**
It is a powerful research tool to display the structure of DNA, proteins, and smaller molecules. Protein Explorer, a derivative of RasMol, is an easier to use program.

**PROSPECT**
PROSPECT (PROtein Structure Prediction and Evaluation Computer ToolKit) is a protein-structure prediction system that employs a computational technique called protein threading to construct a protein's 3-D model.

**PatternHunter**
PatternHunter, based on Java, can identify all approximate repeats in a complete genome in a short time using little memory on a desktop computer. Its features are its advanced patented algorithm and data structures, and the java language used to create it. The Java language version of PatternHunter is just 40 KB, only 1% the size of Blast, while offering a large portion of its functionality.

**COPIA**
COPIA (COnsensus Pattern Identification and Analysis) is a protein structure analysis tool for discovering motifs (conserved regions) in a family of protein sequences. Such motifs can be then used to determine membership to the family for new protein sequences, predict secondary and tertiary structure and function of proteins and study evolution history of the sequences.

## Application of Programmes in Bioinformatics:

### JAVA in Bioinformatics

Since research centers are scattered all around the globe ranging from private to academic settings, and a range of hardware and OSs are being used, Java is emerging as a key player in bioinformatics. Physiome Sciences' computer-based biological simulation technologies and Bioinformatics Solutions' PatternHunter are two examples of the growing adoption of Java in bioinformatics.

### Perl in Bioinformatics

String manipulation, regular expression matching, file parsing, data format interconversion etc are the common text-processing tasks performed in bioinformatics. Perl excels in such tasks and is being used by many developers. Yet, there are no standard modules designed in Perl specifically for the field of bioinformatics. However, developers have designed several of their own individual modules for the purpose, which have become quite popular and are coordinated by the BioPerl project.

## Bioinformatics Projects

**BioJava**
The BioJava Project is dedicated to providing Java tools for processing biological data which includes objects for manipulating sequences, dynamic programming, file parsers, simple statistical routines, etc.

**BioPerl**
The BioPerl project is an international association of developers of Perl tools for bioinformatics and provides an online resource for modules, scripts and web links for developers of Perl-based software.

**BioXML**
A part of the BioPerl project, this is a resource to gather XML documentation, DTDs and XML aware tools for biology in one location.

**Biocorba**
Interface objects have facilitated interoperability between bioperl and other perl packages such as Ensembl and the Annotation Workbench. However, interoperability between bioperl and packages written in other languages requires additional support software. CORBA is one such framework for interlanguage support, and the biocorba project is currently implementing a CORBA interface for bioperl. With biocorba, objects written within bioperl will be able to communicate with objects written in biopython and biojava (see the next subsection). For more information, see the biocorba project website at http://biocorba.org/. The Bioperl BioCORBA server and client bindings are available in the bioperl-corba-server and bioperl-corba-client bioperl CVS repositories respecitively. (see http://cvs.bioperl.org/ for more information).

**Ensembl**

Ensembl is an ambitious automated-genome-annotation project at EBI. Much of Ensembl\'s code is based on bioperl, and Ensembl developers, in turn, have contributed significant pieces of code to bioperl. In particular, the bioperl code for automated sequence annotation has been largely contributed by Ensembl developers. Describing Ensembl and its capabilities is far beyond the scope of this tutorial The interested reader is referred to the Ensembl website at http://www.ensembl.org/.

**bioperl-db**
Bioperl-db is a relatively new project intended to transfer some of Ensembl's capability of integrating bioperl syntax with a standalone Mysql database (http://www.mysql.com/) to the bioperl code-base. More details on bioperl-db can be found in the bioperl-db CVS directory at http://cvs.bioperl.org/cgi-bin/viewcvs/viewcvs.cgi/bioperl-db/?cvsroot=bioperl. It is worth mentioning that most of the bioperl objects mentioned above map directly to tables in the bioperl-db schema. Therefore object data such as sequences, their features, and annotations can be easily loaded into the databases, as in $loader->store($newid,$seqobj) Similarly one can query the database in a variety of ways and retrieve arrays of Seq objects. See biodatabases.pod, Bio::DB::SQL::SeqAdaptor, Bio::DB::SQL::QueryConstraint, and Bio::DB::SQL::BioQuery for examples.

### Biopython and biojava

Biopython and biojava are open source projects with very similar goals to bioperl. However their code is implemented in python and java, respectively. With the development of interface objects and biocorba, it is possible to write java or python objects which can be accessed by a bioperl script, or to call bioperl objects from java or python code. Since biopython and biojava are more recent projects than bioperl, most effort to date has been to port bioperl functionality to biopython and biojava rather than the other way around. However, in the future, some bioinformatics tasks may prove to be more effectively implemented in java or python in which case being able to call them from within bioperl will become more important. For more information, go to the biojava http://biojava.org/ and biopython http://biopython.org/ websites.

# Bioinformatics -Employment Opportunities

**Career Outlook**

Bioinformatics now looks like a hot cake. Prognosis. Good, very good in U.S.A alone. The number of jobs advertised for bioinformatics in Science magazine from 1996 to 1997 increased by 96% bioinformatics job advertisements. Masters students salaries range between $40,000 to $65,000 for persons with top masters training. In addition, Masters students reportedly have received hiring bonuses and relocation assistance. Of course you are ultimately responsible for landing a great job and without your effort nothing will be accomplished. Enormous volume of data has been generated and will be generated in the post genomic era through research including genomics, proteomics, structural biology and dynamics of macromolecular assemblies. All these information will be help better understanding of biology and of disease and an improved ability to design targeted drugs. But manipulation and interpretation of these data demand a lot of skilled personnel.

The need for bioinformatician to make use of these data is well known among the industrialists and academicians in the area. Demand of these trained and skilled personnel, which has opened up a new carrier option as bioinformaticians, is high in academic institution and in the bioindustries. This demand will clearly become exacerbated with advances in genomics and post genomic research. This skill gap is highly visible around the world.

*The carrier in bioinformatics is divided into two parts-developing software and using it.* Most of the bioinformatics companies mostly prefer persons from physics, maths or computer science background rather than biologists-turned programmers. Otherwise, it will create isolated groups that have little interaction with molecular biologists and biochemists which will ultimately fail to achieve promise of better understanding of biology and diseases can be restored.

The solution is to the problem of either by identifying science and biology, which will be difficult or to create interdisciplinary teams, which is much more realistic approach. The other way to tackle this imbalance is to include training in bioinformatics in all undergraduate biology courses.

People working in this field must have knowledge in more than one field-molecular biology, mathematics, statistics and computer science and some elements of machine learning.

**Graduate Employment Opportunities**

There is a growing need nationally and internationally for bioinformaticians, especially graduates with a good grounding in computer science and software engineering, and an appreciation of the biological aspects of the problems to be solved. The activities of such individuals will include working closely with bioinformaticians with a background in the biological and biophysical/biochemical science to:

- elucidate requirements
- develop new algorithms
- implement computer programs and tools for bio data analysis and display of results
- design databases for bio data
- participate in data analysis

Such individuals will gain employment in national and private research centres such as the European Bioinformatics Institue (EBI) at Hinxton Cambridge, and the European Media Lab, academic institutions (eg. the Laboratory for Molecular Biology (LMB) in Cambridge) as well as in pharmaceutical companies (both multinational, for example GlaxoWellcome, smithkline, Astro Zencia, Roche etc, and small but influential companies, for example InPharmatica).

Furthermore, there is a growing demand for graduates with software engineering and distributed systems skills to develop and maintain the powerful and sophisticated computer systems that support research, development and production activities in bioinformatics. There is a heavy use of internet and intranet technology within the sector in order to manage the storage, analysis and retrieval of biodata whose attributes are

- large volume
- rapidly increase in volume
- high complexity

**Career Outlook in India**

According to Confederation of Indian Industry (CII), the global bioinformatics industry clocked an estimated turnover of $2 billion in 2000 and is expected to become $60 billion by 2005. If the industry and government work together it is possible to achieve a five percent global market share by 2005, i.e., a $3 billion opportunity in India.

The past two years has seen many large multinational pharmaceutical companies acquiring other small companies and developing in the biosciences sector. IDC currently forecasts a compound annual growth rate (from 2001-02 to 2004-05) of about 10 percent in the spending on Information Technology by bioscience organizations. Considering the local market is generally less mature than those in the US and Europe, IDC forecasts more aggressive growth beyond 2005, as many of the organizations attempt to play "catch-up". Enterprise applications including data warehousing, knowledge management and storage are being pursued by these companies as priorities.

IDC expects IT spending in biosciences in India will cross $138 million by 2005, mainly in the areas of system clusters, storage, application software, and services. Also the government's life science focus provides a great deal of the necessary backbone to develop and deliver innovative products and technologies. This focus will also help to build fast-growing and lucrative enterprises, attract international investment, and create additional high-value employment opportunities.

Hence the focus of the IT sector should be on products and services that aligns with bioscience needs. Demonstrating a true understanding of the IT requirements of biotechnology processes is the key for IT suppliers to bridge the chasm that currently exists between IT and Science.

**Employment Opportunities in India**

There will be 10% annual growth in the Bioinformatics market for years to come; and the National Science Foundation estimated that 20,000 new jobs in the field would be created in the field in just the next four years.

The market for bioinformatics-based services and databases, experts say, could well surpass $2 billion in a couple of years, though a large part of the global bioinformatics market is centred on US.

Significantly, the growing interest in bioinformatics, industry watchers say, may even lead to an acute shortage of experts in this segment worldwide, which perhaps also explains the poaching that organisations like CCMB have to face. Eight of the dozen-and-odd scientists at the premier biotechnology research body have left for greener pastures forcing it to go on a recruitment drive.

Experts say that the manpower cost of a bioinformatics unit ranges in the region of 50%, while licensing third party databases may account for roughly 25-30% of total operating costs. With its strengths in mathematics, computing and physics and chemistry, the country is positioned ideally to emerge a frontrunner in bioinformatics. The need for new algorithms and new software tools places India in an advantageous position.

**The Future of Bioinformatics Professionals**

In USA, Europe or now in India also if you ask anyone about job prospects in the world of bioinformatics, young scientists will give the same answer. The field is hot!, It is far from being overpopulated and the number of jobs is growing further. Some of the biggest drug firms like SmithKline Beecham, Merck, Johnson & Johnson, and Glaxo Wellcome are hunting for bioinformatics experts while smaller firms have difficulties to get the staffers they want. While traditional scientific job markets are shrinking, here might be the opportunity many young scientists are looking for.

# Bioinformatics Applications

**Molecular medicine**
The human genome will have profound effects on the fields of biomedical research and clinical medicine. Every disease has a genetic component. This may be inherited (as is the case with an estimated 3000-4000 hereditary disease including Cystic Fibrosis and Huntingtons disease) or a result of the body's response to an environmental stress which causes alterations in the genome (e.g. cancers, heart disease, diabetes.). The completion of the human genome means that we can search for the genes directly associated with different diseases and begin to understand the molecular basis of these diseases more clearly. This new knowledge of the molecular mechanisms of disease will enable better treatments, cures and even preventative tests to be developed.

**Personalised medicine**
Clinical medicine will become more personalised with the development of the field of pharmacogenomics. This is the study of how an individual's genetic inheritance affects the body's response to drugs. At present, some drugs fail to make it to the market because a small percentage of the clinical patient population show adverse affects to a drug due to sequence variants in their DNA. As a result, potentially life saving drugs never makes it to the marketplace. Today, doctors have to use trial and error to find the best drug to treat a particular patient as those with the same clinical symptoms can show a wide range of responses to the same treatment. In the future, doctors will be able to analyse a patient's genetic profile and prescribe the best available drug therapy and dosage from the beginning.

**Preventative medicine**
With the specific details of the genetic mechanisms of diseases being unraveled, the development of diagnostic tests to measure a persons susceptibility to different diseases may become a distinct reality. Preventative actions such as change of lifestyle or having treatment at the earliest possible stages when they are more likely to be successful, could result in huge advances in our struggle to conquer disease.

**Gene therapy**
In the not too distant future, the potential for using genes themselves to treat disease may become a reality. Gene therapy is the approach used to treat, cure or even prevent disease by changing the expression of a person's genes. Currently, this field is in its infantile stage with clinical trials for many different types of cancer and other diseases ongoing.

**Drug development**
At present all drugs on the market target only about 500 proteins. With an improved understanding of disease mechanisms and using computational tools to identify and validate new drug targets, more specific medicines that act on the cause, not merely the symptoms, of the disease can be developed. These highly specific drugs promise to have fewer side effects than many of today's medicines.

**Microbial genome applications**
Microorganisms are ubiquitous, that is they are found everywhere. They have been found surviving and thriving in extremes of heat, cold, radiation, salt, acidity and pressure. They are present in the environment, our bodies, the air, food and water. Traditionally, use has been made of a variety of microbial properties in the baking, brewing and food industries. The arrival of the complete genome sequences and their potential to provide a greater insight into the microbial world and its capacities could have broad and far reaching implications for environment, health, energy and industrial applications. For these reasons, in 1994, the US Department of Energy (DOE) initiated the MGP (Microbial Genome Project) to sequence genomes of bacteria useful in energy production, environmental cleanup, industrial processing and toxic waste reduction. By studying the genetic material of these organisms, scientists can begin to understand these microbes at a very fundamental level and isolate the genes that give them their unique abilities to survive under extreme conditions.

**Waste cleanup**

Deinococcus radiodurans is known as the world's toughest bacteria and it is the most radiation resistant organism known. Scientists are interested in this organism because of its potential usefulness in cleaning up waste sites that contain radiation and toxic chemicals.

**Climate change Studies**

Increasing levels of carbon dioxide emission, mainly through the expanding use of fossil fuels for energy, are thought to contribute to global climate change. Recently, the DOE (Department of Energy, USA) launched a program to decrease atmospheric carbon dioxide levels. One method of doing so is to study the genomes of microbes that use carbon dioxide as their sole carbon source.

**Alternative energy sources**

Scientists are studying the genome of the microbe Chlorobium tepidum which has an unusual capacity for generating energy from light

**Biotechnology**

The archaeon Archaeoglobus fulgidus and the bacterium Thermotoga maritima have potential for practical applications in industry and government-funded environmental remediation. These microorganisms thrive in water temperatures above the boiling point and therefore may provide the DOE, the Department of Defence, and private companies with heat-stable enzymes suitable for use in industrial processes.

Other industrially useful microbes include, Corynebacterium glutamicum which is of high industrial interest as a research object because it is used by the chemical industry for the biotechnological production of the amino acid lysine. The substance is employed as a source of protein in animal nutrition. Lysine is one of the essential amino acids in animal nutrition. Biotechnologically produced lysine is added to feed concentrates as a source of protein, and is an alternative to soybeans or meat and bonemeal.

Xanthomonas campestris pv. is grown commercially to produce the exopolysaccharide xanthan gum, which is used as a viscosifying and stabilising agent in many industries. Lactococcus lactis is one of the most important micro-organisms involved in the dairy industry, it is a non-pathogenic rod-shaped bacterium that is critical for manufacturing dairy products like buttermilk, yogurt and cheese. This bacterium, Lactococcus lactis ssp., is also used to prepare pickled vegetables, beer, wine, some bread and sausages and other fermented foods. Researchers anticipate that understanding the physiology and genetic make-up of this bacterium will prove invaluable for food manufacturers as well as the pharmaceutical industry, which is exploring the capacity of L. lactis to serve as a vehicle for delivering drugs.

**Antibiotic resistance**
Scientists have been examining the genome of Enterococcus faecalis-a leading cause of bacterial infection among hospital patients. They have discovered a virulence region made up of a number of antibiotic-resistant genes that may contribute to the bacterium's transformation from a harmless gut bacteria to a menacing invader. The discovery of the region, known as a pathogenicity island, could provide useful markers for detecting pathogenic strains and help to establish controls to prevent the spread of infection in wards.

**Forensic analysis of microbes**
Scientists used their genomic tools to help distinguish between the strain of Bacillus anthryacis that was used in the summer of 2001 terrorist attack in Florida with that of closely related anthrax strains

**The reality of bioweapon creation**
Scientists have recently built the virus poliomyelitis using entirely artificial means. They did this using genomic data available on the Internet and materials from a mail-order chemical supply. The research was financed by the US Department of Defense as part of a biowarfare response program to prove to the world the reality of bioweapons. The researchers also hope their work will discourage officials from ever relaxing programs of immunisation. This project has been met with very mixed feelings

**Evolutionary studies**
The sequencing of genomes from all three domains of life, eukaryota, bacteria and archaea means that evolutionary studies can be performed in a quest to determine the tree of life and the last universal common ancestor.

**Crop improvement**
Comparative genetics of the plant genomes has shown that the organisation of their genes has remained more conserved over evolutionary time than was previously believed. These findings suggest that information obtained from the model crop systems can be used to suggest improvements to other food crops. At present the complete genomes of Arabidopsis thaliana (water cress) and Oryza sativa (rice) are available.

**Insect resistance**
Genes from Bacillus thuringiensis that can control a number of serious pests have been successfully transferred to cotton, maize and potatoes. This new ability of the plants to resist insect attack means that the amount of insecticides being used can be reduced and hence the nutritional quality of the crops is increased.

**Improve nutritional quality**
Scientists have recently succeeded in transferring genes into rice to increase levels of Vitamin A, iron and other micronutrients. This work could have a profound impact in reducing occurrences of blindness and anaemia caused by deficiencies in Vitamin A and iron respectively. Scientists have inserted a gene from yeast into the tomato, and the result is a plant whose fruit stays longer on the vine and has an extended shelf life

**Development of Drought resistance varieties**
Progress has been made in developing cereal varieties that have a greater tolerance for soil alkalinity, free aluminium and iron toxicities. These varieties will allow agriculture to succeed in poorer soil areas, thus adding more land to the global production base. Research is also in progress to produce crop varieties capable of tolerating reduced water conditions

**Vetinary Science**

Sequencing projects of many farm animals including cows, pigs and sheep are now well under way in the hope that a better understanding of the biology of these organisms will have huge impacts for improving the production and health of livestock and ultimately have benefits for human nutrition.

**Comparative Studies**

Analysing and comparing the genetic material of different species is an important method for studying the functions of genes, the mechanisms of inherited diseases and species evolution. Bioinformatics tools can be used to make comparisons between the numbers, locations and biochemical functions of genes in different organisms.

Organisms that are suitable for use in experimental research are termed model organisms. They have a number of properties that make them ideal for research purposes including short life spans, rapid reproduction, being easy to handle, inexpensive and they can be manipulated at the genetic level.

An example of a human model organism is the mouse. Mouse and human are very closely related (>98%) and for the most part we see a one to one correspondence between genes in the two species. Manipulation of the mouse at the molecular level and genome comparisons between the two species can and is revealing detailed information on the functions of human genes, the evolutionary relationship between the two species and the molecular mechanisms of many human diseases.

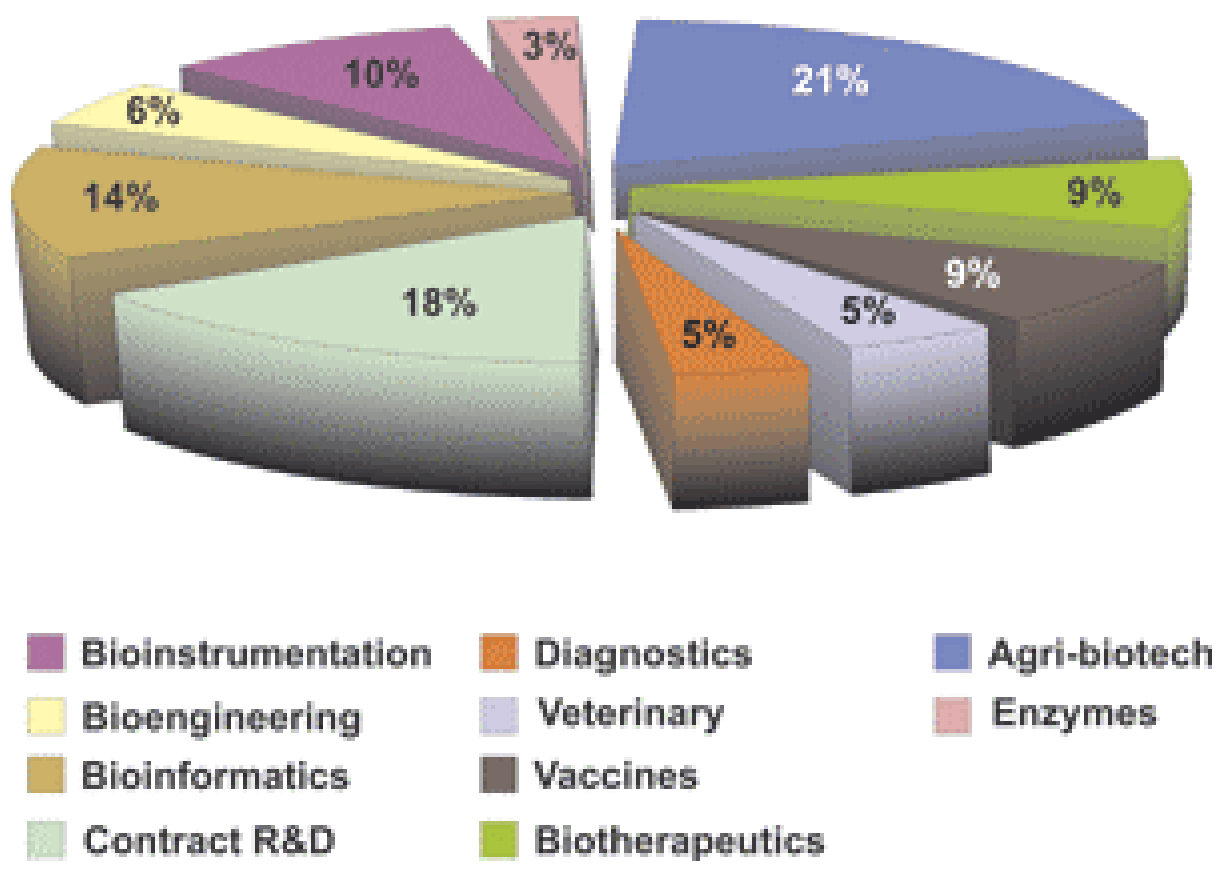

Fig. Applications Areas

# Bioinformatics in India

Studies of IDC points out that India will be a potential star in bioscience field in the coming years after considering the factors like bio-diversity, human resources, infrastructure facilities and government's initiatives. According to IDC, bioscience includes pharma, Bio-IT (bioinformatics), agriculture and R&D. IDC has been reported that the pharmaceutical firms and research institutes in India are looking forward for cost-effective and high-quality research, development, and manufacturing of drugs with more speed.

Bioinformatics has emerged out of the inputs from several different areas such as biology, biochemistry, biophysics, molecular biology, biostatics, and computer science. Specially designed algorithms and organized databases is the core of all informatics operations. The requirements for such an activity make heavy and high level demands on both the hardware and software capabilities.

This sector is the quickest growing field in the country. The vertical growth is because of the linkages between IT and biotechnology, spurred by the human genome project. The promising start-ups are already there in Bangalore, Hyderabad, Pune, Chennai, and Delhi. There are over 200 companies functioning in these places. IT majors such as Intel, IBM, Wipro are getting into this segment spurred by the promises in technological developments.

**Government initiatives**

Informatics is a very essential component in the biotech revolution. Ranging from reference to type-culture collections or comparing gene sequences access to comprehensive up-to-date biological information, all are crucial in almost every aspects of biotechnology. India, as a hub of scientific and academic research, was one of the first countries in the world to establish a nation wide bioinformatics network.

The department of biotechnology (DBT) initiated the program on bioinformatics way back in 1986-87. The Biotechnology Information System Network (BTIS), a division of DBT, has covered the entire country by connecting to the 57 key research centers. BTIS is providing an easy access to huge database to the scientists. Six national facilities on interactive graphics are dedicated to molecular modeling and other related areas. More than 100 databases on biotechnology have been developed. Two major databases namely coconut biotechnology databases and complete genome of white spota syndrome of shrimp has been released for public use. Several major international data bases for application to genomics and proteomics have been established in the form of mirror sites under the National Jai Vigyan Mission.

The BTIS proposes to increase the bandwidth of existing network and provide high-speed internet connectivity to continue with its present activities of training, education mirroring of public utility packages, consideration of R&D projects and support to different research activities in this field. The DBT is planning to set up a National Bioinformatics Institute as an apex body for the bioinformatics network in the country. The DBT also proposes to bolster a proper education in bioinformatics through publication of textbooks and monographs by reputed scientists in the field. Collaboration with the industry is also poised to increase in the coming years.

**Opportunities**
According to Confederation of Indian Industry(CII), the global bioinformatics industry clocked an estimated turnover of $2 billion in 2000 and is expected to become $60 billion by 2005. If the industry and government work together it is possible to achieve a five percent global market share by 2005, i.e., a $3 billion opportunity in India.

The past two years has seen many large multinational pharmaceutical companies acquiring other small companies and developing in the biosciences sector. IDC currently forecasts a compound annual growth rate (from 2001-02 to 2004-05) of about 10 percent in the spending on Information Technology by bioscience organizations. Considering the local market is generally less mature than those in the US and Europe, IDC forecasts more aggressive growth beyond 2005, as many of the organizations attempt to play "catch-up". Enterprise applications including data warehousing, knowledge management and storage are being pursued by these companies as priorities.

IDC expects IT spending in biosciences in India will cross $138 million by 2005, mainly in the areas of system clusters, storage, application software, and services. Also the governments' life science focus provides a great deal of the necessary backbone to develop and deliver innovative products and technologies. This focus will also help to build fast-growing and lucrative enterprises, attract international investment, and create additional high-value employment opportunities.

Hence the focus of the IT sector should be on products and services that align with bioscience needs. Demonstrating a true understanding of the IT requirements of biotechnology processes is the key for IT suppliers to bridge the chasm that currently exists between IT and Science.

**Advantages India has**
India is well placed to take the global leadership in genome analysis, as is in a unique position in terms of genetic resources. India has several ethnic populations that are valuable in providing information about disease predisposition and susceptibility, which in turn will help in drug discovery.

However, as India lacks the records of clinical information about the patients, sequence data without clinical information will have little meaning. And hence partnership with clinicians is essential. The real money is in discovering new drugs for ourselves and not in supplying genetic information and data to the foreign companies, who would then use this information to discover new molecules.

The genomic data provides information about the sequence, but it doesn't give information about the function. It is still not possible to predict the actual 3-D structure of proteins. This is a key area of work as tools to predict correct folding patterns of proteins will help drug design research substantially. India has the potential to lead if it invests in this area.

Looking at this biotech and pharma companies need tremendous software support. Software expertise is required to write algorithms, develop software for existing algorithms, manage databases, and in final process of drug discovery.

Some major opportunity areas for IT companies include:

- Improving content and utility of databases
- Developing better tools for data generation, capture, and annotation
- Developing and improving tools and databases for comprehensive functional studies
- Developing and improving tools for representing and analyzing sequence similarity and variation
- Creating mechanisms to support effective approaches for producing robust, software that can be widely shared.

As major pharmaceutical and genome-based biotech companies invest heavily in software, Indian IT companies have a great business opportunity to offer complete database solutions to major pharmaceutical and genome-based biotech companies in the world.

Pure cost benefits for the biotech companies will definitely drive the bioinformatics industry in the country. The biotech industry in 2000 has spent an estimated 36 percent on R & D. Success for many will mean a drastic reduction in R&D costs. Thus biotech companies will be forced to outsource software rather than developing propriety software like in the past. Since the cost of programs for handling this data is extremely high in the west, Indian IT companies have a great business opportunity to offer complete database solutions to major pharmaceutical and genome-based biotech companies in the world.

The IT industry can also focus more on genomics through different levels of participation areas such as hardware, database product and packages, implementation and customization of software, and functionality enhancement of database.

Abraham Thomas, managing director, IBM India Ltd, says, "the alignment of a vast pool of scientific talent, a world-class IT industry, a vigorous generic pharmaceutical sector and government initiatives in establishment of public sector infrastructure and research labs are positioning India to emerge as a significant participant on the global biotech map."

With an objective to help and rise bioinformatics sector to the world map the Bioinformatics Society of India (Inbios) has been working since August 2001. The Inbios already has over 270 members in a short span of one and half years. It has become a common informal platform for the younger generation to learn and contribute to this sun rising field in India.

**Problems in the sector**

The major issue for India is its transition from a recognized global leader in software development to areas of real strength upon which it can capitalize in the biosciences. The identifiable areas are in computation biology and bioinformatics, where a substantial level of development skills are required to develop custom applications to knot together and integrate disparate databases (usually from several global locations), simulations, molecular images, docking programs etc.

The industry people, meanwhile, say that the mushrooming of bioinformatic institutes is creating a problem of finding talented and trained individuals in this industry. While many of them have a superficial knowledge and a certificate, India lacks true professionals in this area.

Most people, who opt for bioinformatics are from the life sciences areas that do not have exposure to the IT side of bioinformatics, which is very important. Another issue is that some companies face shortage of funds and infrastructure. The turn around time for an average biotech industry to breakeven would be around three to five years.

Most of the venture capitals and other sources of funding would not be very supportive, especially if the company is not part of a larger group venture. It would help if the government would take an active role in building infrastructure and funding small and medium entrepreneurs.

**Biotechnology Information Systems (BTIS)**

The objective of setting up BTIs

- To provide a national bioinformation network designed to bridge the inter-disciplinary gaps in biotechnology information and to establish link among scientists in organisations involved in R&D and manufacturing activities in biotechnology.
- To build up information resources, prepares databases on biotechnology and to develop relevant information handling tools and techniques.
- To continuously access information requirements, organize creation of necessary infrastructure and to provide information and computer support services to the national community of users.
- To establish linkages with international resources in Biotechnology information (eg. Databanks on genetic materials, published literature, patents and other information of scientific and commercial value).
- To evolve and implement programmes on education of users and training of information scientists responsible for handling of biotechnology information and its applications to biotechnology research and development.
- To develop, support and enhance public information resources for biotechnology in India, eg.Genbanks, molecular biology data and related research information resources.
- To undertake preparing and publishing surveys, state of the art reports and forecasts for several branches of the sector.
- To provide a national backbone for exchange and analysis of information covering aspects of mult-disciplinary Biotechnology amongst the national community comprising scientists, research scholars, students, industry and planning personnel.

# Proteomics-Introduction

**Definition:** "The analysis of complete complements of proteins. Proteomics includes not only the identification and quantification of proteins, but also the determination of their localization, modifications, interactions, activities, and, ultimately, their function. Initially encompassing just two- dimensional (2D) gel electrophoresis for protein separation and identification, proteomics now refers to any procedure that characterizes large sets of proteins. The explosive growth of this field is driven by multiple forces - genomics and its revelation of more and more new proteins; powerful protein technologies such as newly developed mass spectrometry approaches, global [yeast] two - hybrid techniques, and spin-offs from DNA arrays; and innovative computational tools and methods to process, analyze, and interpret prodigious amounts of data."

The theme of molecular biology research, in the past, has been oriented around the gene rather than the protein. This is not to say that researchers have neglected to study proteins, but rather that the approaches and techniques most commonly used have looked primarily at the nucleic acids and then later at the protein(s) implicated.

The main reason for this has been that the technologies available, and the inherent characteristics of nucleic acids, have made the genes the low hanging fruit. This situation has changed recently and continues to change as larger scale, higher throughput methods are developed for both nucleic acids and proteins. The majority of processes that take place in a cell are not performed by the genes themselves, but rather by the proteins that they code for.

A disease can arise when a gene/protein is over- or under-expressed, or when a mutation in a gene results in a malformed protein, or when post translational modifications alter a protein's function. Thus to truly understand a biological process, the relevant proteins must be studied directly. But there are more challenges when studying proteins compared to studying genes, due to their complex 3-D structure which is related to the function, analogous to a machine.

***Proteomics*** *is defined as the systematic large-scale analysis of protein expression under normal and perturbed (stressed, diseased, and/or drugged) states, and generally involves the separation, identification, and characterization of all of the proteins in a cell or tissue sample.* The meaning of the term has also been expanded, and is now used loosely to refer to the approach of analyzing which proteins a particular type of cell synthesizes, how much the cell synthesizes, how cells modify proteins after synthesis, and how all of those proteins interact.

There are orders of magnitude more proteins than genes in an organism - based on alternative splicing (several per gene) and post translational modifications (over 100 known), there are estimated to be a million or more.

Fortunately there are features such as folds and motifs, which allow them to be categorized into groups and families, making the task of studying them more tractable. There is a broad range of technologies used in proteomics, but the central paradigm has been the use of 2-D gel electrophoresis (2D-GE) followed by mass spectrometry (MS). 2D-GE is used to first separate the proteins by isoelectric point and then by size.

The individual proteins are subsequently removed from the gel and prepared, then analyzed by MS to determine their identity and characteristics. There are various types of mass analyzers used in proteomics MS including quadrupole, time-of-flight (TOF), and ion trap, and each has its own particular capabilities. Tandem arrangements are often used, such as quadrupole-TOF, to provide more analytical power. The recent development of soft ionization techniques, namely matrix-assisted laser desorption ionization (MALDI) and electro-spray ionization (ESI), has allowed large biomolecules to be introduced into the mass analyzer without completely decomposing their structures, or even without breaking them at all, depending on the design of the experiment.

There are techniques which incorporate liquid chromatography (LC) with MS, and others that use LC by itself. Robotics has been applied to automate several steps in the 2DGE-MS process such as spot excision and enzyme digests. To determine a protein's structure, XRD and NMR techniques are being improved to reach higher throughput and better performance.

For example, automated high-throughput crystallization methods are being used upstream of XRD to alleviate that bottleneck. For NMR, cryo-probes and flow probes shorten analysis time and decrease sample volume requirements. The hope is that determining about 10,000 protein structures will be enough to characterize the estimated 5,000 or so folds, which will feed into more reliable in silico structural prediction methods.

Structure by itself does not provide all of the desired information, but is a major step in the right direction. Protein chips are being developed for many of the processes in proteomics. For example, researchers are developing protocols for protein microarrays at institutions such as Harvard and Stanford as well as at several companies. These chips - grids of attached peptide fragments, attached antibodies, or gel "pads" with proteins suspended inside - will be used for various experiments such as protein-protein interaction studies and differential expression analysis.

They can also be used to filter out high abundance proteins before further experiments; one of the major challenges in proteomics is isolating and analyzing the low abundance proteins, which are thought to be the most important. There are many other types of protein chips, and the number will continue to grow. For example, microfluidics chips can combine the sample preparation steps prior to MS, such as enzyme digests, with nanoelectrospray ionization, all on the one chip. Or, the samples can be ionized directly off of the surface of the chip, similar to a MALDI target. Microfluidics chips are also being combined with NMR.

In the next few years, various protein chips will be used increasingly in diagnostic applications as well. The bioinformatics side of proteomics includes both databases and analysis software. There are many public and private databases containing protein data ranging from sequences, to functions, to post translational modifications. Typically, a researcher will first perform 2D-GE followed by MS; this will result in a fingerprint, molecular weight, or even sequence for each protein of interest, which can then be used to query databases for similarities or other information.

Swiss-Prot and TrEMBL, developed in collaboration between the Swiss Institute of Bioinformatics and the European Bioinformatics Institute, are currently the major databases dedicated to cataloging protein data, but there are dozens of more specialized databases and tools. New bioinformatics approaches are constantly being introduced. Recent customized versions of PSI-BLAST can, for example, utilize not only the curated protein entries in Swiss-Prot but also linguistic analyses of biomedical journal articles to help determine protein family relationships. Publicly available databases and tools are popular, but there are also several companies offering subscriptions to proprietary databases, which often include protein-protein interaction maps generated using the yeast two-hybrid (Y2H) system.

The proteomics market is comprised of instrument manufacturers, bioinformatics companies, laboratory product suppliers, service providers, and other biotech related companies which can defy categorization. A given company can often overlap more than one of these areas. Many of the companies involved in the proteomics market are actually doing drug discovery as their major focus, while partnering, or providing services or subscriptions, to other companies to generate short term revenues. The market for proteomics products and services was estimated to be $1.0B in 2000, growing at a CAGR of 42% to about $5.8B in 2005.

The major drivers will continue to be the biopharmaceutical industry's pursuit of blockbuster drugs and the recent technological advances, which have allowed large-scale studies of genes and proteins. Alliances are becoming increasingly important in this field, because it is challenging for companies to find all of the necessary expertise to cover the different activities involved in proteomics. Synergies must be created by combining forces. For example, many companies working with mass spectrometry, both the manufacturers and end user labs, are collaborating with protein chip related companies. The technologies are a natural fit for many applications, such as microfluidic chips, which provide nanoelectrospray ionization into a mass spectrometer.

There are many combinations of diagnostics, instrumentation, chip, and bioinformatics companies which create effective partnerships. In general, proteomics appears to hold great promise in the pursuit of biological knowledge. There has been a general realization that the large-scale approach to biology, as opposed to the strictly hypothesis-driven approach, will rapidly generate much more useful information.

The two approaches are not mutually exclusive, and the happy medium seems to be the formation of broad hypotheses, which are subsequently investigated by designing large-scale experiments and selecting the appropriate data. Proteomics and genomics, and other varieties of 'omics', will all continue to complement each other in providing the tools and information for this type of research.

# Microarrays

**Microarray:** A 2D array, typically on a glass, filter, or silicon wafer, upon which genes or gene fragments are deposited or synthesized in a predetermined spatial order allowing them to be made available as probes in a high-throughput, parallel manner.

Microarrays that consist of ordered sets of DNA fixed to solid surfaces provide pharmaceutical firms with a means to identify drug targets. In the future, the emerging technology promises to help physicians decide the most effective drug treatments for individual patients.

*Microarrays are simply ordered sets of DNA molecules of known sequence.* Usually rectangular, they can consist of a few hundred to hundreds of thousands of sets. Each individual feature goes on the array at precisely defined location on the substrate. The identity of the DNA molecule fixed to each feature never changes. Scientists use that fact in calculating their experimental results.

Microarray analysis permits scientists to detect thousands of genes in a small sample simultaneously and to analyze the expression of those genes. As a result, it promises to enable biotechnology and pharmaceutical companies to identify drug targets - the proteins with which drugs actually interact. Since it can also help identify individuals with similar biological patterns, microarray analysis can assist drug companies in choosing the most appropriate candidates for participating in clinical trials of new drugs. In the future, this emerging technology has the potential to help medical professionals select the most effective drugs, or those with the fewest side effects, for individual patients.

**Potential of Microarray analysis**

The academic research community stands to benefit from microarray technology just as much as the pharmaceutical industry. The ability to use it in place of existing technology will allow researchers to perform experiments faster and more cheaply, and will enable them to concentrate on analyzing the results of microarray experiments rather than simply performing the experiments. This research could then lead to a better understanding of the disease process. That will require many different levels of research. While the field of expression has received most attention so far, looking at the gene copy level and protein level is just as important. Microarray technology has potential applications in each of these three levels.

Identifying drug targets provided the initial market for the microarrays. A good drug target has extraordinary value for developing pharmaceuticals. By comparing the ways in which genes are expressed in a normal and diseased heart, for example, scientists might be able to identify the genes and hence the associated proteins -- that are part of the disease process. Researchers could then use that information to synthesize drugs that interact with these proteins, thus reducing the disease's effect on the body.

Gene sequences can be measured simultaneously and calculated instantly when an ordered set of DNA molecules of known sequence a microarray is used. Consequently, scientists can evaluate an entire set of genes at once, rather than looking at physiological changes one gene at a time. For example, Genetics Institute, a biotechnology company in Cambridge, Massachusetts, built an array consisting of genes for cytokines, which are proteins that affect cell physiology during the inflammatory response, among other effects. The full set of DNA molecules contained more than 250 genes. While that number was not large by current standards of microarrays, it vastly outnumbered the one or two genes examined in typical pre-microarray experiments. The Genetics Institute scientists used the array to study how changes experienced by cells in the immune system during the inflammatory response are reflected in the behavior of all 250 genes at the same time. This experiment established the potential for using the patterns of response to help locate points in the body at which drugs could prove most effective.

**Microarray Products**

Within that basic technological foundation, microarray companies have created a variety of products and services. They range in price, and involve several different technical approaches. A kit containing a simple array with limited density can cost as little as $1,100, while a versatile system favored by R&D laboratories in pharmaceutical and biotechnology companies costs more than $200,000. The differences among products lie in the basic components and the precise nature of the DNA on the arrays.

The type of molecule placed on the array units also varies according to circumstances. The most commonly used molecule is cDNA, or complementary DNA, which is derived from messenger RNA and cloned. Since they are derived from a distinct messenger RNA, each feature represents an expressed gene.

**Microarray-Identifying interactions**

To detect interactions at microarray features, scientists must label the test sample in such a way that an appropriate instrument can recognize it. Since the minute size of microarray features limits the amount of material that can be located at any feature, detection methods must be extremely sensitive.

Other than a few low-end systems that use radioactive or chemiluminescent tagging, most microarrays use fluorescent tags as their means of identification. These labels can be delivered to the DNA units in several different ways. One simple and flexible approach involves attaching a fluorophore such as fluorescein or Cy3 to the oligonucleotide layer. While relatively simple, this approach has low sensitivity because it delivers only one unit of label per interaction. Technologists can achieve more sensitivity by multiplexing the labeled entity -- that is, delivering more than one unit of label per interaction.

# Bioinformatics career guidance

- What is Bioinformatics & why it is needed?

Biotech research generates such massive volumes of information, so quickly that we need newer and swifter ways of crunching the deluge of data (much of it hopelessly jumbled) churned out by the research labs. Mining, digitising and indexing this enormous quantum of data requires high-end computational technology and sophisticated software solutions.

- Is Bioinformatics industry is really potential?

Representing a marriage of IT and biotechnology, bioinformatics is poised to be one of the most prodigious growth areas in the next two decades. Currently valued at around $ 1 billion, the industry is expected to grow exponentially over the next 10 years stretching the very boundaries of biotechnology —transforming an essentially lab-based science into an information science yielding rare biological insights and hastening the discovery of new drugs. Globally, the biotech computing sector is estimated to touch a whopping $30 billion by 2003 and $ 60 billion in 2005. This in turn will create a corresponding boom in job opportunities.

- What type of people did bioinformatics require?

Companies need cross-functional manpower at all levels — biologists with IT skills, or IT professionals with a serious interest in biology (just one of the skills is not enough) who can offer complete cost-effective database solutions to pharma and genome-based biotech companies all over the world.

- What type of work available in Bioinformatics?

There is no such thing as a typical career path in this field. Bioinformaticians need to perform two critical roles: develop IT tools embodying novel algorithms and analytical techniques, and apply existing tools to achieve new insights into molecular biology. However, you must remember that although powerful and highly specialised in itself, bioinformatics is only a part of biotechnology. Getting a DNA sequence coding for a new protein does not automatically make it useful. Unless this information is converted into useful processes and products, it serves no purpose. You can not, for instance, have a virtual drug or a virtual vaccine. We need real products. And we need to develop our own new molecules (particularly if we have to survive in the new IPR regime).

- What type of career opportunities available in Bioinformatics?

There are different types of career opportunities available for different stream students,

**Life Sciences:**
Scientific Curator, Gene Analyst, Protein Analyst, Phylogenitist, Research Scientist / Associate.
**Computer Science / Engineering:**
Data base programmer, Bio-informatics software developer, Computational biologist, Network Administrator / Analyst.

**Applied Science:**
Structural analyst, Molecular Modeler, Bio-statistician, Bio-mechanics, Database programmer.
**Pharmaceutical Science:**
Cheminformatician, Pharmacogenetician, Pharmacogenomics, Research Scientist / Associate.

- What skills should a Bioinformatician have?

According to the scientist working at companies such as Celera Genomics and Eli Lilly, the following "core requirements" for bioinformaticians:

>> Fairly deep background in some aspect of molecular biology. It can be biochemistry, molecular biology, molecular biophysics, or even molecular modeling, but without a core of knowledge of molecular biology is like, "run into brick walls too often."

>>Understanding the central dogma of molecular biology, how and why DNA sequence is transcribed into RNA and translated into protein is vital.

>>Should have substantial experience with at least one or two major molecular biology software packages, either for sequence analysis or molecular modeling. The experience of learning one of these packages makes it much easier to learn to use other software quickly.

>>Should be comfortable working in a command-line computing environment. Working in Linux or UNIX will provide this experience.

>>Should have experience with programming in a computer language such as Java, Unix, C, C++, RDBMS such as Oracle and Sybase, CORBA, Perl or Python, CGI and web scripting.

***Source:****Extracted from the book "Developing Bioinformatics Computer Skills" by Cynthia Gibbs & Per Jambeck, O'Reilly & Associates, Inc*

- **How much salary did Bioinformatician get?**

**India:**

Starting with a package of Rs 12,000 to Rs 15,000, you can expect Rs 20,000 with a couple of years of experience under your belt. In fact, the acute shortage of experts in the field has given rise to active poaching of scientists from premier research institutions. The going price for bioinformaticians with a year's experience is upwards of Rs 50,000 per month.

**Abroad Countries:**
Starting salaries in the USA range between $60,000 and $ 90,000 for professionals with a couple of years of experience.

| Average salaries in biotech and pharmaceutical companies are as follows: | | |
|---|---|---|
| **Clinical Research**<br><br>* Associate: $51,500<br>*Senior associate: $65,000<br>*Manager: $85,000<br>*Clinical research physician: $90,000200,000<br>*Senior laboratory technician: $34,000<br>*Junior laboratory technician: $21,715 | **Biostatistics**<br><br>*MS entry-level: $74,500<br>*PhD entry-level: $110,000 | **Regulatory Affairs**<br><br>*Associate: $52,000<br>*Senior associate: $76,000 |
| **Quality Assurance**<br><br>* Specialist: $54,500<br>*Engineer: $58,000<br>*President/General manager: $94,000$400,000 | In 1999, average earnings of scientists employed by the federal government were:<br><br>*General biologists: $56,000<br>*Microbiologists: $62,600<br>*Physiologists: $71,300<br>*Geneticists: $68,200 | Average salary offers in 1999 for those with degrees in biological science were:<br><br>*BS: $29,000<br>*MS: $34,450<br>*PhD: $45,700 |
| Median earnings in industries employing the greatest number of biological and medical scientists in 1997 were:<br><br>*Federal government: $48,600<br>*Pharmaceuticals: $46,300<br>*Research and testing services: $40,800<br>*State government: $38,000 | | |
| Sources: Stax Research (1999), Abbott, Langer & Associates (1999), and U.S. Bureau of Labor Statistics | | |

- What are the real industry requirements?

There are distinct categories of professionals that the industry needs:

1. Computer Programmers, Mathematicians and people trained in Physics, Statistics etc. who develop software tools and applications for biotechnology and life science companies. They are cross trained in life sciences, such as molecular biology, DNA sequence analysis and in addition that they would need skills in writing algorithms and codes for developing such programs. A very specific training is required for such professionals to meet the need of life sciences companies.

2. People with a background in life sciences who are the end users of such programs and packages and they use these tools to translate the information into tangible products such as new molecules, drugs, enzymes etc. They can conduct their R&D program more effectively if they are cross-trained in computing skills. They can also be Business Analysts for life science companies.

- Is it easier to move from biology to computers or the reverse?

The answer depends on whether you are talking to a computer scientist who ‘does’ biology or a molecular biologist who ‘does’ computing. Most of what you will read in the popular press is that the importance of interdisciplinary scientists cannot be over-stressed and that the young people getting the top jobs in the next few years will be those graduating from truly interdisciplinary programs.

However, there are many types of bioinformatics jobs available, so no one background is ideal for all of them. The fact is that many of the jobs available currently involve the design and implementation of programs and systems for the storage, management and analysis of vast amounts of DNA sequence data. Such positions require in-depth programming and relational database skills which very few biologists possess and so it is largely the computational specialists who are filling these roles.

This is not to say the computer-savvy biologist doesn’t play an important role. As the bioinformatics field matures there will be a huge demand for outreach to the biological community as well as the need for individuals with the in-depth biological background necessary to sift through gigabases of genomic sequence in search of novel targets. It will be in these areas that biologists with the necessary computational skills will find their niche.

- How to become a bioinformatics expert?

Bioinformatics combines the tools and techniques of mathematics, computer science and biology in order to understand the biological significance of a variety of data. So if you like to get into this new scientific field you should be fond of these ‘classic’ disciplines. Because the field is so new, almost everyone in it did something else before. Some biologist went into bioinformatics by picking up programming but others entered via the reverse route.

# History of Bioinformatics

Bioinformatics is the application of computer technology to the management of biological information. Computers are used to gather, store, analyze and integrate biological and genetic information which can then be applied to gene-based drug discovery and development. The need for Bioinformatics capabilities has been precipitated by the explosion of publicly available genomic information resulting from the Human Genome Project. The goal of this project - determination of the sequence of the entire human genome (approximately three billion base pairs) - will be reached by the year 2002. The science of Bioinformatics, which is the melding of molecular biology with computer science, is essential to the use of genomic information in understanding human diseases and in the identification of new molecular targets for drug discovery. In recognition of this, many universities, government institutions and pharmaceutical firms have formed bioinformatics groups, consisting of computational biologists and bioinformatics computer scientists. Such groups will be key to unraveling the mass of information generated by large scale sequencing efforts underway in laboratories around the world.

The Modern bioinformatics is can be classified into two broad categories, Biological Science and computational Science. Here is the data of historical events for both biology and computer science.

The history of biology in general, B.C. and before the discovery of genetic inheritance by G. Mendel in 1865, is extremely sketch and inaccurate. This was the start of Bioinformatics history. **Gregor Mendel.** is known as the "Father of Genetics". He did experiment on the cross-fertilization of different colors of the same species. He carefully recorded the data and analyzed the data. Mendel illustrated that the inheritance of traits could be more easily explained if it was controlled by factors passed down from generation to generation.

The understanding of genetics has advanced remarkably in the last thirty years. In 1972, Paul berg made the first recombinant DNA molecule using ligase. In that same year, Stanley Cohen, Annie Chang and Herbert Boyer produced the first recombinant DNA organism. In 1973, two important things happened in the field of genomics. The advancement of computing in 1960-70s resulted in the basic methodology of bioinformatics. However, it is the 1990s when the INTERNET arrived when the full fledged bioinformatics field was born.

Here are some of the major events in bioinformatics over the last several decades. The events listed in the list occurred long before the term, "bioinformatics", was coined.

| BioInformatics Events | |
|---|---|
| 1665 | Robert Hooke published Micrographia, described the cellular structure of cork. He also described microscopic examinations of fossilized plants and animals, comparing their microscopic structure to that of the living organisms they resembled. He argued for an organic origin of fossils, and suggested a plausible mechanism for their formation. |
| 1683 | Antoni van Leeuwenhoek discovered bacteria. |
| 1686 | John Ray, John Ray's in his book "Historia Plantarum" catalogued and described 18,600 kinds of plants. His book gave the first definition of species based upon common descent. |
| 1843 | Richard Owen elaborated the distinction of **homology** and **analogy.** |
| 1864 | Ernst Haeckel (Häckel) outlined the essential elements of modern zoological classification. |
| 1865 | Gregory Mendel (1823-1884), Austria, established the theory of genetic inheritance. |
| 1902 | The chromosome theory of heredity is proposed by Sutton and Boveri, working independently. |
| 1962 | Pauling's theory of molecular evolution |
| 1905 | The word "genetics" is coined by William Bateson. |
| 1913 | First ever linkage map created by Columbia undergraduate Alfred Sturtevant (working with T.H. Morgan). |
| 1930 | Tiselius, Uppsala University, Sweden, A new technique, electrophoresis, is introduced by Tiselius for separating proteins in solution. "The moving-boundary method of studying the electrophoresis of proteins" (published in *Nova Acta Regiae Societatis Scientiarum Upsaliensis*, Ser. IV, Vol. 7, No. 4) |
| 1946 | Genetic material can be transferred laterally between bacterial cells, as shown by Lederberg and Tatum. |
| 1952 | Alfred Day Hershey and Martha Chase proved that the DNA alone carries genetic information. This was proved on the basis of their bacteriophage research. |
| 1961 | Sidney Brenner, François Jacob, Matthew Meselson, identify messenger RNA, |
| 1965 | Margaret Dayhoff's Atlas of Protein Sequences |
| 1970 | Needleman-Wunsch algorithm |
| 1977 | DNA sequencing and software to analyze it (Staden) |
| 1981 | Smith-Waterman algorithm developed |
| 1981 | The concept of a sequence motif (Doolittle) |
| 1982 | GenBank Release 3 made public |
| 1982 | Phage lambda genome sequenced |
| 1983 | Sequence database searching algorithm (Wilbur-Lipman) |
| 1985 | FASTP/FASTN: fast sequence similarity searching |
| 1988 | National Center for Biotechnology Information (NCBI) created at NIH/NLM |
| 1988 | EMBnet network for database distribution |
| 1990 | BLAST: fast sequence similarity searching |
| 1991 | EST: expressed sequence tag sequencing |
| 1993 | Sanger Centre, Hinxton, UK |
| 1994 | EMBL European Bioinformatics Institute, Hinxton, UK |
| 1995 | First bacterial genomes completely sequenced |
| 1996 | Yeast genome completely sequenced |
| 1997 | PSI-BLAST |
| 1998 | Worm (multicellular) genome completely sequenced |
| 1999 | Fly genome completely sequenced |
| 2000 | Jeong H, Tombor B, Albert R, Oltvai ZN, Barabasi AL. **The large-scale organization of metabolic networks.** Nature 2000 Oct 5;407(6804):651-4, PubMed |
| 2000 | The genome for *Pseudomonas aeruginosa* (6.3 Mbp) is published. |
| 2000 | The A. thaliana genome (100 Mb) is sequenced. |
| 2001 | The human genome (3 Giga base pairs) is published. |

# Protein Folding

Proteins are the biological molecules that are the building blocks of cells and organs, and the biochemical processes required to keep living organisms alive are catalyzed and regulated by proteins called enzymes. Proteins are linear polymers of amino acids that fold into conformations dictated by the physical and chemical properties of the amino acid chain. *The biological function of a protein is dependent on the protein folding into the correct, or "native", state. Protein folding is usually a spontaneous process, and often when a protein unfolds because of heat or chemical denaturation, it will be capable of refolding into the correct conformation as soon as it is removed from the environment of the denaturant.*

Protein folding can go wrong for many reasons. When an egg is boiled, the proteins in the white unfold and misfold into a solid mass of protein that will not refold or redissolve. This type of irreversible misfolding is similar to the insoluble protein deposits found in certain tissues that are characteristic of some diseases, such as Alzheimer's disease. These protein deposits are aggregates of proteins folded into the wrong shapes.

*Determining the process by which proteins fold into particular shapes, characteristic of their amino acid sequence, is commonly called "the protein folding problem", an area of study at the forefront of computational biology.* One approach to studying the protein folding process is the application of statistical mechanics techniques and molecular dynamics simulations to the study of protein folding.

The Stanford University Folding@home project has simulated protein folding with atomistic detail in an implicit solvent model by using a large scale distributed computing project that allows timescales thousands to millions of times longer than previously achievable with a model of this detail. Look at the menu on the left border of the Stanford Folding@home web page. Click on the "Science" link to read the scientific background behind the protein folding distributed computing project. (Q1) What are the 3 functions of proteins that are mentioned in the "What are proteins?" section of the scientific background? (Q2) What are 3 diseases that are believed to result from protein misfolding? (Q3) What are typical timescales for molecular dynamics simulations? (Q4) What are typical timescales at which the fastest proteins fold? (Q5) How does the Stanford group break the microsecond barrier with their simulations?

Return to the Stanford Folding@home home page. Click on the "Results" link in the left border of the web page. Look at the information on the folding simulations of the villin headpiece. (Q6) How many amino acids are in the simulated villin headpiece? (Q7) How does this compare with the number of amino acids in a typical protein? (Q8) Taking into consideration the size of the biological molecules in these simulations and the requirements that necessitated using large scale distributed computing methods for the simulations, what are the biggest impediments to understanding the protein folding problem?

With the publication of entire genomes that contain sequences to many unknown proteins, it is possible to imagine someday having the ability to predict the final structure of a protein based on its sequence. This would require an understanding of the fundamental rules that govern protein folding. Also, inroads into the mechanisms behind protein folding would provide important knowledge for fighting disease states where misfolded proteins are implicated.

# BioInformatics - Introduction

To build a house you need bricks and mortar and something else -- the "know-how", or "***information***" as to how to go about your business. The Victorians knew this. But when it came to the "building" of animals and plants, -- the word, and perhaps the concept, of "information" is difficult to trace in their writings.

Classical scholars tell us that Aristotle did not have this problem. The "eidos", the form-giving essence that shapes the embryo "contributes nothing to the the material body of the embryo but only communicates its program of development" (see Delbrück's "Aristotle-totle-totle" in ***Of Microbes and Life*** 1971, pp. 50-55).

William Bateson spoke of a "**factor**" (gene) having the "**power**" to bring about the building of the characters which make up an organism. He used the "information" concept, but not the word. He was prepared to believe his factors were molecules of the type we would today call macromolecules, but he did not actually call them "***informational*** macromolecules".

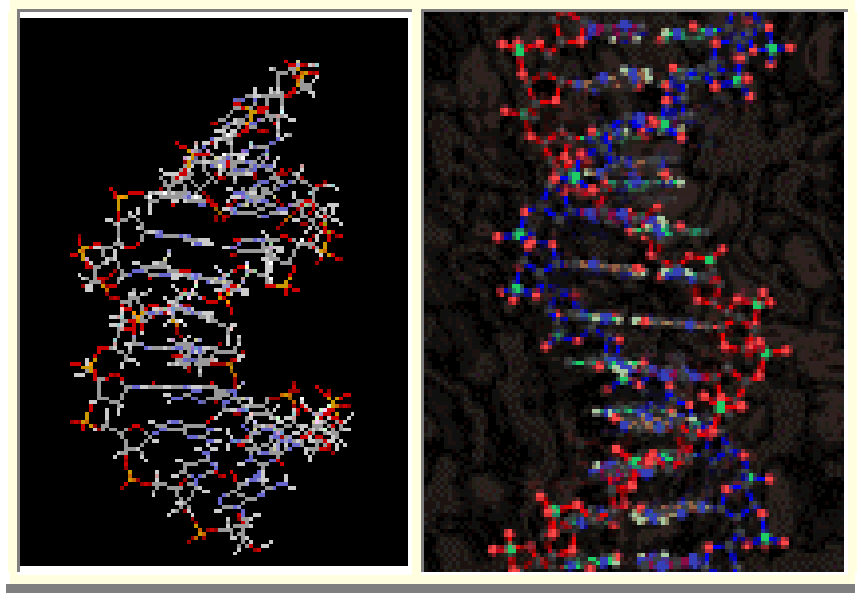

Information has many forms. If you turn down the corner of a page of a book to remind you where you stopped reading ("book-mark"), then you have left information on the page. In future you read ("decode") the bookmark with the knowledge that it means "continue here". A future historian might be interested in where you paused in your reading. Coming across the book, he/she would notice creases suggesting that a flap had been turned down. Making assumptions about the code you were employing, a feasible map of the book could then be made with your pause sites. It might be discovered that you paused at particular sites, say at the ends of chapters. In this case pauses would be correlated with the distribution of the book's "primary information". Or perhaps there was a random element to your pausing ... perhaps when your partner wanted the light out. In this case pausing would be influenced by your pairing relationship.

A more familiar form of information is the linear form you are now decoding (reading), which is similar to the form you might decode on the page of a book. If a turned-down flap on a page is a large one, it might cover up some of the information. Thus, **one form of information might *interfere* with another form of information**. To read the text you would have to correct (fold back) the "secondary structure" of the page (the flap) so that it no longer overlapped the text. Thus, there is a ***conflict***. You can either retain the flap and not read the text, or get rid of the flap and read the text.

In the case of a book page, the text is imposed on an underlying flat two dimensional base, the paper. The text (message) and the medium are different. Similarly, in the case of our genetic material, DNA, the "medium" is a chain of two units (phosphate and ribose) and the most easily recognized "message" is provided by a sequence of "letters" (bases) attached, like beads, to the chain. As in the case of a written text on paper, "flaps" in DNA (secondary structure) can conflict with the base sequence (primary structure). Thus the pressures to convey information (messages) encoded in a particular sequence, ***and*** to convey information encoded in a "flap", may be in conflict. The "hand" of evolution has to resolve these apparently ***intrinsic*** conflicts while dealing with other pressures (***extrinsic***) from the environment.

The stunning novelty of the Watson-Crick model of DNA was not only that it was beautiful, but that it also explained so much of the biology of heredity. There was not just one sequence of letters, but two. These were wrapped round each other in the form of a double helix. One was the complement of the other, so that the sequence of one string (strand) could be inferred from the sequence of the other. If there were damage to one strand of DNA, then that strand could potentially be repaired on the basis of the text of the opposite strand. When the cell divided the two strands would part and separate. New "daughter" strands, synthesized from nucleotide "building blocks" (each consisting of phosphate, ribose and a base) , would replace those which had separated, so that duplexes identical to the parental duplex would be created.

There were two main types of bases, purines (**R**) and pyrimidines (**Y**). Thus, disregarding the phosphate-ribose chain, the first nucleic acids to appear in evolution could accurately be represented as a binary sequence such as **RYRRYRYYRYR**.... Each base would be a "binary digit". Conventionally, we represent binary digits in computer language as strings of **0**s and **1**s. If a **Y** and an **R** were equally likely alternatives in a sequence position, then each could be quantitated as one "**bit**" of information.

Each base came to acquire two flavours. There are two main types of purines, adenine (**A**) and guanine (**G**), and two main types of pyrimidines, cytosine (**C**) and thymine (**T**). Thus, the above sequence might now be represented as (say) **ACGATGCCGTA.... Chargaff's first parity rule** is that purines pair with pyrimidines, specifically **A** with **T** and **C** with **G**. Thus, a duplex containing this sequence, with pairing between ***complementary bases*** in the "top" and "bottom" strands could be written as:

```
ACGATGCCGTAGCATCGT
TGCTACGGCATCGTAGCA
```

It was later realized that, under certain circumstances, this double helix could form "flaps". Thus each of the above two strands ("top" and "bottom") can form stem-loop secondary structures, of the following type, due to pairing with complementary bases in the ***same*** strand.

```
                C
ACGATGC         G
TGCTACG         T
       A
```

For this to happen there have to be matching (complementary) bases. Only the bases in the loop (**CGTA**) are unpaired in this structure. The stem consists of paired bases. Thus Chargaff's parity rule has to apply, to a close approximation, to ***single strands*** of DNA. When one examines DNAs from whatever biological source, one invariably finds that the rule applies. We refer to this as **Chargaff's second parity rule**.

Returning to our own written textual form of information, the sentence "**Mary had a little lamb its fleece was white as snow**" contains the ***information*** that a person called Mary is in possession of an immature sheep. The same information might be written in Chinese or Greek. Thus, the sentence contains not only its ***primary*** information, but ***secondary*** information about its origin -- e.g. it is likely that the author is more familiar with English than other languages. Some believe that English is on the way to displacing other languages, so that eventually it (or the form it evolves to) will constitute the only language used by human beings on this planet. Similarly, in the course of early evolution it is likely that a prototypic nucleic acid language displaced contenders.

It would be difficult to discern a relationship between the English, Chinese and Greek versions of the above sentence, because these languages diverged from primitive root languages thousands of years ago. However, in England, if a person with a Cockney accent were to speak the sentence it would sound like "**Miree ader liawl laimb sfloyce wors woyt ers snaa**". Cockney English and "regular" English diverged more recently and it is easy to discern similarities.

Now look at the following text:

**yewas htbts llem ws arifea ac wMhitte alidsnoe la**
**irsnwwis aee ar lal larfoMyce b sos woilmyt erdea**

One line of text is the regular English version with the letters shuffled. The other line is the cockney version with the letters shuffled. Can you tell which is which? If the shuffling was thorough, the primary information has been destroyed. However, there is still ***some*** information left. With the knowledge that cockneys tend to "drop" their **H**s, it can be deduced that the upper text is more likely to be from someone who spoke regular English. With a longer text, this could be more precisely quantitated. Languages have characteristic letter frequencies. You can take a segment ("window") and count the various letters in that segment.

In this way you can identify a text as English, Cockney, Chinese or Greek, without too much trouble. We can call this information "secondary information". There may be various other ***levels*** of information in a sequence of symbols. To evaluate the secondary information in DNA (with only four "letters"), you select a "window" (say 1000 bases) and counts the number of bases in that window. You can apply the same window to another section of the DNA, or to another DNA molecule from a different biological species, and repeat the count. Then you can compare DNA "accents".

The best understood type of primary information in DNA is the information for proteins. The DNA sequence of bases (one type of "letter") encodes another type of "letter", the "**amino acids**". There are 20 amino acids, with names such as **aspartate**, **glycine**, **phenylalanine**, **serine** and **valine** (which are abbreviated as **Asp**, **Gly**, **Phe**, **Ser** and **Val**). Under instructions received from DNA, amino acids are joined together in the same order as they are encoded in DNA, to form **proteins**. The latter, chains of amino acids which fold in complicated ways, play a major role in determining how we interact with our environment. The proteins determine our "phenotype". For example, in an organism of a particular species ("***A***") the twenty one base DNA sequence:

TTTTCATTAGTTGGAGATAAA

read in sets of three bases ("codons"), conveys primary information for a seven amino acid protein fragment (**PheSerLeuValGlyAspLys**). All members of the species will tend to have the same DNA sequence, and differences between members of the species will tend to be rare and of minor degree. If the protein is fundamental to cell function it is likely that organisms of ***another*** species ("***B***") will have DNA which encodes the ***same*** protein fragment. However, when we examine their DNA we might find major differences compared with the DNA of the first species (the similarities are emphasized in red):

**TT**CAGCC**T**C**GT**G**GG**G**GA**C**AA**G

This sequence ***also*** encodes the above protein fragment, showing that the DNA contains the ***same*** primary information as in the first DNA sequence, but it is "spoken" with a different "accent". This ***secondary*** information might have some biological role. It is theoretical possible (but unlikely) that all the genes in an organism of species ***B*** would have this "accent", yet otherwise encode the ***same*** proteins. In this case, organisms of species ***A*** and ***B*** would be both anatomically and functionally (physiologically) ***identical***, while differing dramatically with respect to secondary information.

On the other hand, consider a single change in the sequence of species ***A*** to:

TTTTCATTAGTTGGAG**T**TAAA

Here the difference would change one of the seven amino acids. It is likely that such ***minor*** changes in a ***very small*** number of genes affecting development would be sufficient to cause anatomical and morphological differentiation ***within*** species ***A*** (e.g. compare a bulldog and a poodle, as "varieties" of dogs, which are able to breed with each other). Yet, in this case the secondary information would be hardly changed.

The view developed in these pages is that, like the Cockney's dropped H's, the role of secondary information is to ***initiate,*** and, for a while, ***maintain***, reproductive isolation. This can occur because the genetic code is a "**redundant**" or "**degenerate**" code; for example, the amino acid **serine** is not encoded by just one codon; there are six possible codons (**TCT**, **TCC**, **TCA**, **TCG**, **AGT**, **AGC**). In the first of the above DNA sequences (*A*) the amino acid **serine** (**Ser**) is encoded by TCA, whereas AGC is used in the second (***B***). On the other hand, the change in species ***A*** from GAT (first sequence) to GTT (third sequence) changes the encoded amino acid from **aspartic acid** (**Asp**) to **valine** (**Val**), and this should be sufficient to change the properties of the corresponding protein, and hence change the phenotype.

Thus, the biological interest of linguistic barriers is that they also tend to be **reproductive barriers**. Even if a Chinese person and an English person are living in the same territory ("sympatrically"), if they do not speak the same language they are unlikely to marry. The Chinese tend to marry Chinese and produce more Chinese. The English tend to marry English and produce more English. Even in England, because of the "class" barriers so colourfully portrayed by George Bernard Shaw, Cockneys tend to marry Cockneys, and the essence of the barrier from people speaking "regular" English is the difference in accent. Because of other ("blending") factors at work in our society it is unlikely that this linguistic speciation will continue to the extent that Cockney will become an independent language. However, the point is that when there is "incipient" linguistic speciation, it is the ***secondary*** information (dropped H's) , not the primary information, which constitutes the barrier.

Before the genetic code was deciphered in the early 1960s, researchers such as Wyatt (1952) and Sueoka (1961) studied the base composition of DNAs with a major interest in the primary information -- how a sequence of bases might be related to a sequence of amino acids. However, their results have turned out to be of greater interest with respect to the secondary information in DNA.

# Biology in the Computer Age

From the interaction of species and populations, to the function of tissues and cells within an individual organism, biology is defined as the study of living things. In the course of that study, biologists collect and interpret data. Now, at the beginning of the 21st century, we use sophisticated laboratory technology that allows us to collect data faster than we can interpret it. We have vast volumes of DNA sequence data at our fingertips. But how do we figure out which parts of that DNA control the various chemical processes of life? We know the function and structure of some proteins, but how do we determine the function of new proteins? And how do we predict what a protein will look like, based on knowledge of its sequence? We understand the relatively simple code that translates DNA into protein. But how do we find meaningful new words in the code and add them to the DNA-protein dictionary?

*Bioinformatics* is the science of using information to understand biology; it's the tool we can use to help us answer these questions and many others like them. Unfortunately, with all the hype about mapping the human genome, bioinformatics has achieved buzzword status; the term is being used in a number of ways, depending on who is using it. Strictly speaking, bioinformatics is a subset of the larger field of *computational biology,* the application of quantitative analytical techniques in modeling biological systems. In this book, we stray from bioinformatics into computational biology and back again. The distinctions between the two aren't important for our purpose here, which is to cover a range of tools and techniques we believe are critical for molecular biologists who want to understand and apply the basic computational tools that are available today.

The field of bioinformatics relies heavily on work by experts in statistical methods and pattern recognition. Researchers come to bioinformatics from many fields, including mathematics, computer science, and linguistics. Unfortunately, biology is a science of the specific as well as the general. Bioinformatics is full of pitfalls for those who look for patterns and make predictions without a complete understanding of where biological data comes from and what it means. By providing algorithms, databases, user interfaces, and statistical tools, bioinformatics makes it possible to do exciting things such as compare DNA sequences and generate results that are potentially significant. "Potentially significant" is perhaps the most important phrase. These new tools also give you the opportunity to overinterpret data and assign meaning where none really exists. We can't overstate the importance of understanding the limitations of these tools. But once you gain that understanding and become an intelligent consumer of bioinformatics methods, the speed at which your research progresses can be truly amazing.

### *How Is Computing Changing Biology?*

An organism's hereditary and functional information is stored as DNA, RNA, and proteins, all of which are linear chains composed of smaller molecules. These macromolecules are assembled from a fixed alphabet of well-understood chemicals: DNA is made up of four deoxyribonucleotides (adenine, thymine, cytosine, and guanine), RNA is made up from the four ribonucleotides (adenine, uracil, cytosine, and guanine), and proteins are made from the 20 amino acids. Because these macromolecules are linear chains of defined components, they can be represented as sequences of symbols. These sequences can then be compared to find similarities that suggest the molecules are related by form or function.

Sequence comparison is possibly the most useful computational tool to emerge for molecular biologists. The World Wide Web has made it possible for a single public database of genome sequence data to provide services through a uniform interface to a worldwide community of users. With a commonly used computer program called fsBLAST, a molecular biologist can compare an uncharacterized DNA sequence to the entire publicly held collection of DNA sequences. In the next section, we present an example of how sequence comparison using the BLAST program can help you gain insight into a real disease.

## The Eye of the Fly

Fruit flies (*Drosophila melanogaster*) are a popular model system for the study of development of animals from embryo to adult. Fruit flies have a gene called *eyeless*, which, if it's "knocked out (i.e., eliminated from the genome using molecular biology methods), results in fruit flies with no eyes. It's obvious that the *eyeless* gene plays a role in eye development.

Researchers have identified a human gene responsible for a condition called *aniridia*. In humans who are missing this gene (or in whom the gene has mutated just enough for its protein product to stop functioning properly), the eyes develop without irises.

If the gene for *aniridia* is inserted into an eyeless drosophila "knock out," it causes the production of normal drosophila eyes. It's an interesting coincidence. Could there be some similarity in how *eyeless* and *aniridia* function, even though flies and humans are vastly different organisms? Possibly. To gain insight into how *eyeless* and *aniridia* work together, we can compare their sequences. Always bear in mind, however, that genes have complex effects on one another. Careful experimentation is required to get a more definitive answer.

As little as 15 years ago, looking for similarities between *eyeless* and *aniridia* DNA sequences would have been like looking for a needle in a haystack. Most scientists compared the respective gene sequences by hand-aligning them one under the other in a word processor and looking for matches character by character. This was time-consuming, not to mention hard on the eyes.

In the late 1980s, fast computer programs for comparing sequences changed molecular biology forever. Pairwise comparison of biological sequences is the foundation of most widely used bioinformatics techniques. Many tools that are widely available to the biology community--including everything from multiple alignment, phylogenetic analysis, motif identification, and homology-modeling software, to web-based database search services--rely on pairwise sequence-comparison algorithms as a core element of their function.

These days, a biologist can find dozens of sequence matches in seconds using sequence-alignment programs such as BLAST and FASTA. These programs are so commonly used that the first encounter you have with bioinformatics tools and biological databases will probably be through the National Center for Biotechnology Information's (NCBI) BLAST web interface. Figure 1-1 shows a standard form for submitting data to NCBI for a BLAST search.

**Figure 1-1. Form for submitting a BLAST search against nucleotide databases at NCBI**

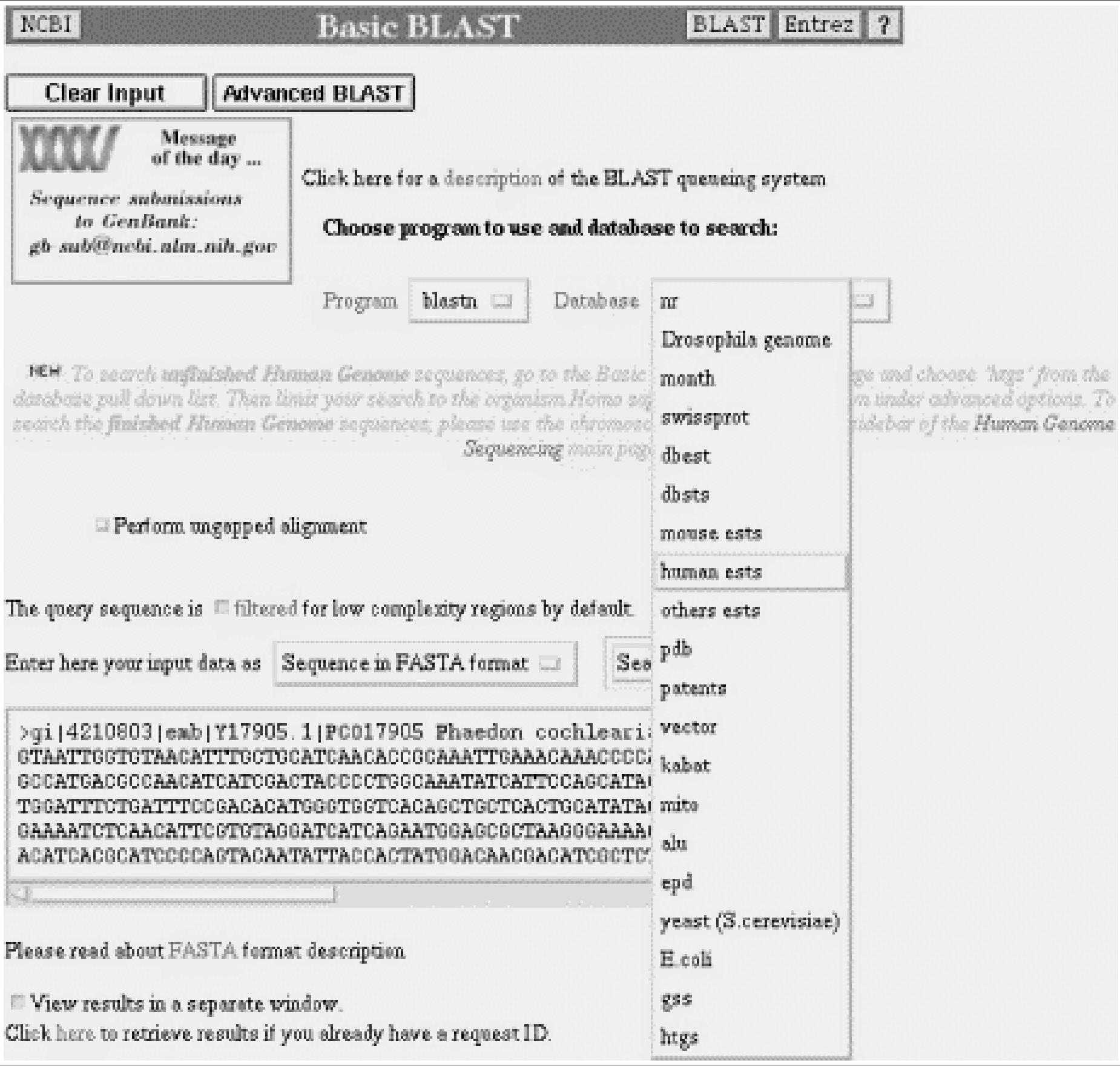

## Labels in Gene Sequences

Before you rush off to compare the sequences of *eyeless* and *aniridia* with BLAST, let us tell you a little bit about how sequence alignment works.

It's important to remember that biological sequence (DNA or protein) has a chemical function, but when it's reduced to a single-letter code, it also functions as a unique label, almost like a bar code. From the information technology point of view, sequence information is priceless. The sequence label can be applied to a gene, its product, its function, its role in cellular metabolism, and so on. The user searching for information related to a particular gene can then use rapid pairwise sequence comparison to access any information that's been linked to that sequence label.

The most important thing about these sequence labels, though, is that they don't just uniquely identify a particular gene; they also contain biologically meaningful patterns that allow users to compare different labels, connect information, and make inferences. So not only can the labels connect all the information about one gene, they can help users connect information about genes that are slightly or even dramatically different in sequence.

If simple labels were all that was needed to make sense of biological data, you could just slap a unique number (e.g., a GenBank ID) onto every DNA sequence and be done with it. But biological sequences are related by evolution, so a partial pattern match between two sequence labels is a significant find. BLAST differs from simple keyword searching in its ability to detect partial matches along the entire length of a protein sequence.

## Comparing eyeless and aniridia with BLAST

When the two sequences are compared using BLAST, you'll find that *eyeless* is a partial match for *aniridia*. The text that follows is the raw data that's returned from this BLAST search:

```
pir||A41644 homeotic protein aniridia - human
      Length = 447

 Score = 256 bits (647), Expect = 5e-67
 Identities = 128/146 (87%), Positives = 134/146 (91%), Gaps = 1/146 (0%)

Query: 24 IERLPSLEDMAHKGHSGVNQLGGVFVGGRPLPDSTRQKIVELAHSGARPCDISRILQVSN 83
          I R P+  M + HSGVNQLGGVFV GRPLPDSTRQKIVELAHSGARPCDISRILQVSN
Sbjct: 17 IPRPPARASMQNS-HSGVNQLGGVFVNGRPLPDSTRQKIVELAHSGARPCDISRILQVSN 75

Query: 84 GCVSKILGRYYETGSIRPRAIGGSKPRVATAEVVSKISQYKRECPSIFAWEIRDRLLQEN 143
          GCVSKILGRYYETGSIRPRAIGGSKPRVAT EVVSKI+QYKRECPSIFAWEIRDRLL E
Sbjct: 76 GCVSKILGRYYETGSIRPRAIGGSKPRVATPEVVSKIAQYKRECPSIFAWEIRDRLLSEG 135

Query: 144 VCTNDNIPSVSSINRVLRNLAAQKEQ 169
           VCTNDNIPSVSSINRVLRNLA++K+Q
Sbjct: 136 VCTNDNIPSVSSINRVLRNLASEKQQ 161


 Score = 142 bits (354), Expect = 1e-32
 Identities = 68/80 (85%), Positives = 74/80 (92%)

Query: 398 TEDDQARLILKRKLQRNRTSFTNDQIDSLEKEFERTHYPDVFARERLAGKIGLPEARIQV 457
           +++ Q RL LKRKLQRNRTSFT +QI++LEKEFERTHYPDVFARERLA KI LPEARIQV
Sbjct: 222 SDEAQMRLQLKRKLQRNRTSFTQEQIEALEKEFERTHYPDVFARERLAAKIDLPEARIQV 281

Query: 458 WFSNRRAKWRREEKLRNQRR 477
           WFSNRRAKWRREEKLRNQRR
Sbjct: 282 WFSNRRAKWRREEKLRNQRR 301
```

The output shows local alignments of two high-scoring matching regions in the protein sequences of the *eyeless* and *aniridia* genes. In each set of three lines, the query sequence (the *eyeless* sequence that was submitted to the BLAST server) is on the top line, and the *aniridia* sequence is on the bottom line. The middle line shows where the two sequences match. If there is a letter on the middle line, the sequences match exactly at that position. If there is a plus sign on the middle line, the two sequences are different at that position, but there is some chemical similarity between the amino acids (e.g., D and E, aspartic and glutamic acid). If there is nothing on the middle line, the two sequences don't match at that position.

In this example, you can see that, if you submit the whole *eyeless* gene sequence and look (as standard keyword searches do) for an exact match, you won't find anything. The local sequence regions make up only part of the complete proteins: the region from 24-169 in *eyeless* matches the region from 17-161 in the human *aniridia* gene, and the region from 398-477 in *eyeless* matches the region from 222-301 in *aniridia*. The rest of the sequence doesn't match! Even the two regions shown, which match closely, don't match 100%, as they would have to, in order to be found in a keyword search.

However, this partial match is significant. It tells us that the human *aniridia* gene, which we don't know much about, is substantially related in sequence to the fruit fly's *eyeless* gene. And we do know a lot about the *eyeless* gene, from its structure and function (it's a DNA binding protein that promotes the activity of other genes) to its effects on the phenotype--the form of the grown fruit fly.

BLAST finds local regions that match even in pairs of sequences that aren't exactly the same overall. It extends matches beyond a single-character difference in the sequence, and it keeps trying to extend them in all directions until the overall score of the sequence match gets too small. As a result, BLAST can detect patterns that are imperfectly replicated from sequence to sequence, and hence distant relationships that are inexact but still biologically meaningful.

Depending on the quality of the match between two labels, you can transfer the information attached to one label to the other. A high-quality sequence match between two full-length sequences may suggest the hypothesis that their functions are similar, although it's important to remember that the identification is only tentative until it's been experimentally verified. In the case of the *eyeless* and *aniridia* genes, scientists hope that studying the role of the *eyeless* gene in Drosophila eye development will help us understand how *aniridia* works in human eye development.

## Isn't Bioinformatics Just About Building Databases?

Much of what we currently think of as part of bioinformatics--sequence comparison, sequence database searching, sequence analysis--is more complicated than just designing and populating databases. Bioinformaticians (or computational biologists) go beyond just capturing, managing, and presenting data, drawing inspiration from a wide variety of quantitative fields, including statistics, physics, computer science, and engineering. Figure 1-2 shows how quantitative science intersects with biology at every level, from analysis of sequence data and protein structure, to metabolic modeling, to quantitative analysis of populations and ecology.

Bioinformatics is first and foremost a component of the biological sciences. The main goal of bioinformatics isn't developing the most elegant algorithms or the most arcane analyses; the goal is finding out how living things work. Like the molecular biology methods that greatly expanded what biologists were capable of studying, bioinformatics is a tool and not an end in itself. Bioinformaticians are the tool-builders, and it's critical that they understand biological problems as well as computational solutions in order to produce useful tools.

Research in bioinformatics and computational biology can encompass anything from abstraction of the properties of a biological system into a mathematical or physical model, to implementation of new algorithms for data analysis, to the development of databases and web tools to access them.

**Figure 1-2. How technology intersects with biology**

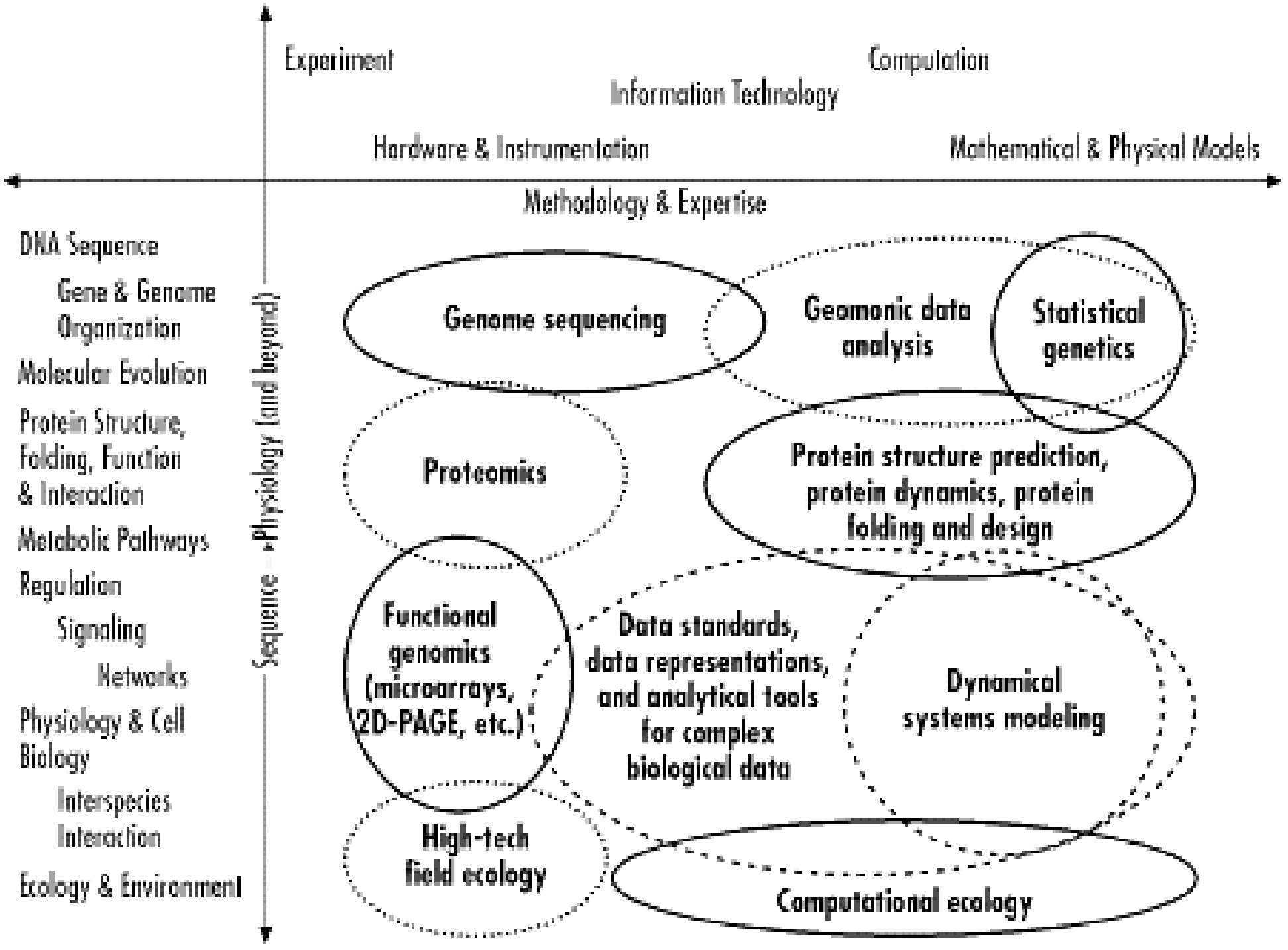

## The First Information Age in Biology

Biology as a science of the specific means that biologists need to remember a lot of details as well as general principles. Biologists have been dealing with problems of information management since the 17*th* century.

The roots of the concept of evolution lie in the work of early biologists who catalogued and compared species of living things. The cataloguing of species was the preoccupation of biologists for nearly three centuries, beginning with animals and plants and continuing with microscopic life upon the invention of the compound microscope. New forms of life and fossils of previously unknown, extinct life forms are still being discovered even today.

All this cataloguing of plants and animals resulted in what seemed a vast amount of information at the time. In the mid-16*th* century, Otto Brunfels published the first major modern work describing plant species, the *Herbarium vitae eicones*. As Europeans traveled more widely around the world, the number of catalogued species increased, and botanical gardens and herbaria were established. The number of catalogued plant types was 500 at the time of Theophrastus, a student of Aristotle. By 1623, Casper Bauhin had observed 6,000 types of plants. Not long after John Ray introduced the concept of distinct species of animals and plants, and developed guidelines based on anatomical features for distinguishing conclusively between species. In the 1730s, Carolus Linnaeus catalogued 18,000 plant species and over 4,000 species of animals, and established the basis for the modern taxonomic naming system of kingdoms, classes, genera, and species. By the end of the 18*th* century, Baron Cuvier had listed over 50,000 species of plants.

It was no coincidence that a concurrent preoccupation of biologists, at this time of exploration and cataloguing, was classification of species into an orderly taxonomy. A botany text might encompass several volumes of data, in the form of painstaking illustrations and descriptions of each species encountered. Biologists were faced with the problem of how to organize, access, and sensibly add to this information. It was apparent to the casual observer that some living things were more closely related than others. A rat and a mouse were clearly more similar to each other than a mouse and a dog. But how would a biologist know that a rat was like a mouse (but that rat was not just another name for mouse) without carrying around his several volumes of drawings? A nomenclature that uniquely identified each living thing and summed up its presumed relationship with other living things, all in a few words, needed to be invented.

The solution was relatively simple, but at the time, a great innovation. Species were to be named with a series of one-word names of increasing specificity. First a very general division was specified: animal or plant? This was the kingdom to which the organism belonged. Then, with increasing specificity, came the names for class, genera, and species. This schematic way of classifying species, as illustrated in Figure 1-3, is now known as the "Tree of Life."

**Figure 1-3. The "Tree of Life" represents the nomenclature system that classifies species**

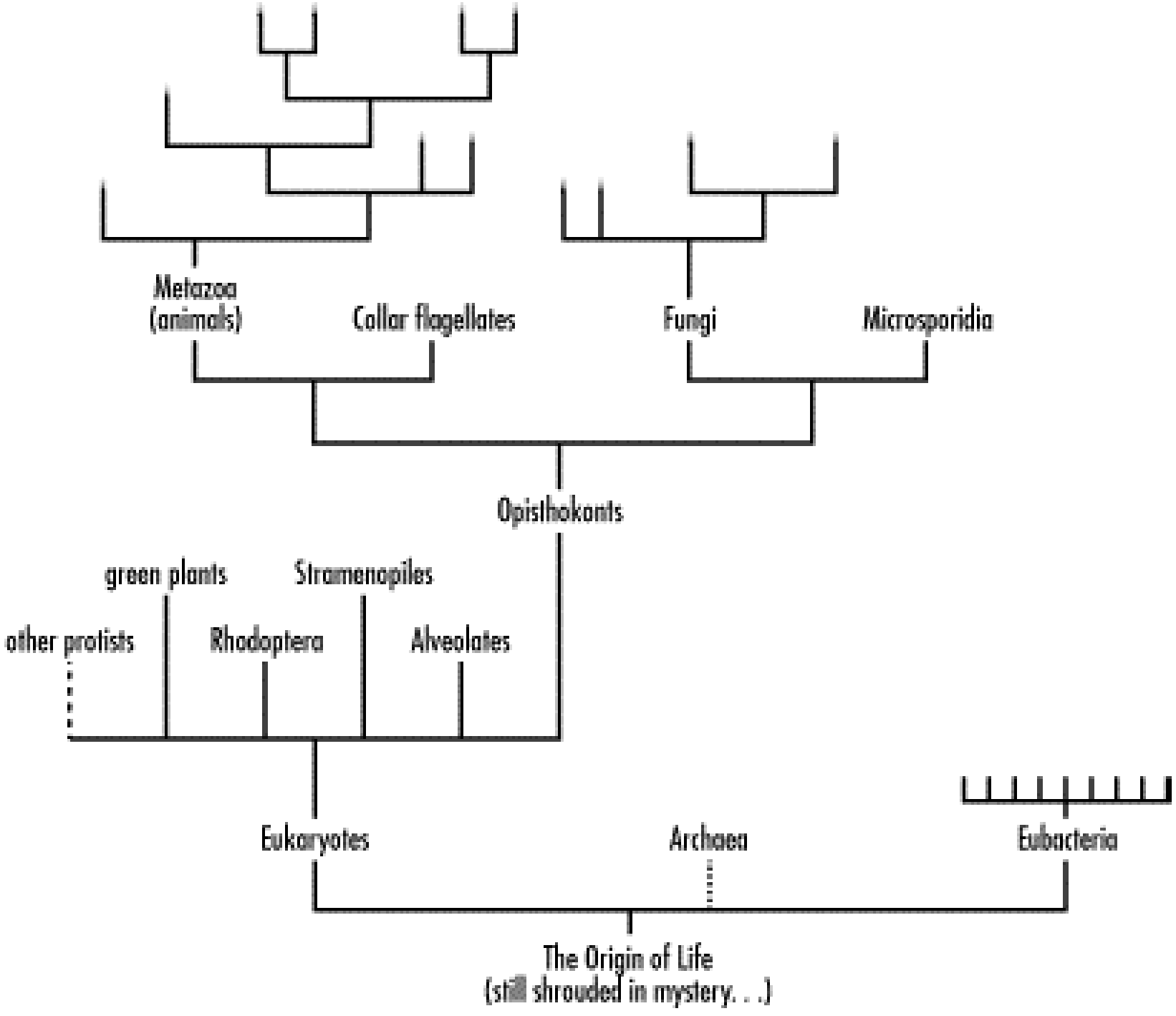

A modern taxonomy of the earth's millions of species is too complicated for even the most zealous biologist to memorize, and fortunately computers now provide a way to maintain and access the taxonomy of species. The University of Arizona's Tree of Life project and NCBI's Taxonomy database are two examples of online taxonomy projects.

Taxonomy was the first informatics problem in biology. Now, biologists have reached a similar point of information overload by collecting and cataloguing information about individual genes. The problem of organizing this information and sharing knowledge with the scientific community at the gene level isn't being tackled by developing a nomenclature. It's being attacked directly with computers and databases from the start.

The evolution of computers over the last half-century has fortuitously paralleled the developments in the physical sciences that allow us to see biological systems in increasingly fine detail. Figure 1-4 illustrates the astonishing rate at which biological knowledge has expanded in the last 20 years.

**Figure 1-4. The growth of GenBank and the Protein Data Bank has been astronomical**

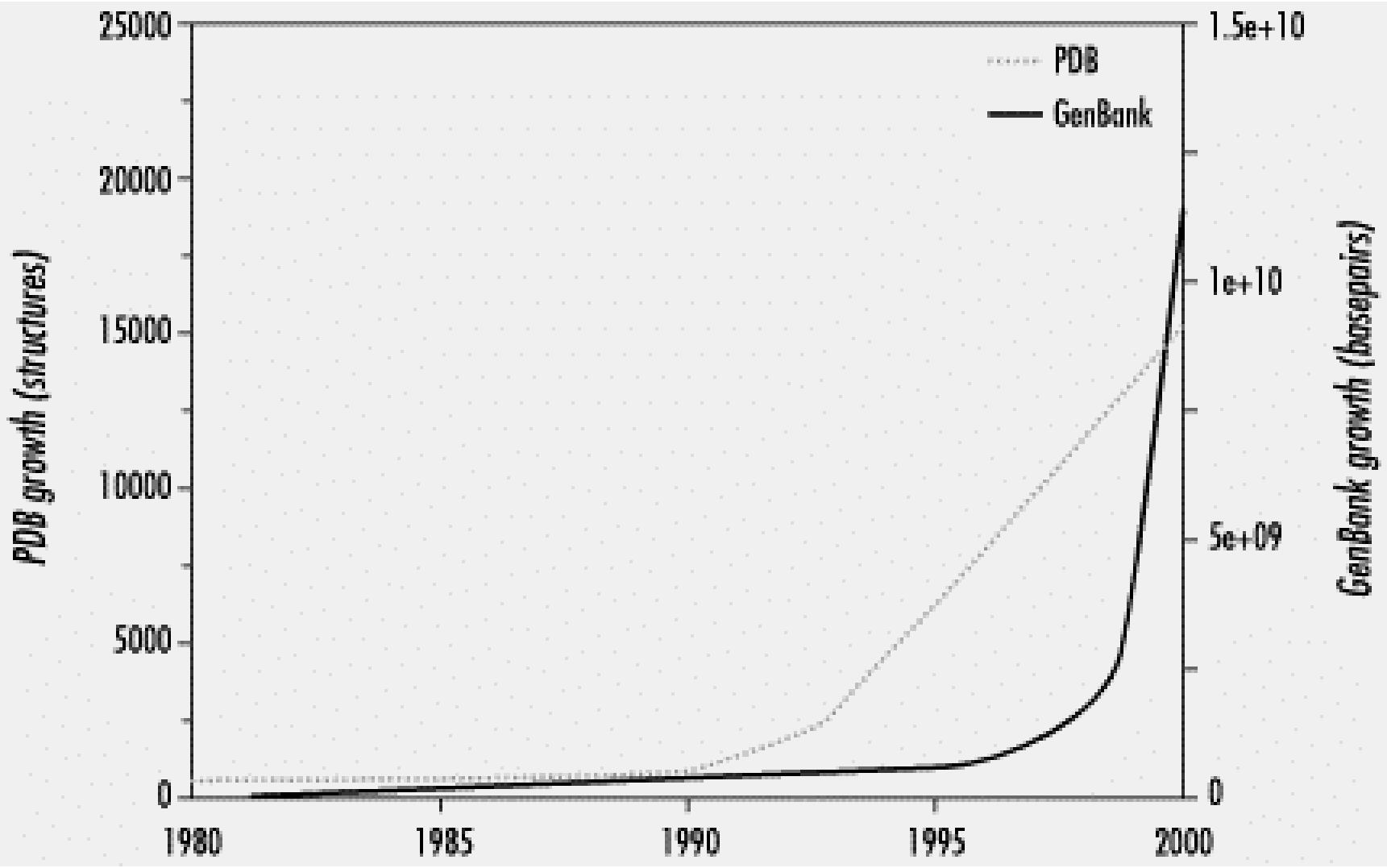

Simply finding the right needles in the haystack of information that is now available can be a research problem in itself. Even in the late 1980s, finding a match in a sequence database was worth a five-page publication. Now this procedure is routine, but there are many other questions that follow on our ability to search sequence and structure databases. These questions are the impetus for the field of bioinformatics.

## *What Does Informatics Mean to Biologists?*

The science of informatics is concerned with the representation, organization, manipulation, distribution, maintenance, and use of information, particularly in digital form. There is more than one interpretation of what bioinformatics--the intersection of informatics and biology--actually means, and it's quite possible to go out and apply for a job doing bioinformatics and find that the expectations of the job are entirely different than you thought.

The functional aspect of bioinformatics is the representation, storage, and distribution of data. Intelligent design of data formats and databases, creation of tools to query those databases, and development of user interfaces that bring together different tools to allow the user to ask complex questions about the data are all aspects of the development of bioinformatics infrastructure.

Developing analytical tools to discover knowledge in data is the second, and more scientific, aspect of bioinformatics. There are many levels at which we use biological information, whether we are comparing sequences to develop a hypothesis about the function of a newly discovered gene, breaking down known 3D protein structures into bits to find patterns that can help predict how the protein folds, or modeling how proteins and metabolites in a cell work together to make the cell function. The ultimate goal of analytical bioinformaticians is to develop predictive methods that allow scientists to model the function and phenotype of an organism based only on its genome sequence. This is a grand goal, and one that will be approached only in small steps, by many scientists working together.

## *What Challenges Does Biology Offer Computer Scientists?*

The goal of biology, in the era of the genome projects, is to develop a quantitative understanding of how living things are built from the genome that encodes them.

Cracking the genome code is complex. At the very simplest level, we still have difficulty identifying unknown genes by computer analysis of genomic sequence. We still have not managed to predict or model how a chain of amino acids folds into the specific structure of a functional protein.

Beyond the single-molecule level, the challenges are immense. The sheer amount of data in GenBank is now growing at an exponential rate, and as data types beyond DNA, RNA, and protein sequence begin to undergo the same kind of explosion, simply managing, accessing, and presenting this data to users in an intelligible form is a critical task. Human-computer interaction specialists need to work closely with academic and clinical researchers in the biological sciences to manage such staggering amounts of data.

Biological data is very complex and interlinked. A spot on a DNA array, for instance, is connected not only to immediate information about its intensity, but to layers of information about genomic location, DNA sequence, structure, function, and more. Creating information systems that allow biologists to seamlessly follow these links without getting lost in a sea of information is also a huge opportunity for computer scientists.

Finally, each gene in the genome isn't an independent entity. Multiple genes interact to form biochemical pathways, which in turn feed into other pathways. Biochemistry is influenced by the external environment, by interaction with pathogens, and by other stimuli. Putting genomic and biochemical data together into quantitative and predictive models of biochemistry and physiology will be the work of a generation of computational biologists. Computer scientists, mathematicians, and statisticians will be a vital part of this effort.

## *What Skills Should a Bioinformatician Have?*

There's a wide range of topics that are useful if you're interested in pursuing bioinformatics, and it's not possible to learn them all. However, in our conversations with scientists working at companies such as Celera Genomics and Eli Lilly, we've picked up on the following "core requirements" for bioinformaticians:

- You should have a fairly deep background in some aspect of molecular biology. It can be biochemistry, molecular biology, molecular biophysics, or even molecular modeling, but without a core of knowledge of molecular biology you will, as one person told us, "run into brick walls too often."
- You must absolutely understand the central dogma of molecular biology. Understanding how and why DNA sequence is transcribed into RNA and translated into protein is vital. (In *Chapter 2, Computational Approaches to Biological Questions*, we define the central dogma, as well as review the processes of transcription and translation.)

- You should have substantial experience with at least one or two major molecular biology software packages, either for sequence analysis or molecular modeling. The experience of learning one of these packages makes it much easier to learn to use other software quickly.
- You should be comfortable working in a command-line computing environment. Working in Linux or Unix will provide this experience.
- You should have experience with programming in a computer language such as C/C++, as well as in a scripting language such as Perl or Python.

There are a variety of other advanced skill sets that can add value to this background: molecular evolution and systematics; physical chemistry--kinetics, thermodynamics and statistical mechanics; statistics and probabilistic methods; database design and implementation; algorithm development; molecular biology laboratory methods; and others.

## *Why Should Biologists Use Computers?*

Computers are powerful devices for understanding any system that can be described in a mathematical way. As our understanding of biological processes has grown and deepened, it isn't surprising, then, that the disciplines of computational biology and, more recently, bioinformatics, have evolved from the intersection of classical biology, mathematics, and computer science.

### A New Approach to Data Collection

Biochemistry is often an anecdotal science. If you notice a disease or trait of interest, the imperative to understand it may drive the progress of research in that direction. Based on their interest in a particular biochemical process, biochemists have determined the sequence or structure or analyzed the expression characteristics of a single gene product at a time. Often this leads to a detailed understanding of one biochemical pathway or even one protein. How a pathway or protein interacts with other biological components can easily remain a mystery, due to lack of hands to do the work, or even because the need to do a particular experiment isn't communicated to other scientists effectively.

The Internet has changed how scientists share data and made it possible for one central warehouse of information to serve an entire research community. But more importantly, experimental technologies are rapidly advancing to the point at which it's possible to imagine systematically collecting all the data of a particular type in a central "factory" and then distributing it to researchers to be interpreted.

In the 1990s, the biology community embarked on an unprecedented project: sequencing all the DNA in the human genome. Even though a first draft of the human genome sequence has been completed, automated sequencers are still running around the clock, determining the entire sequences of genomes from various life forms that are commonly used for biological research. And we're still fine-tuning the data we've gathered about the human genome over the last 10 years. Immense strings of data, in which the locations of only a relatively few important genes are known, have been and still are being generated. Using image-processing techniques, maps of entire genomes can now be generated much more quickly than they could with chemical mapping techniques, but even with this technology, complete and detailed mapping of the genomic data that is now being produced may take years.

Recently, the techniques of x-ray crystallography have been refined to a degree that allows a complete set of crystallographic reflections for a protein to be obtained in minutes instead of hours or days. Automated analysis software allows structure determination to be completed in days or weeks, rather than in months. It has suddenly become possible to conceive of the same type of high-throughput approach to structure determination that the Human Genome Project takes to sequence determination. While crystallization of proteins is still the limiting step, it's likely that the number of protein structures available for study will increase by an order of magnitude within the next 5 to 10 years.

Parallel computing is a concept that has been around for a long time. Break a problem down into computationally tractable components, and instead of solving them one at a time, employ multiple processors to solve each subproblem simultaneously. The parallel approach is now making its way into experimental molecular biology with technologies such as the DNA microarray. Microarray technology allows researchers to conduct thousands of gene expression experiments simultaneously on a tiny chip. Miniaturized parallel experiments absolutely require computer support for data collection and analysis. They also require the electronic publication of data, because information in large datasets that may be tangential to the purpose of the data collector can be extremely interesting to someone else. Finding information by searching such databases can save scientists literally years of work at the lab bench.

The output of all these high-throughput experimental efforts can be shared only because of the development of the World Wide Web and the advances in communication and information transfer that the Web has made possible.

The increasing automation of experimental molecular biology and the application of information technology in the biological sciences have lead to a fundamental change in the way biological research is done. In addition to anecdotal research--locating and studying in detail a single gene at a time--we are now cataloguing all the data that is available, making complete maps to which we can later return and mark the points of interest. This is happening in the domains of sequence and structure, and has begun to be the approach to other types of data as well. The trend is toward storage of raw biological data of all types in public databases, with open access by the research community. Instead of doing preliminary research in the lab, scientists are going to the databases first to save time and resources.

## Why Use Unix or Linux?

Setting up your computer with a Linux operating system allows you to take advantage of cutting-edge scientific-research tools developed for Unix systems. As it has grown popular in the mass market, Linux has retained the power of Unix systems for developing, compiling, and running programs, networking, and managing jobs started by multiple users, while also providing the standard trimmings of a desktop PC, including word processors, graphics programs, and even visual programming tools. This book operates on the assumption that you're willing to learn how to work on a Unix system and that you'll be working on a machine that has Linux or another flavor of Unix installed. For many of the specific bioinformatics tools we discuss, Unix is the most practical choice.

On the other hand, Unix isn't necessarily the most practical choice for office productivity in a predominantly Mac or PC environment. The selection of available word processing and desktop publishing software and peripheral devices for Linux is improving as the popularity of the operating system increases. However, it can't (yet) go head-to-head with the consumer operating systems in these areas. Linux is no more difficult to maintain than a normal PC operating system, once you know how, but the skills needed and the problems you'll encounter will be new at first.

## *What Information and Software Are Available?*

Only a few years ago, biologists had to know how to do literature searches using printed indexes that led them to references in the appropriate technical journals. Modern biologists search web-based databases for the same information and have access to dozens of other information types as well. Knowing how to navigate these resources is a vital skill for every biologist, computational or not.

We then introduce the basic tools you'll need to locate databases, computer programs, and other resources on the Web, to transfer these resources to your computer, and to make them work once you get them there. In Chapters through we turn to particular types of scientific questions and the tools you will need to answer them. In some cases, there are computer programs that are becoming the standard for solving a particular type of problem (e.g., BLAST and FASTA for amino acid and nucleic acid sequence alignment). In other areas, where the method for solving a problem is still an open research question, there may be a number of competing tools, or there may be no tool that completely solves the problem.

## Why Do I Need to Install a Program from the Web?

Handling large volumes of complex data requires a systematic and automated approach. If you're searching a database for matches to one query, a web form will do the trick. But what if you want to search for matches to 10,000 queries, and then sort through the information you get back to find relationships in the results? You certainly don't want to type 10,000 queries into a web form, and you probably don't want your results to come back formatted to look nice on a web page. Shared public web servers are often slow, and using them to process large batches of data is impractical. *Chapter 12, Automating Data Analysis with Perl*, contains examples of how to use Perl as a driver to make your favorite program process large

## *How Do I Understand Sequence Alignment Data?*

It's hard to make sense of your data, or make a point, without visualization tools. The extraction of cross sections or subsets of complex multivariate data sets is often required to make sense of biological data. Storing your data in structured databases, which are discussed in Chapter 13, creates the infrastructure for analysis of complex data.

Once you've stored data in an accessible, flexible format, the next step is to extract what is important to you and visualize it. Whether you need to make a histogram of your data or display a molecular structure in three dimensions and watch it move in real time, there are visualization tools that can do what you want. *Chapter 14, Visualization and Data Mining*, covers data-analysis and data-visualization tools, from generic plotting packages to domain-specific programs for marking up biological sequence alignments, displaying molecular structures, creating phylogenetic trees, and a host of other purposes.

## *How Do I Write a Program to Align Two Biological Sequences?*

An important component of any kind of computational science is knowing when you need to write a program yourself and when you can use code someone else has written. The efficient programmer is a lazy programmer; she never wastes effort writing a program if someone else has already made a perfectly good program available. If you are looking to do something fairly routine, such as aligning two protein sequences, you can be sure that someone else has already written the program you need and that by searching you can probably even find some source code to look at. Similarly, many mathematical and statistical problems can be solved using standard code that is freely available in code libraries. Perl programmers make code that simplifies standard operations available in modules; there are many freely available modules that manage web-related processes, and there are projects underway to create standard modules for handling biological-sequence data.

## *How Do I Predict Protein Structure from Sequence?*

There are some questions we can't answer for you, and that's one of them; in fact, it's one of the biggest open research questions in computational biology. What we can and do give you are the tools to find information about such problems and others who are working on them, and even, with the proper inspiration, to develop approaches to answering them yourself. Bioinformatics, like any other science, doesn't always provide quick and easy answers to problems.

## *What Questions Can Bioinformatics Answer?*

The questions that drive (and fund) bioinformatics research are the same questions humans have been working away at in applied biology for the last few hundred years. How can we cure disease? How can we prevent infection? How can we produce enough food to feed all of humanity? Companies in the business of developing drugs, agricultural chemicals, hybrid plants, plastics and other petroleum derivatives, and biological approaches to environmental remediation, among others, are developing bioinformatics divisions and looking to bioinformatics to provide new targets and to help replace scarce natural resources.

The existence of genome projects implies our intention to use the data they generate. The implicit goals of modern molecular biology are, simply stated, to read the entire genomes of living things, to identify every gene, to match each gene with the protein it encodes, and to determine the structure and function of each protein. Detailed knowledge of gene sequence, protein structure and function, and gene expression patterns is expected to give us the ability to understand how life works at the highest possible resolution. Implicit in this is the ability to manipulate living things with precision and accuracy.

# Watching a Protein Fold

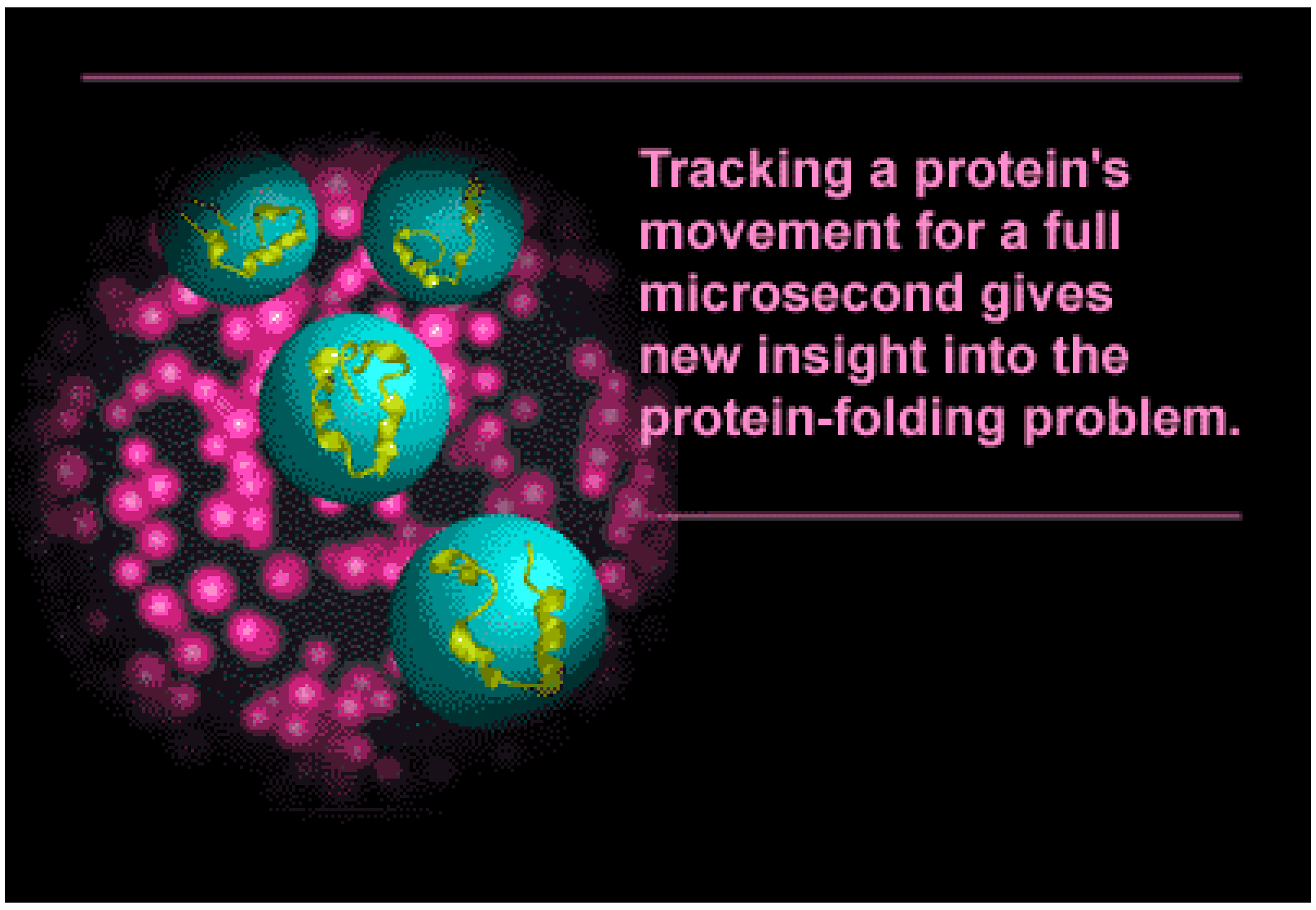

The protein-folding problem — one of the major challenges of molecular biology in the 1990s — could be thought of as a version of cryptography. Scientists like Peter Kollman are the code breakers, trying to uncover a set of rules somehow embedded in a protein's sequence of amino acids. This chemical alphabet of 20 characters, strung like beads on a chain along the peptide backbone of a newborn protein, carries a blueprint that specifies the protein's mature folded shape.

Within a matter of seconds — or less — after rolling off the protein assembly line (in the cellular ribosome), the stretched-out chain wraps into a bundle, with twists and turns, helices, sheets and other 3D features. For proteins, function follows from form — the grooves and crevices of its complex folds are what allow it to latch onto other molecules and carry out its biological role.

But what are the rules? How is it that a particular sequence of amino acids results in a particular folded shape? "The protein-folding problem is still the single most exciting problem in computational biochemistry," says Kollman, professor of pharmaceutical chemistry at the University of California, San Francisco, and a leader in using a research tool called "molecular dynamics," a method of computational simulation that tracks the minute shifts of a protein's structure over time. "To be able to predict the structure of the protein from just the amino-acid sequence would have tremendous impact in all of biotechnology and drug design."

In 1997, Kollman and his coworkers Yong Duan and Lu Wang used the CRAY T3D at Pittsburgh Supercomputing Center to develop molecular dynamics software that exploits parallel systems like the T3D and CRAY T3E much more effectively than before. Using this improved software on the T3D and, later, on a T3E at Cray Research, the researchers tracked the folding of a small protein in water for a full microsecond, 100 times longer in time than previous simulations. The result is a more complete view of how one protein folds — in effect, a look at what hasn't been seen before, and it offers precious new insight into the folding process.

## Exploiting Massive Parallelism

Simulating a millionth of a second of protein movement may sound less than impressive, until you realize that the longest prior simulations of similar proteins extended only 10 to 20 nanoseconds (billionths of a second). The limitation holding back this critical work has been the tremendous computational demand of the simulations, which must account for interactions between each atom in a protein and all the other atoms and surrounding water molecules. To capture protein movement at a useful level of detail, the full set of these interactions must be recalculated every femtosecond (a millionth of a nanosecond) of protein time. Even with the most advanced systems, it's a daunting, costly computational challenge.

The Kollman team's recent effort focused on a small, 36 amino-acid protein called the villin headpiece sub-domain. With surrounding water molecules, the computation involved about 12,000 atoms. To capture the first 200 nanoseconds of folding took 40 days of dedicated computing using all 256 processors of the T3D. With a 256-processor T3E, four times faster, the next 800 nanoseconds took about another two months.

This big leap in folding simulation was made possible by Duan's ability to exploit the parallelism of 256 processors running simultaneously. Using Pittsburgh's T3D, he devised and tested software-doctoring manipulations to the molecular dynamics part of AMBER (a widely used package for modeling proteins and DNA, developed by Kollman's research group). The changes boosted single-processor performance about 70% and greatly improved the "load balancing" and communication among processors, resulting overall in 256-processor performance six times faster than before.

"The real challenge for molecular dynamics," says Kollman, "is to effectively use massively parallel machines. If you have 256 processors, you have to divide the computation so each processor does the same amount of work, which is difficult because each particle has to keep track of every other." Juggling information back and forth between processors and memory involves a high communication overhead, which has seriously limited the usefulness of parallel processing for these calculations. "Virtually all other molecular dynamics codes in the literature level off at 40, 50 or 60 processors," says Kollman. "In other words, even if you use all the processors, you don't get any faster because the communication among processors is rate limiting."

Duan broke key parts of the task — distribution of updated position coordinates for the atoms and "force collection" for the atom-to-atom interactions — into small pieces that each processor does on an as-needed basis. In earlier versions of AMBER, each processor kept a complete set of coordinates and forces. He also implemented a "spatial decomposition" scheme, breaking the entire protein and water system into rectangular blocks, each block assigned to a processor. This reduces redundant communications and "latency," time required to open communication pathways between processors.

With these changes and others, the software showed significantly improved parallel "scaling" — it now runs 170 times faster on 256 processors than on one alone. "This was a tour-de-force of parallel programming," says Kollman, "and it wouldn't have been possible except for Pittsburgh making the T3D available to us."

"The dedicated T3D allowed me to conduct extensive tests on a variety of plausible schemes," says Duan, who cites training at a PSC parallel programming workshop and discussions with PSC scientist Michael Crowley as also being instrumental to his work on this project.

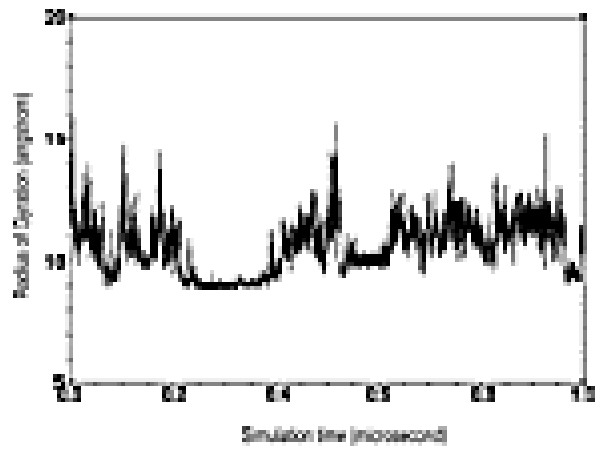

**Radius of Gyration over Time -** The Quiet Time for Protein Folding

What did the researchers learn from viewing a full simulated microsecond of protein folding? A burst of folding in the first 20 nanoseconds quickly collapses the unfolded structure, suggesting that initiation of folding for a small protein can occur within the first 100 nanoseconds. Over the first 200 nanoseconds, the protein moves back and forth between compact states and more unfolded forms. The researchers capture this behavior by plotting the protein's radius of gyration — how much the structure spreads out from its center — as a function of time. "If you look at those curves," notes Kollman, "they're very noisy — the structure is moving, wiggling and jiggling a lot."

The folded structures, often called "molten globules," have 3D features, such as partially formed helices loosely packed together, that bear resemblance to the final folded form. They are only marginally stable, notes Kollman, and unfold again before settling into other folded structures.

The next 800 nanoseconds reveal an intriguing "quiet period" in the folding. From about 250 nanoseconds until 400 nanoseconds the fluctuating movement back and forth between globules and unfolding virtually ceases. "For this period in the later part of the trajectory," says Kollman, "everything becomes quiet. And that's where the structure gets closest to the native state. It's quite happy there for awhile, then it eventually drifts off again for the rest of the period out to a microsecond."

For Kollman, this behavior suggests that folding may be characterized as a searching process. "It's a tantalizing idea — that the mechanism of protein folding is to bounce around until it finds something close, and stay there for a period, but if it isn't good enough it eventually leaves and keeps searching. It might have 10 of these quiet periods before it arrives at a period where enough of the amino-acid sidechains are in a good enough environment that it locks into the final structure."

Although only a partial glimpse — even the fastest proteins need 10 to 100 microseconds to fully fold — these results represent a major step forward in protein-folding simulation. New experimental methods are also providing more detailed looks at the process, offering the possibility of direct comparison between experiment and simulation, which will further advance understanding.

The challenge of the protein-folding problem is to have the ability to predict protein structure more accurately. For the pharmaceutical industry, this holds the prospects of greatly reducing the cost and expense of developing new therapeutic drugs. Recent research, furthermore, suggests that certain diseases, such as "mad cow disease" and possibly Alzheimer's can be understood as malfunctions in protein folding.

Ultimately, for Kollman and others using molecular dynamics, the goal is to follow the entire folding process. With the promise of more powerful computing and higher level parallelism, Kollman sees that goal as within reach. "We're getting some new insights that weren't available before because the simulations weren't in the right time-scale. Being able to visualize the folding process of even a small protein in a realistic environment has been a goal of many researchers. We believe our work marks the beginning of a new era of the active participation of full-scale simulations in helping to understand the mechanism of protein folding."

*CONFIDENTIAL GOVERNMENT DOCUMENT*

**NOT FOR GENERAL CIRCULATION**

# BIOINFORMATICS POLICY OF INDIA (BPI – 2004)

*"Long years ago we had made a tryst with destiny… and now the time comes when we shall redeem our pledge…not wholly or in full measure, but very substantially…….."*

*Jawaharlal Nehru*

सत्यमेव जयते

# NATIONAL BIOINFORMATICS POLICY

## 1. OVERVIEW

Growth of biotechnology has accelerated particularly during the last decade due to accumulation of vast sequence and structure information as a result of sequencing of genomes and solving of crystal structures. This, coupled with advances in information technology, has made biotechnology increasingly dependent on computationally intensive approaches. This has led to the emergence of a super- speciality discipline, called bioinformatics.

Bioinformatics has become a frontline applied science and is of vital importance to the study of new biology, which is widely recognised as the defining scientific endeavour of the twenty-first century. The genomic revolution has underscored the central role of bioinformatics in understanding the very basics of life processes.

India's predominantly agrarian economy, the vast biodiversity and ethnically diverse population makes biotechnology a crucial determinant in achieving national development. As India's population crossed one billion figure, the country is faced with newer challenges of **conservation of biodiversity to ensure** food security, healthcare, **tackling bio-piracy and safe guarding IPR of Plant Genetic Resources (PGR) and associated knowledge systems,** environment protection and education. The liberalisation and globalisation of the economy pose further challenge to society and the government to modernise and respond to the increasingly competitive international environment. As rapid technological advancements and innovation impact the basic activities like agriculture, industry, environment and services, the country has to evolve programmes that would aid in economic development driven by science and technology. It is therefore of utmost importance that India participates in and contributes to the ensuing global bioinformatics revolution.

In recognition of its importance, the Department of Biotechnology, Government of India has identified bioinformatics as an area of high priority during the tenth plan period in order to ensure that this sector attains levels demanded in the international arena. This can be achieved through organisational and functional restructuring; integration and optimal utilisation of the available resources; planned expansion based on actual market demand; increasing autonomy of the system; transfer of technology from laboratory to the industry; sustainable development of human resources; and finally, enhancing accountability of the participating institutions.

Beginning early last decade, India has endeavoured to create an infrastructure that would enable it to harness biotechnology through the application of bioinformatics. The government took a major step forward in establishing a national grid of bioinformatics centres as the Biotechnology Information System Network **(BTISNet)**. The network has presently grown to sixty one centres covering all parts of the country. While this is an appreciable achievement, much more needs to be done to move forward. It is in this context that this policy paper is being drawn up.

The **Bioinformatics Policy of India (BPI–2004)** has been formulated against a backdrop of our experience; building up on the successes and learning from the shortcomings. The primary objective is to make India competitive in the changing global scenario. Over the last fifteen years, bioinformatics, because of the concerted efforts of DBT, has made significant progress particularly in the areas of infrastructure development, orientation of the scientific workforce to use computational approaches and providing training to practising scientists. At the same time, there have been serious pitfalls as to the goal-oriented approach of the programme, integration of resources and generating manpower to suit the growing national and international market. The policy paper is envisaged to provide a framework for the national strategies to be adopted over the coming years to promote growth of bioinformatics in India and to encourage its application.

# 2. PROGRAMME OBJECTIVES

The principal aim of the bioinformatics programme was to ensure that India emerged a key international player in the field of bioinformatics; enabling a greater access to information wealth created during the post-genomic era and catalysing the country's attainment of lead position in medical, agricultural, animal and environmental biotechnology. India should make a niche in Bioinformatics industry and would work to create bioinformatics industry with US$10 billion by the end of 10$^{th}$ Plan period. It was felt that these could be achieved through a focussed approach in terms of information acquisition, storage, retrieval and distribution.

The Department of Biotechnology had adopted the following strategies to achieve these objectives:

- Develop the programme as an array of distributed resource repositories in areas of specialisation pertinent to the needs of India's economic development.
- Coordination of the network through an Apex Secretariat

- System design and implementation in terms of **computing and** communication infrastructure, bioinformatics utilities etc.
- Facilitate and **enhance** application of bioinformatics
- Support and **promote organisation of** long-term, short-term and continued training/education in Bioinformatics. National level testing for quality assurance on human resource on Bioinformatics shall also be conducted through reputed universities.
- Establish linkages with international resources in biotechnology information
- An international institute on Bioinformatics shall also be established to promote various activities on bioinformatics particularly international and entrepreneurial participation in these activities.

# 3. CRITICAL EVALUATION OF THE PROGRAMME

### *3.1 Achievements*

- *Infrastructure Development:-* The programme has over the years developed a significant infrastructure through computer hardware, software, communication facilities and ancillary facilities such as mirror sites and interactive graphics capabilities. The present infrastructure is well placed for higher intellectual development.
- *Orientation of the Scientific Community to Computational Approaches: -* The programme has succeeded in orienting the scientific community to increasingly use computational approaches in solving biological problems.
- *Training:* - There has been a significant success in training scientists and researchers in diverse application of bioinformatics and computational biology.

### *3.2 Pitfalls*

- *Inadequate Focus:-* Despite the establishment of the programme as a coordinated activity, there is a lack of focus in programmes of the different centres that constitute the network. Activities undertaken under the scheme are more in tune with the individual specialisation and thrust of the centres. This is partly due to absence of a clearly defined framework for operation and partly due to the non-homogenous pattern of setting up of the centres. There is need for a more central theme around which the entire network should function.

- *Lack of Resource Integration:-* **A considerable amount of** resources exist within the network; ranging from databases, software, educational packages to human resources. Nevertheless, the resource remains largely fragmentary without proper access provisions. This has not only led to the high under-utilisation of the resource, but has led to an unacceptable amount of redundancy thereby increasing the financial burden.
- *Inadequate Demand-Supply Forecasts:-* Development of the bioinformatics initiatives has been largely chance based and dependant on the individual investigator's specialisation and liking. There has been no attempt to undertake adequate techno-market surveys and appropriate opinion polls involving academia, industries and the people for determining the programmes to be undertaken.
- *Inadequate Control Mechanisms:-* The present system of centrally administered network through an apex secretariat has resulted in a dilution of accountability of the constituent centres. It has also become difficult for the secretariat to undertake close monitoring and management of the network given the large geographical distances among the centres.

# 4. BENEFICIARIES

- Agriculture, Health and Environment Sector
- Industry
- **National resources: Capacity building in conservation and sustainable utilization (bio prospecting) of biodiversity including protection of IPR and prevention of biopiracy.**
- National resources in higher education

# 5. STRATEGIES FOR THE FUTURE

DEFINED GOALS

- Generate a national level resource strength in bioinformatics
- Promote R& D programmes **in bioinformatics**
- Promote entrepreneurial development
- Globalisation of the national bioinformatics initiatives
- Encourage development of quality human resources
- Restructuring the BTIS network for optimised performance

# *POLICY RECOMMENDATIONS*

This policy formulation suggests the following programmes for fulfilling the above-mentioned objectives.

**5.1 Generate National Level Resource Strength in Bioinformatics:**

***5.1.1 Improvement of Bioinformatics Capabilities of India***

Towards development of strong bioinformatics and bio-computational capabilities the programme should focus heavily upon the following activities:

- ***Modernisation of infrastructure: Internet bandwidth for resource sharing:*** All the centres of the network should have access to the Internet and should have sufficient bandwidth as demanded by the applications. The Internet connectivity might be in the form of leased lines so as to optimise cost constraints.
- *Uninterrupted Network Access:* The computer infrastructure of the centre should preferably be arranged into a network so as to optimise usage and internal sharing of information and resources.
- ***Popularise Use/ Access to Public Domain Utilities:*** Presently, a large number of public domain utilities are availables for bioinformatics applications. The centres are to be encouraged to use these utilities rather than acquiring costly packages for this purpose. The apex body should ensure that the centre has the necessary system configuration to use the concerned packages.

- ***Enhanced Solutions/ tools Development Capabilities:*** Successful implementation of an operational bioinformatics programme requires adequate emphasis on solution development. As such, activities such as development of databases and software and management of databases constitute important components of the programme. The centres within the network should be encouraged to develop these required capabilities indigenously.
- *Research*: The centres of the network need to be encouraged to develop and inculcate a research culture and undertake active research in frontier areas of bioinformatics and computational biology.
- ***Human Resource Development:*** Lack of adequate trained manpower is a major bottleneck for the bioinformatics initiatives. The centres of the network would be encouraged to take steps that would result in generation of trained manpower in bioinformatics. These training may be conducted either at the entry level or at on-job level. To meet the requirement of qualified manpower at least 20 M.Tech. and 10 M.Sc. courses to be introduced at various prominent institutions.

#### *5.1.2 Resource Integration*

Being a distributed network and knowledge resource, it is expected that the total knowledge component developed out of the programme be adequately utilised for the benefit of the country. To this end, all resources constituting the **BTISNet** should be considered as National Resources in Higher Education and should be made available to all members of the scientific community of India. In this regard, the following integration protocol is proposed:

- The resources of the **BTISNet** should be brought under a common Internet portal maintained at the Apex Centre. The portal shall consist, apart from factual data, utility services of the sort that are required for research, teaching and practice in bioinformatics.
- All members of the network should have free and unhindered access to this portal and would be free to use the utility packages contained therein.
- It should be mandatory for all Bioinformatics Centres to have their own URL, which will be used as the platform for integrating the resources as stated above.

- The above resource sharing notwithstanding, the Apex Secretariat of the BTIS should take adequate measures to ensure safeguard of strategic components of the information such as biodiversity data, software and so on in order to prevent usage that would be detrimental to the national interest.

### 5.2 Promote R& D Strengths

#### *5.2.1 Provision of Extramural Support for Bioinformatics Research Projects*

Completion of the genome projects and progress in structure elucidation has opened a new vista for downstream research in bioinformatics. These studies range from modelling of cellular function, metabolic pathways, validation of drug targets to understanding gene function in health and disease. Currently, a number of scientists throughout the country – both within the BTIS programme and outside – are involved in *in-silico* studies of biological processes. The BTIS programme should encourage more such studies. Extramural support to bioinformatics research projects should include:

- Funding of bioinformatics research projects
- Provide financial support for resource creation and maintenance
- Enable access to bioinformatics utilities required for research
- Provide financial support for travel/training of scientists

#### *5.2.2 Generation of a Bioinformatics Research Support System*

Downstream research in bioinformatics relies heavily on the availability of curetted secondary and tertiary databases and knowledge resources. In this regard, the BTIS programme should undertake the process of acquiring and mining useful information from primary databases and compiling secondary, tertiary and quaternary databases covering specialised areas of biology. Such databases and knowledge repositories, would serve as research support systems for augmenting the bioinformatics research activities.

#### *5.2.3 Adequate Cover to Intellectual Property*

In its attempts to promote R&D activities, the BTIS programme should take adequate measures for protection of intellectual property generated out of the projects.

## 5.3 Promote Entrepreneurial Development

### *5.3.1 Industry Participation and Building Academia-Industry Interface*

Transformation of a knowledge resource to economic development is dependent on the rate at which the technology developed is absorbed by the industry. This push-pull synergy between technology and market is the hallmark for sustainability of any technologically intensive programme. Synchronisation of this approach requires progressive yet sustained liaison with the industry. Fostering a closer cooperation with the industry and building a meaningful academia-industry interface could optimise the success of the bioinformatics. The growth of a sustained academia-industry relation might be implemented through the following processes:

- **Understanding the industry environment and market environment through techno-economic market surveys**
- **Inventorying the transferable technologies available within the BTIS network and making them available on the public domain**
- **Developing an Industry-Institution Partnership programme at national, regional and international level.**
- **Promoting the growth of business incubators in the field of bioinformatics**
- **Provision for industrial and entrepreneurial consultancy services**
- **Encourage enterprise creation through liberalised flow of foreign capital, outsourcing, infrastructure generation etc.**
- **Adequate emphasis on human resource development that suits the requirement of the industries.**

Apart from these adequate importance should be given to the promotion and protection of intellectual property rights.

The Industry and academia cooperation can be of two ways- (i) The academic institutions can outsource the expenditure for the finishing and packaging of any databases & software development for which the academic institutions should retain the copy right/ patent, and (ii) The collaboration can be for the entire project in which the academic institutions and the industry shall share the copy right/patent.

### *5.3.2 Exploration of Marketing Channels for the Bioinformatics Products & Services*

Technologically intensive products and services have significant business potential. However, it requires intensive marketing efforts through positioning, promotion and distribution. Among these, marketing channels seems to hold the key to the success of the endeavour. The current state-of-art of the bioinformatics resources generated out of the BTIS programme indicates existence of a number of potentially marketable products and services in the form of software, databases, knowledge bases and human resources. The BTIS programme should undertake the task of identifying the proper marketing channels, and align these channels with the programme mandate on one hand and market trends on the other. Exploration of marketing channels might be undertaken by the following proposed mechanisms:

- ***Development of Web Directory:-*** **Maintaining an inventory of development of web directory of relevant technical products in different institutes, universities and industries and their manner of acquiring and transfer. This should have controlled access.**
- *Establishing Liaison with Agencies:-* Identification of marketing agencies who might be entrusted the channel function
- ***Formulation of a Pricing Strategy:-*** **Formulation of** a realistic pricing strategy of the products and monitoring the same keeping in view the market trends.

### *5.3.3 Capital Funding for the Public-Private partnership in Bioinformatics*

The bioinformatics industry being a high technology area requires venture capital funding for active participation of private & public sector organisation. There is a strong need to create such a fund. A separate centre shall be established for the promotion of public-private partnership with sufficient autonomy in decision making.

## 5.4 Globalisation the National Bioinformatics Initiatives

Globalisation refers to the process that enables integration of the national economy with the world economy. In technologically intensive sectors, this is achieved by free flow of technology across state boundaries and mutual resource sharing that caters to the interest of the international community. The Indian bioinformatics programme is likely to become more competitive through technology diffusion, assistance in capital formation and innovation from its overseas brethrens.

### *5.4.1 International Collaboration on Resource Sharing*

The global bioinformatics initiatives are progressively focussing upon collaborative endeavours. Examples include the International Sequence Database Collaboration among the NCBI, USA; EBI, UK and DDBJ, Japan. Other international collaborative groups also exist like the International Consortium for Collaboration in Bioinformatics [ICCBnet], International Centre for Genetic Engineering & Biotechnology [ICGEBnet], Asia-Pacific Bioinformatics Network [APBioNet] and so on. The functioning and operation of the BTIS network would be greatly optimised and enhanced if there is a closer link with the major international bioinformatics initiatives. In this regard, the following measures might be adopted:

- ***Establishment of Mirror Sites of International Servers**:-* Establishing mirror sites of major international bioinformatics servers not only enable the Indian scientific community to access these resources, but this would promote greater cooperation with the global community.
- *Establishing Nodes of International Bioinformatics Initiatives:-* The centres of the BTISnet might be encouraged to become national nodes of noteworthy international initiatives so as to become a part of a bigger network. Notable examples include initiatives like the S*.org of the National University of Singapore, **ILDIS, U.K.**
- *Establishment of Virtual Libraries of Bioinformatics:-* Digitising the existing collection of library resources in bioinformatics present in the different BTIS centres, should be encouraged to enable the centres to form a part of international virtual library networks. This would enable resource sharing and sharing of bibliographic and other information at an enhanced rate.
- *National Genome Net and Trace Archives:-* The BTISnet servers should undertake the task of compiling the sequence and structure information generated *de novo* within the country and share it on a mutual resource sharing protocol with other countries of the world. Apart from this, the BTIS centres should maintain a curetted archive of the major genome information for facilitating development of secondary/tertiary databases and downstream research. The entire genome repository generated in the process, should be integrated into a single-window platform through a National Genome Information Network.

### *5.4.2 International Participation in Bioinformatics Capital Formation*

In the current international collaboration trends of the bioinformatics programme, the Indian counterpart is essentially at the level of a user. The basic resource base, utility platform and the intellectual property rest with the foreign component of the collaboration. Minimal effort has till date been given to involve the developed countries to take part in generation of bioinformatics capital resources in the country.

The liberalised economy has opened a new vista in the field of foreign capital inflow and foreign technology agreements. The programme should utilise this regime judiciously to strengthen the country's bioinformatics infrastructure. Apart from accelerating the present pace of progress of the programme, it would also enable job creation, resource generation, and output maximisation and provide the BTIS with a competitive edge over other countries.

International participation in bioinformatics capital formation might be undertaken in the following proposed ways:

- To allow foreign institutional investors, MNCs and academic establishments to invest in any chosen Bioinformatics Centre with a ceiling of 51% equity.
- To allow foreign pharmaceutical companies to outsource from the Bioinformatics Centres in all areas other than those of strategic relevance
- To encourage major international bioinformatics institutions like the NCBI, EBI, DDBJ, SIB etc to open offshore centres in India using the infrastructure of the BTIS centres.
- To allow leading technology companies to invest in technology upgradation schemes with the Bioinformatics Centres.

### *5.4.3 Establishment of International Institute on Bioinformatics:*

In the recent past, almost every government agencies working on science has shown keen interest in Bioinformatics and there is a need to establish an apex body or institute where sufficiently large number of people can interact with each other towards promoting bioinformatics activities such as R&D, service and education. Similar to NCBI in USA, EBI in Cambridge, Singapore National University and South African National Bioinformatics Institute (SANBI), India needs to establish an international institute to support the growth of bioinformatics. This institute shall evolve criteria for quality training and evaluate bioinformatics training programmes offered by various organisations in the country. Besides these activities the institute shall promote international cooperation in this area.

### *5.4.4 National Eligibility Test on Bioinformatics (NETBI) for certifying quality human resource in Bioinformatics:*

In India several government and non-government agencies including small private organisations are conducting bioinformatics courses in different levels starting from graduation to post graduation levels. However, the students those are passing out through these courses are unable to attract job opportunities because of the lowest standard of teaching programmes. In order to select high quality human resources in Bioinformatics the department shall conduct a national eligibility test (NETBI) for certification of quality human resource in bioinformatics. On the basis of performance on such tests some fellowships may be awarded for pursuing higher studies such as M.Tech., Ph.D. in bioinformatics.

### 5.5 Encourage Human Resource Development, Education, & Awareness

The dearth of adequate trained manpower in bioinformatics is a major global problem. The case is no different for India. The programme should lay adequate emphasis on manpower development to address this problem. These should be long-term, short-term and continued training programmes. More M.Tech. & Ph.D. courses shall be introduced at various institutions.

### *5.5.1 Short-term Training*

As in the past, the programme should continue organising workshops, seminars and symposia all over the year covering as many numbers of centres as possible. Adequate emphasis should be given to understanding theoretical principles as well as hands-on experience during the training. The subject area of the training modules should be such that it covers the frontier areas of bioinformatics and computational biology.

### *5.5.2 Long-term Training*

The long term training programmes should be tailored to generate domain knowledge in the field of bioinformatics. These should include Diploma level courses, similar to the ones already in operation, as well as Degree courses at the level of M.Sc., M.Tech. and Ph.D. The individuals trained up in these schemes would develop the expertise to perform as group leaders in bioinformatics.

### *5.5.3 Continued Education in Bioinformatics*

Bioinformatics is a fast growing discipline and therefore, the domain knowledge gathered by an individual, tends to become obsolete within a very short period of time. As such there is an acute requirement of continuous training of the scientists, teachers and researchers in the newly emerging areas and concepts. In view of this, the BTIS programme should take necessary measures to provide continued education to the practicing bioinformaticians of the country. This may be undertaken in the following proposed ways:

- Organising summer and winter schools in bioinformatics
- Organising refresher courses
- Institution of Travel Fellowships for training abroad

### *5.5.4 Awareness of Bioinformatics*

- **Provide opportunity for web-based learning through BTIS portal**
- **Publication of news updates**
- **Content creation in bioinformatics e.g. through generation of multimedia and e-learning packages.**
- **Generation of a compendium of experts in bioinformatics, updated regularly.**

## 5.6 Restructuring of the BTIS Organisation for Optimised Performance

### *5.6.1 Enhanced Autonomy of the Apex Centre*

The Apex Bioinformatics Centre should be thoroughly restructured and reoriented to develop it as a Product Management Centre of the bioinformatics programme. It is suggested that the centre be made independent and declared an autonomous institution under the Department of Biotechnology, specialising in bioinformatics research, application and management.

### 5.6.2 Compilation of Work Areas for the Programme

The Apex Centre should compile a list of work areas for general guidance of the centres, keeping in view the national requirements, expertise available, funds available and basic objectives of the Government of India. These would include:

- **Identification of high priority research areas that need to be addressed.**
- **Identification of core interest areas of the Bioinformatics programme.**
- **Identification of high priority databases and software that need to be developed.**
- **Standardisation of course curriculum for bioinformatics education.**
- **Monitoring, analysing and publishing the market demand of different bioinformatics applications (tools) and utilities**

### 5.6.3 Generation of Work Groups

The Apex Centre, based on the reports of the individual centres, should evolve small functional work groups and foster closer interaction within these individual groups. Each of the work groups should be under the supervision of Group Coordinators, who would oversee the overall functioning of the group. On the basis of the current trends of the network, suggested work groups include:

- Medical Science
- Commercial Biotechnology & Intellectual Property Management
- Computational Biology & Algorithms
- Biodiversity & Environment
- Plant Science, Agriculture & Veterinary Science
- Molecular Biology, Cell Biology & Structural Biology

5.6.4 Clustering of National Computational Resources

In various organisations the high end computer system are not fully utilised and there is a need to bring such computational resources on national Biogrid. Similar to the super computing facility at IIT, Delhi three more facility on regional basis shall be established and networked with the Biogrid. Currently the country is in need of a minimum of 10 teraflops of computational power.

## *5.6.5 Restructuring the BTIS into a Product Organisation*

The BTISnet should be converted into a product organisation with a common mandate and a common minimum programme. Functional modulation should be carried out so as to make optimal use of the individual areas of specialisation of the constituent centres. This can be achieved through re-classification of centres based on functional areas. Thus the traditional classification as DICs and sub-DICs might be replaced by more realistic classifications based on actual function. Suggested functional categories might include:

- Information **Dissemination/** Extension Centres
- Solution Developer Centres
- Research & Development Centres
- Education and Human Resource Development Centres

# 6. CONCLUSIONS

The chief feature of this policy proposal is to **facilitate** a paradigm shift for the **DBT's**/ bioinformatics programme from infrastructure generation to resource building **at national, regional and international level.** In this regard, the functional protocol might be redefined as follows:

- To **focus** on resource building in bioinformatics using the infrastructure already generated

- To ensure balanced and integrated development in terms of information dissemination, application development, R&D **in Bioinformatics.**
- Leadership quality human resource development shall be strengthen by introducing more post graduate/ doctoral programme in bioinformatics.
- Venture capital funding shall be created to provide public-private partnership in bioinformatics.
- Suitable incentive applicable for the industries shall be extended similar to the IT sector for their participation in bioinformatics.
- To encourage and enhance international exchanges/ partnership with global organizations.
- All the above participations/ investments etc. should be implemented after getting a report of Steering Committee on any investment keeping a watch on national interest.
- Furthermore, the qualifications in Bioinformatics awarded by various agencies should be recognised through the National Eligibility Test in Bioinformatics(NETBI).

Implementation of these policy recommendations with adequate follow-up might result in the BTIS being developed into a National Virtual Centre for Bioinformatics, making use of the power of distributed resources and individual expertise in the right combination.

*****

# Biocomputing in a Nutshell

## A Short View onto the Development of Biology

The success of modern molecular biology might be considered a cartesian dream. Reductionism, Rene Descartes' belief of understanding complex phenomena by reducing them to their constituent parts - despite all its limitations - has turned out to be a home run in molecular biology.

The developments in modern biology have their roots in the interdisciplinary work of scientists from many fields. This was a crucial element in the breaking of the code of life; Max Delbrück, Francis Crick and Maurice Wilkins all had backgrounds in physics. In fact, it was the physicist Erwin Schrödinger (ever heard about Schrödinger's cat ?), who in "What is life" was the first to suggest that the "gene" could be viewed as an information carrier whose physical structure corresponds to a succession of elements in a hereditary code script. This later turned out to be the DNA, one of the two types of molecules "on which life is built".

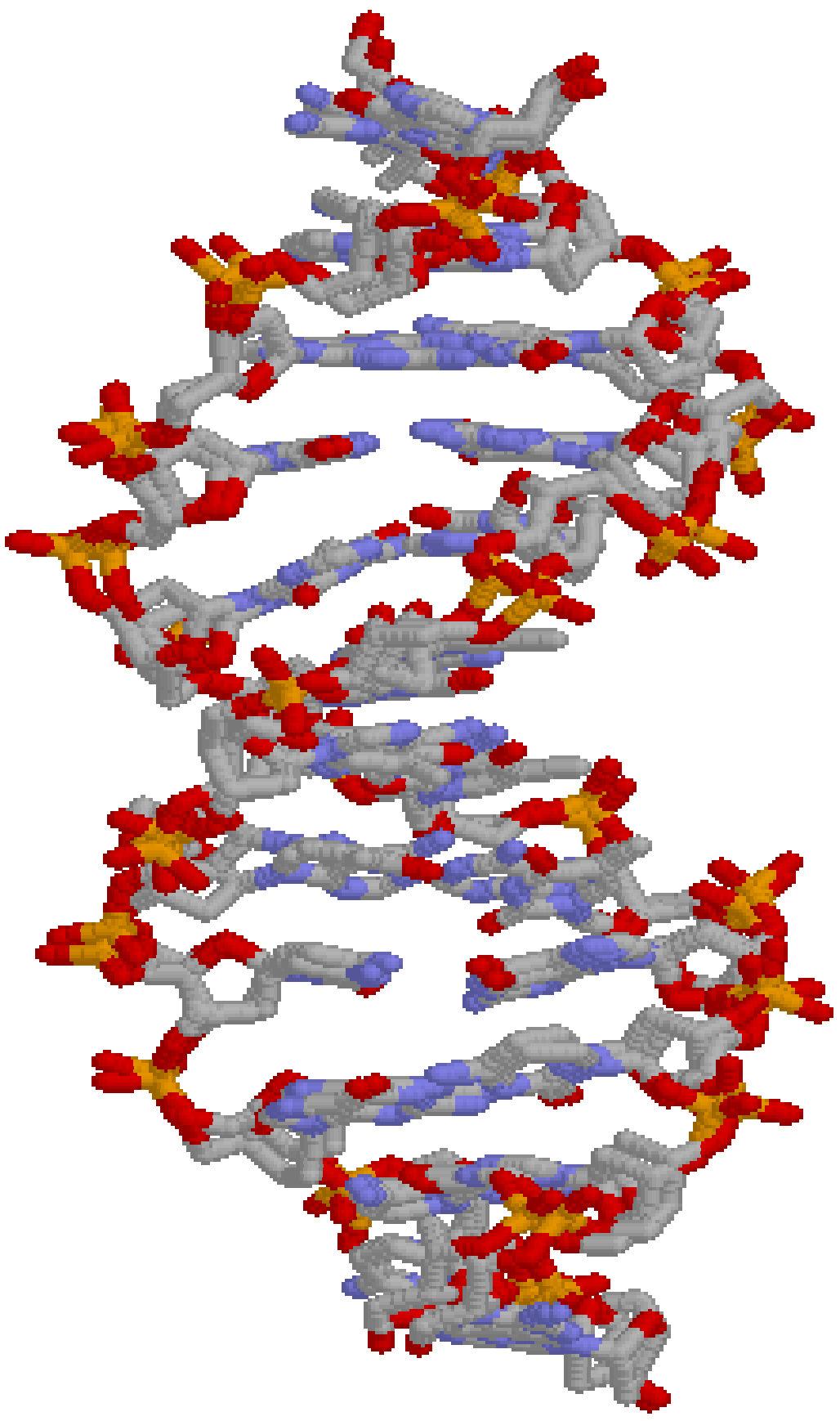

*The 3-dimensional structure of the DNA.*

Linus Pauling, the chemist, vitamin C-ist and anti atom-bombist determined the structure of the other type of molecule, the protein molecule - that is chains made up of things called amino acids.

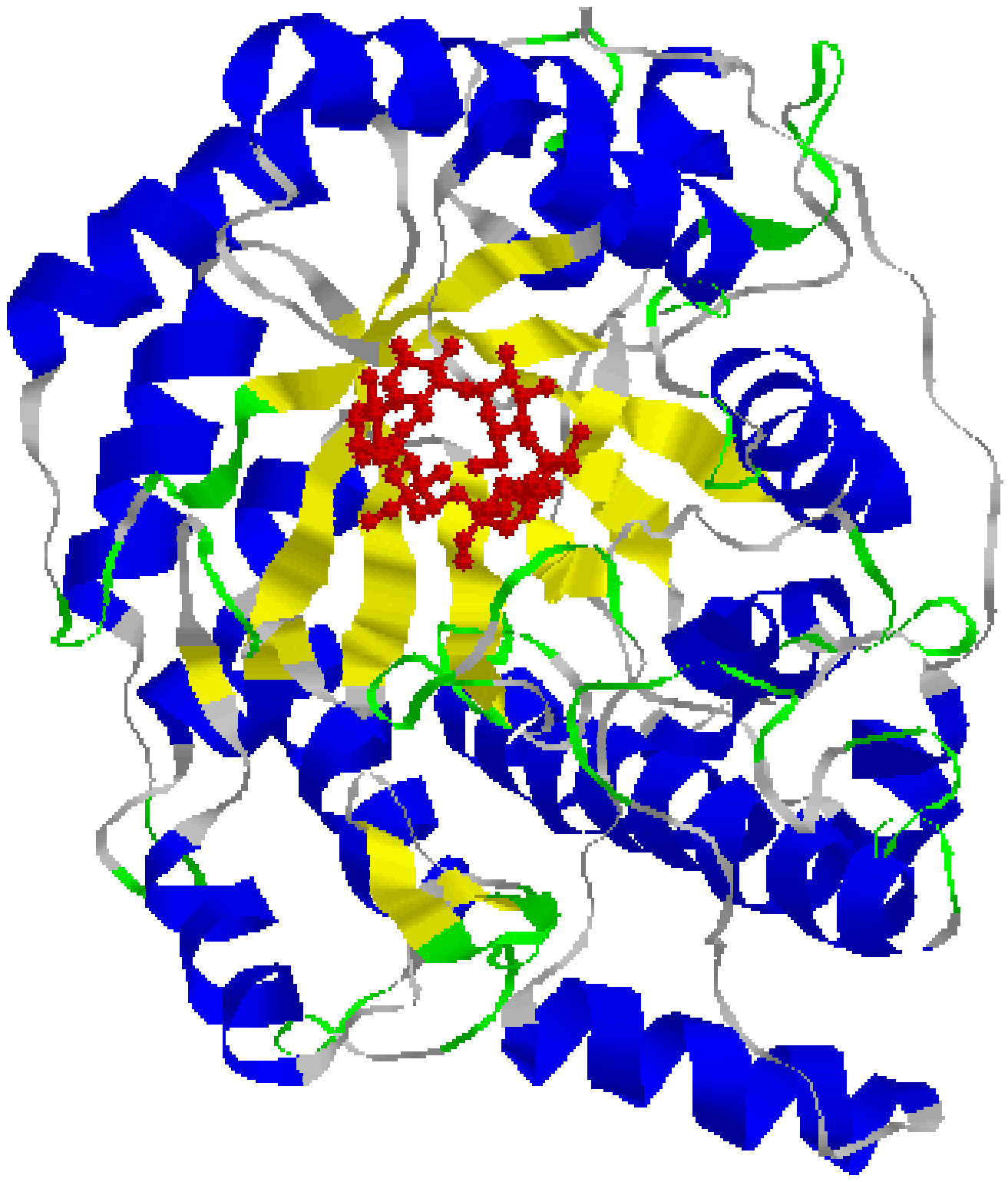

*The 3-dimensional structure of a protein, Beta-amylase. The main structural units of the protein, which are made up of just a few amino acids each, are differently coloured.*

This work inspired James Watson and Francis Crick in 1953 to elucidate the structure of DNA - the ABC of all known living matter. To cut a long story short over the next years many people pieced the puzzle together: The building blocks of life are the 20 amino acids that make up proteins; DNA contains the blueprints for these structures in its own structure. It is a long strand made of 4 nucleotides - this is the code of life. It goes *ACGTTCCTCCCGGGCTCC*, and so on, and so on, and so on. If you know the code you know the structure of all living things, at least in theory.

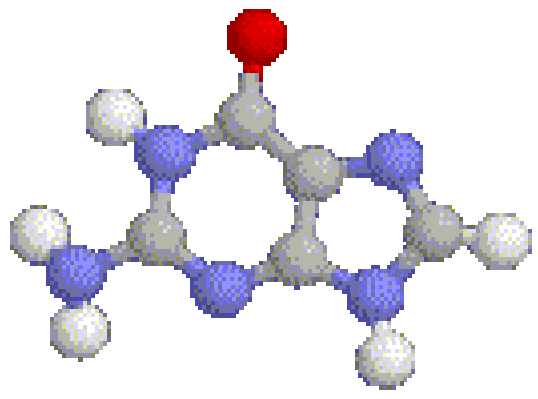

*An animation of Guanine (G), one of the 4 standard nucleotide bases. The colored balls represent the atoms from which it is made. Similar ball-and-stick models can be constructed for the 20 amino acids. (Click here if you'd like to `animate' the Guanine.)*

Here is a summary of the relationship between DNA and protein:

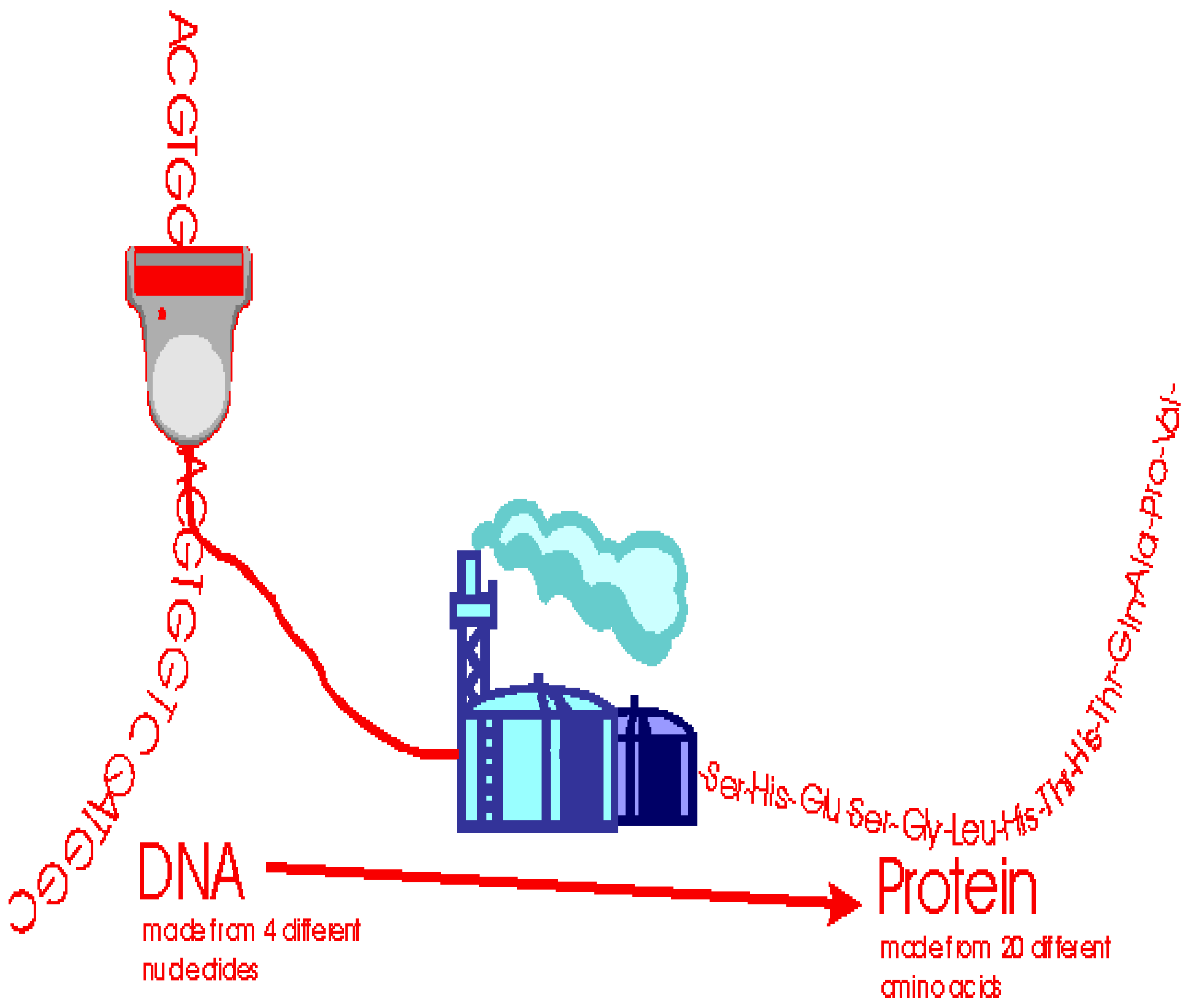

**An Enormous Flood of Data**

Restless technology has produced means of reading genes (DNA) almost like bar - code. The problem is that life is a complicated business, and therefore the code to describe even the smallest of God's creatures would fill many books. But scientists are very ambitious people and do lots of over-time. They have started to decode "themselves" in the Human Genome Project - HUGO for short. In fact, a sort of "average" human is decoded sampling DNA from unknown donors. But the difference in DNA between any human, and another one (or a scientist...) is almost null. Nevertheless, an average human scientist is made up of about 2.9 billion (2.9*109) nucleotides !

This orgy of reductionism presents problems which only big brother can solve: How do I store all this information in a form which is universally accessible and retrievable? What started as a cartesian dream is turning out to Bill Gates' satisfaction: Computers are needed !
Vast computer data banks accessible to you and me store this vast quantity of information. There are a lot of different data banks where DNA and protein sequence information are stored. Three examples are listed in the table below.

| **Name of data bank** | **Type of sequences stored** | **Number of sequences (1996)** |
|---|---|---|
| EMBL / GENBANK | Nucleotide sequences | 827174 |
| SWISSPROT | Protein sequences | 52205 |
| PDB | Protein structures | 4525 |

The growth of one typical data bank is shown in below, the increasing number of sequences in the SWISSPROT data bank as time goes by.

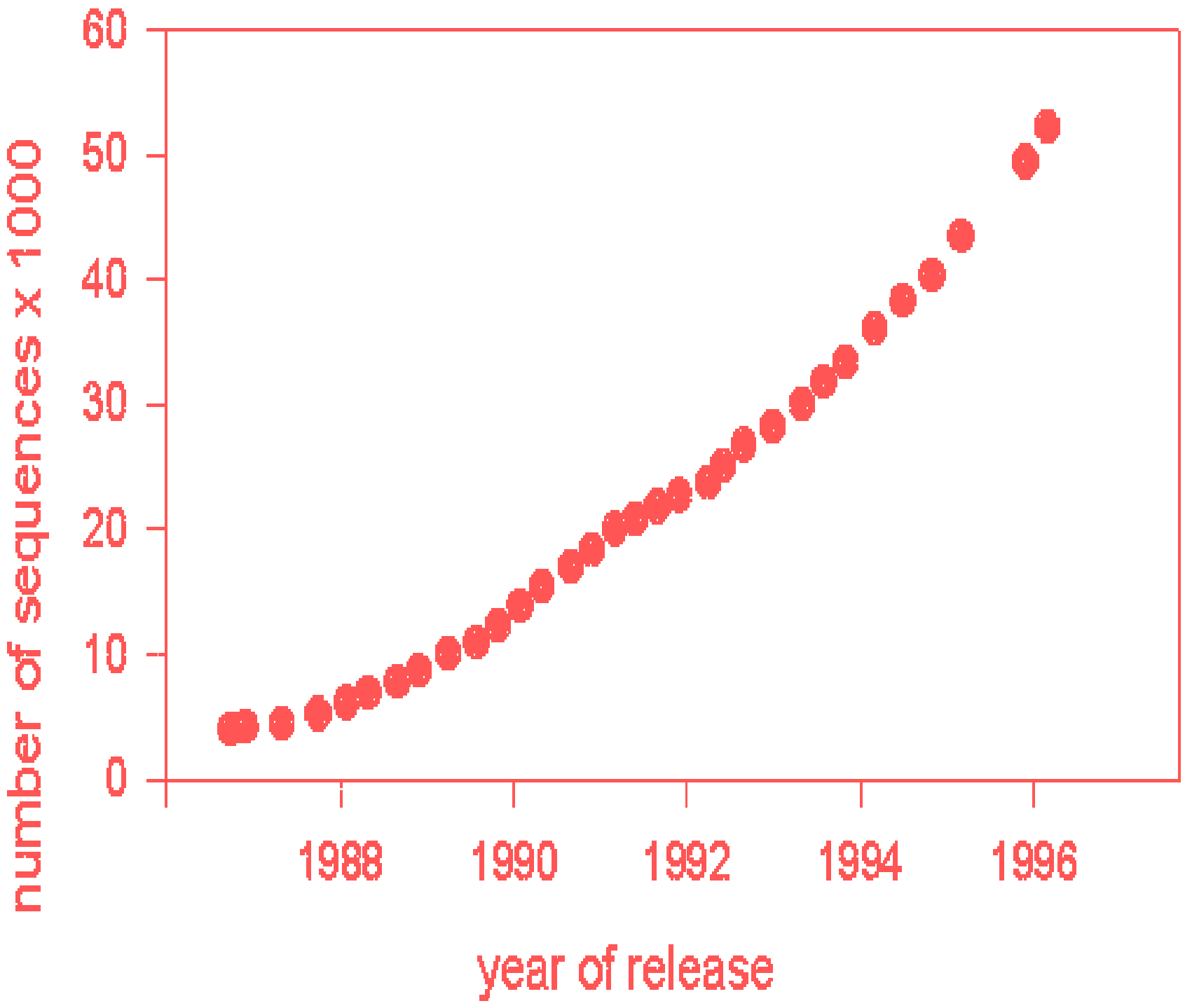

*Growth of the SWISSPROT data bank.*

## How can we analyze the Flood of Data?

An advantage of these data banks is their flexibility. All this information can be ordered and combined according to different patterns and tell us an awful lot. The motto goes: don't just store it, analyze it! By comparing sequences, one can find out about things like

- ancestors of organisms
- phylogenetic trees
- protein structures
- protein function

**Phylogenetic trees** are genealogical trees which are built up with information gained from the comparison of the amino acid sequences of a protein like cytochrome C, sampled from different species. Proteins like Beta-amylase or Hemoglobin cannot be chosen to get the "full picture", that is the full tree, because they don't occur throughout the living matter. Due to Darwinian Evolution, the protein has a slightly different amino acid sequence for each of the species. One phylogenetic tree was created for instance with the sequences of cytochrome C from several plants, animals and fungi. Below, part of this phylogenetic tree is shown.

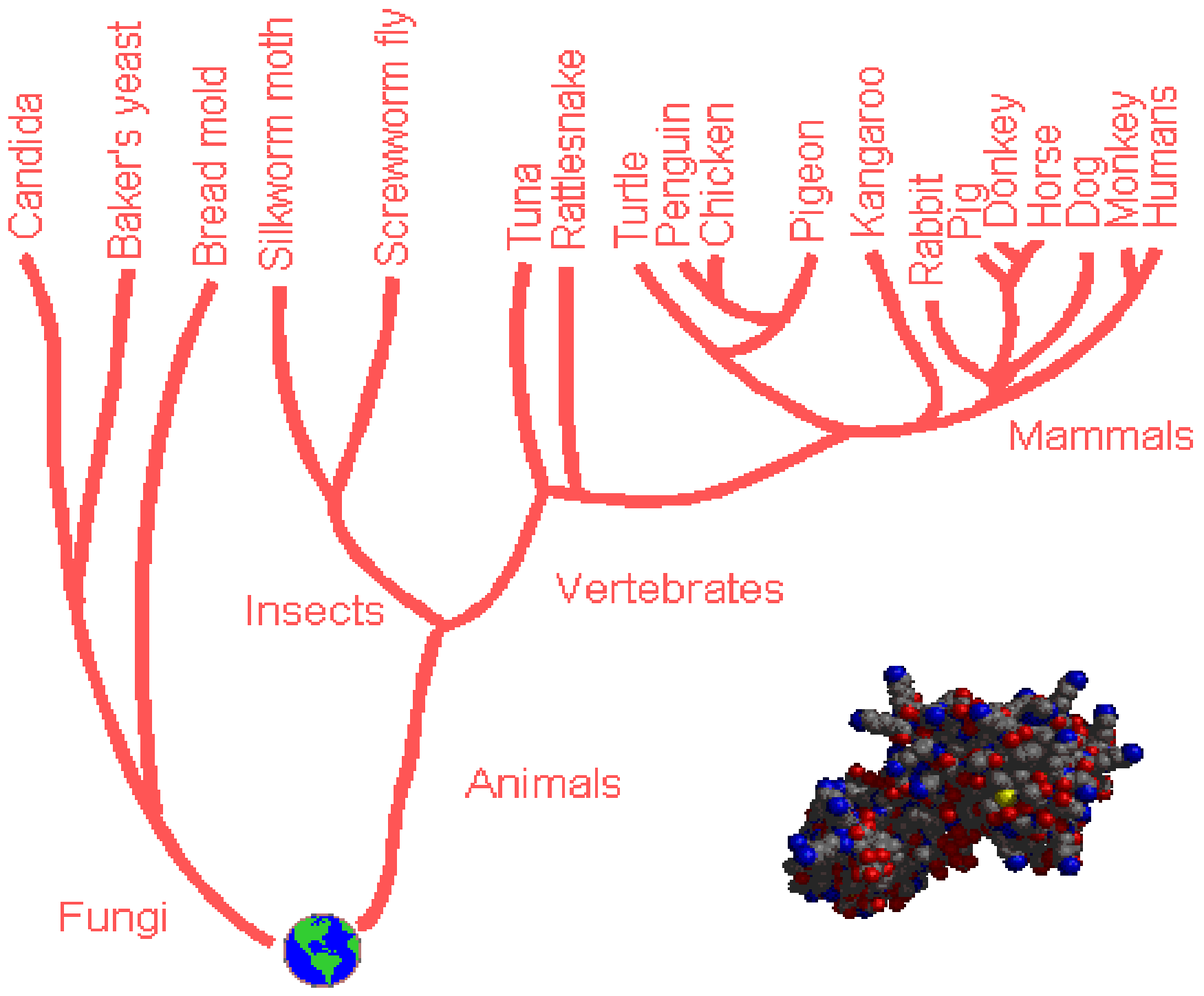

*Drawing of a phylogenetic tree based on the amino acid sequence data of cytocrome C (see inset).*

**Prediction of protein structure** from sequence is one of the most challenging tasks in today's computational biology. More or less, the task is to calculate an image like the one in the second figure of this text. Although most information of 3-dimensional structure is encoded in the amino acid sequence it is still unknown which information controls the process of protein folding. Among millions of possible folding products, proteins take up one working, native structure. Since it is very difficult and expensive to evaluate structures by methods like X-ray diffraction or NMR spectroscopy, there is a big need for the unfailing prediction of 3-dimensional structures of proteins from sequence data. Today there are methods which are able to give a quite reliable result from available sequence data, the odds to get this "right" are about 65%.

**Sequence comparison** is a very powerful tool in molecular biology, genetics and protein chemistry. Frequently it is unknown for which proteins a new DNA sequence codes or if it codes for any protein at all. If you compare a new coding sequence with all known sequences there is a high probability to find a similiar sequence. Often it is already known which role the protein in the data bank plays in the cell. If you assume that a similar sequence implies a similar function, you now have much more knowledge about your new sequence than before

Proteins of one class often show a few amino acids that always occur at the same positions in the amino acid sequence. By looking for "patterns" you will be able to gain information about the activity of a protein of which only the gene (DNA) is known. Evaluation of such patterns yields information about the architecture of proteins. Often these patterns are involved in active sites, which are the workbenchs of proteins.

# Introduction to Bioinformatics

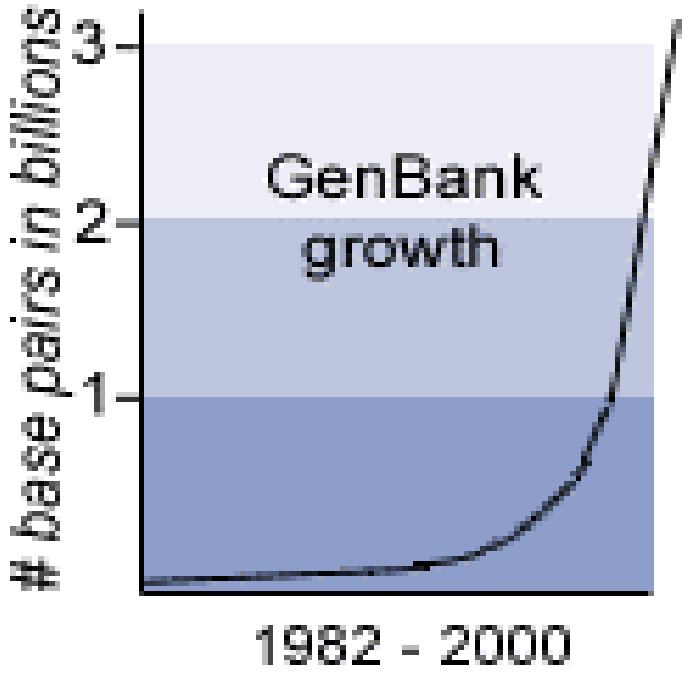

**What is Bioinformatics?**
In the last few decades, advances in molecular biology and the equipment available for research in this field have allowed the increasingly rapid sequencing of large portions of the genomes of several species. In fact, to date, several bacterial genomes, as well as those of some simple eukaryotes (e.g., *Saccharomyces cerevisiae*, or baker's yeast) and more complex eukaryotes (C. elegans and Drosophila) have been sequenced in full. The Human Genome Project, designed to sequence all 24 of the human chromosomes, is also progressing and a rough draft was completed in the spring of 2000.

Popular sequence databases, such as GenBank and EMBL, have been growing at exponential rates. This deluge of information has necessitated the careful storage, organization and indexing of sequence information. Information science has been applied to biology to produce the field called bioinformatics.

**LANDMARK SEQUENCES COMPLETED**

- tRNA - (1964) - 75 bases (old, slow, complicated method)
- First complete DNA genome: X174 DNA (1977) - 5386 bases
- human mitochondrial DNA (1981) - 16,569 bases
- tobacco chloroplast DNA (1986) - 155,844 bases
- First complete bacterial genome (*H. Influenzae*)(1995) - 1.9 x $10^6$ bases
- Yeast genome (eukaryote at ~ 1.5 x $10^7$) completed in 1996
- Several archaebacteria
- *E. coli* -- 4 x $10^6$ bases [1997 & 1998]
- Several pathogenic bacterial genomes sequenced
  - Helicobacter pyloris (ulcers)
  - Treponema pallidium (Syphilis)
  - Borrelia burgdorferi (Lyme disease)
  - Chlamydia trachomatis (trachoma - blindness)
  - Rickettsia prowazekii (epidemic typhus)
  - Mycobacterium tuberculosis (tuberculosis)
- Nematode C. elegans ( ~ 4 x $10^8$) - December 1998
- Drosophila (fruit fly) (2000)
- Human genome (rough draft completed 5/00) - 3 x $10^9$ base

**Bioinformatics** is the recording, annotation, storage, analysis, and searching/retrieval of nucleic acid sequence (genes and RNAs), protein sequence and structural information. This includes databases of the sequences and structural information as well methods to access, search, visualize and retrieve the information.

Sequence data can be used to make predictions of the functions of newly identified genes,estimate evolutionary distance in phylogeny reconstruction, determine the active sites of enzymes, construct novel mutations and characterize alleles of genetic diseases to name just a few uses. Sequence data facilitates:

- Analysis of the organization of genes and genomes and their evolution
- Protein sequence can be predicted from DNA sequence which further facilitates possible prediction of protein properties, structure, and function (proteins rarely sequenced in entirety today)
- Identification of regulatory elements in genes or RNAs
- Identification of mutations that lead to disease, etc.

Bioinformatics is the field of science in which biology, computer science, and information technology merge into a single discipline. The ultimate goal of the field is to enable the discovery of new biological insights as well as to create a global perspective from which unifying principles in biology can be discerned.

There are three important sub-disciplines within bioinformatics involving computational biology:

- the development of new algorithms and statistics with which to assess relationships among members of large data sets;
- the analysis and interpretation of various types of data including nucleotide and amino acid sequences, protein domains, and protein structures; and
- the development and implementation of tools that enable efficient access and management of different types of information.

One of the simpler tasks used in bioinformatics concern the creation and maintenance of databases of biological information. Nucleic acid sequences (and the protein sequences derived from them) comprise the majority of such databases. While the storage and or ganization of millions of nucleotides is far from trivial, designing a database and developing an interface whereby researchers can both access existing information and submit new entries is only the beginning.

The most pressing tasks in bioinformatics involve the analysis of sequence information. **Computational Biology** is the name given to this process, and it involves the following:

- Finding the genes in the DNA sequences of various organisms
- Developing methods to predict the structure and/or function of newly discovered proteins and structural RNA sequences.
- Clustering protein sequences into families of related sequences and the development of protein models.
- Aligning similar proteins and generating phylogenetic trees to examine evolutionary relationships.

Data-mining is the process by which testable hypotheses are generated regarding the function or structure of a gene or protein of interest by identifying similar sequences in better characterized organisms. For example, new insight into the molecular basis of a disease may come from investigating the function of homologs of the disease gene in model organisms. Equally exciting is the potential for uncovering phylogenetic relationships and evolutionary patterns.The process of evolution has produced DNA sequences that encode proteins with very specific functions. It is possible to predict the three-dimensional structure of a protein using algorithms that have been derived from our knowledge of physics, chemistry and most importantly, from the analysis of other proteins with similar amino acid sequences.

## Sequence Analysis: Sequence to Potential Function

### Sequence to Potential Function

- ORF prediction and gene identification
- Search databases for potential protein function or homologue
- Protein structure prediction and multiple sequence alignment (conserved regions)
- Analysis of potential gene regulatory elements
- Gene knockout or inhibition (RNA interference) for phenotypic analysis

## Overview of Sequence Analysis

### Sequence data facilitates:

- Analysis of the organization of genes and genomes
- Prediction of protein properties, functions, and structure from gene sequence or cDNA (proteins rarely sequenced in entirety today) -- Cystic fibrosis example
- Identification of regulatory elements
- Identification of mutations that lead to disease, etc.

### The Creation of Sequence Databases

Most biological databases consist of long strings of nucleotides (guanine, adenine, thymine, cytosine and uracil) and/or amino acids (threonine, serine, glycine, etc.). Each sequence of nucleotides or amino acids represents a particular gene or protein (or section thereof), respectively. Sequences are represented in shorthand, using single letter designations. This decreases the space necessary to store information and increases processing speed for analysis.

While most biological databases contain nucleotide and protein sequence information, there are also databases which include taxonomic information such as the structural and biochemical characteristics of organisms. The power and ease of using sequence information has however, made it the method of choice in modern analysis.

In the last three decades, contributions from the fields of biology and chemistry have facilitated an increase in the speed of sequencing genes and proteins. The advent of cloning technology allowed foreign DNA sequences to be easily introduced into bacteria. In this way, rapid mass production of particular DNA sequences, a necessary prelude to sequence determination, became possible. Oligonucleotide synthesis provided researchers with the ability to construct short fragments of DNA with sequences of their own choosing. These oligonucleotides could then be used in probing vast libraries of DNA to extract genes containing that sequence. Alternatively, these DNA fragments could also be used in polymerase chain reactions to amplify existing DNA sequences or to modify these sequences. With these techniques in place, progress in biological research increased exponentially.

For researchers to benefit from all this information, however, two additional things were required: 1) ready access to the collected pool of sequence information and 2) a way to extract from this pool only those sequences of interest to a given researcher. Simply collecting, by hand, all necessary sequence information of interest to a given project from published journal articles quickly became a formidable task. After collection, the organization and analysis of this data still remained. It could take weeks to months for a researcher to search sequences by hand in order to find related genes or proteins.

Computer technology has provided the obvious solution to this problem. Not only can computers be used to store and organize sequence information into databases, but they can also be used to analyze sequence data rapidly. The evolution of computing power and storage capacity has, so far, been able to outpace the increase in sequence information being created. Theoretical scientists have derived new and sophisticated algorithms which allow sequences to be readily compared using probability theories. These comparisons become the basis for determining gene function, developing phylogenetic relationships and simulating protein models. The physical linking of a vast array of computers in the 1970's provided a few biologists with ready access to the expanding pool of sequence information. This web of connections, now known as the Internet, has evolved and expanded so that nearly everyone has access to this information and the tools necessary to analyze it.

**Databases of protein and nucleic acid sequences**

- In the US, the repository of this information is The National Center for Biotechnology Information (NCBI)
- The database at the NCBI is a collated and interlinked dataset known as the Entrez Databases
  - Description of the Entrez Databases
  - Examples of a selected database files
    - Protein
      - Most protein sequence is derived from conceptual translation
    - Chromosome with genes and predicted proteins (Accession #D50617, Yeast Chromosome VI: Entrez)
    - Genome (C. elegans)
    - Protein Structure (TPI database file or Chime structure)
    - Expressed sequence tags (Ests)(Summary of current data)
  - Neighboring

 **Searching databases** to identify sequences and predicting functions or properties of **predicted** proteins

- Searching by keyword, accession #, etc.
- Searching for homologous sequences
  - see Blast at the NCBI
    - BLAST (Basic Local Alignment Search Tool) is a set of similarity search programs designed to explore all of the available sequence databases regardless of whether the query is protein or DNA.

## Searching for Genes

The collecting, organizing and indexing of sequence information into a database, a challenging task in itself, provides the scientist with a wealth of information, albeit of limited use. The power of a database comes not from the collection of information, but in its analysis. A sequence of DNA does not necessarily constitute a gene. It may constitute only a fragment of a gene or alternatively, it may contain several genes.

Luckily, in agreement with evolutionary principles, scientific research to date has shown that all genes share common elements. For many genetic elements, it has been possible to construct consensus sequences, those sequences best representing the norm for a given class of organisms (e.g, bacteria, eukaroytes). Common genetic elements include promoters, enhancers, polyadenylation signal sequences and protein binding sites. These elements have also been further characterized into further subelements.

Genetic elements share common sequences, and it is this fact that allows mathematical algorithms to be applied to the analysis of sequence data. A computer program for finding genes will contain at least the following elements.

| **Elements of a Gene-seeking Computer Program** | |
|---|---|
| **Algorithms for pattern recognition** | Probability formulae are used to determine if two sequences are statistically similar. |
| **Data Tables** | These tables contain information on consensus sequences for various genetic elements. More information enables a better analysis. |
| **Taxonomic Differences** | Consensus sequences vary between different taxonomic classes of organisms. Inclusion of these differences in an analysis speeds processing and minimizes error. |
| **Analysis rules** | These programming instructions define how algorithms are applied. They define the degree of similarity accepted and whether entire sequences and/or fragments thereof will be considered in the analysis. A good program design enables users to adjust these variables. |

## The Challenge of Protein Prediction and Modelling

There are a myriad of steps following the location of a gene locus to the realization of a three-dimensional model of the protein that it encodes.

**Identification of the location of the transcription start and stop sites.** A proper analysis to locate a genetic locus will usually have already pinpointed at least the approximate sites of the transcriptional start and stop. Such an analysis is usually insufficient in determining protein structure. It is the start and end codons for translation that must be determined with accuracy for prediction of the protein encoded.

**Identification of the translation initiation and stop sites.** The first codon in a messenger RNA sequence is almost always AUG. While this reduces the number of candidate codons, the reading frame of the sequence must also be taken into consideration.

There are six reading frames possible for a given DNA sequence, three on each strand, that must be considered, unless further information is available. Since genes are usually transcribed away from their promoters, the definitive location of this element can reduce the number of possible frames to three. There is not a strong concensus between different species surrounding translation start codons. Therefore, location of the appropriate start codon will include a frame in which they are not apparent abrupt stop codons. Knowledge of a proteinÕs predicted molecular mass can assist this analysis. Incorrect reading frames usually predict relatively short peptide sequences. Therefore, it might seem deceptively simple to ascertain the correct frame. In bacteria, such is frequently the case. However, eukaryotes add a new obstacle to this process: INTRONS!

**Prediction and Identification of the Exon/Intron Splice Sites**. In eukaryotes, the reading frame is discontinuous at the level of the DNA because of the presence of introns. Unless one is working with a cDNA sequence in analysis, these introns must be spliced out and the exons joined to give the sequence that actually codes for the protein. Intron/exon splice sites can be predicted on the basis of their common features. Most introns begin with the nucleotides GT and end with the nucleotides AG. There is a branch sequence near the downstream end of each intron involved in the splicing event. There is a moderate concensus around this branch site.

**Prediction of Protein 3-D Structure**. With the completed primary amino acid sequence in hand, the challenge of modelling the three-dimensional structure of the protein awaits. This process uses a wide range of data and CPU-intensive computer analysis. Most often, one is only able to obtain a rough model of the protein, and several conformations of the protein may exist that are equally probable. The best analyses will utilize data from all the following sources.

| | |
|---|---|
| Pattern Comparison | Alignment to known homologues whose conformation is more secure |
| X-ray Diffraction Data | Most ideal when some data is available on the protein of interest. However, diffraction data from homologous proteins is also very valuable. |
| Physical Forces/Energy States | Biophysical data and analyses of an amino acid sequence can be used to predict how it will fold in space. |

All of this information is used to determine the most probable locations of the atoms of the protein in space and bond angles. Graphical programs can then use this data to depict a three-dimensional model of the protein on the two-dimensional computer screen.

## Introduction to Bioinformatic Databases

Protein sequence and DNA and RNA sequence data have been growing since the development of methods for sequence determination *(While methods for protein sequence determinations were developed prior to those for DNA and RNA, most protein sequence information in the databases today is not derived from direct analysis of proteins, but from conceptual translation of DNA sequence data)*. Similarly three-dimensional structures have been growing since the development of X-ray crystallography and newer NMR methods. These data are stored in databases which are typically publicly accessible (exceptions include proprietary databases from pharmaceutical and genomics companies - to be discussed later).

Most biological databases consist of long strings of nucleotides (guanine, adenine, thymine, cytosine and uracil) and/or amino acids (threonine, serine, glycine, etc.). Each sequence of nucleotides or amino acids represents a genome, chromosomes, particular gene, mRNA, RNA or protein (or section thereof), respectively. Sequences are represented in shorthand, using single letter designations. This decreases the space necessary to store information and increases processing speed for analysis. Structural databases contain the nucleotide or amino acid strings and the X, Y, and Z spatial coordinates of all sequences to define and represent the 3-dimensional structure of the polymer, and in some cases any associated ligands or cofactors.

While most biological databases contain nucleotide and protein sequence information, there are also databases which include taxonomic information, structural information, biochemical or metabolic characteristics of organisms, genome sequence, organized and curated collections of subtypes of proteins or protein domains or motifs, nucleic acids, families of protein structural types, etc.. The power and ease of using sequence information has however, made it the method of choice in modern analysis. More recently, the addition of mRNA and protein expression data has being growing rapidly and has led to the necessity for the need to store, interpret, and access this type of data.

In the last three decades, contributions from the fields of biology and chemistry have facilitated an increase in the speed of sequencing genes and proteins. The advent of cloning technology allowed foreign DNA sequences to be easily introduced into bacteria. In this way, rapid mass production of particular DNA sequences, a necessary prelude to sequence determination, became possible. Oligonucleotide synthesis provided researchers with the ability to construct short fragments of DNA with sequences of their own choosing. These oligonucleotides could then be used in probing vast libraries of DNA to extract genes containing that sequence. Alternatively, these DNA fragments could also be used in polymerase chain reactions to amplify existing DNA sequences or to modify these sequences. With these techniques in place, and particularly the success and use of shotgun sequencing of genomes, both data and progress in biological research increased exponentially.

For researchers to benefit from all this information, however, two additional things were required:

1) ready access to the collected pool of sequence information and
2) a way to extract from this pool only those sequences of interest to a given researcher.

Simply collecting, by hand, all necessary sequence information of interest to a given project from published journal articles quickly became a formidable task. After collection, the organization and analysis of this data still remained. It could take weeks to months for a researcher to search sequences by hand in order to find related genes or proteins.

Computer technology has provided the obvious solution to this problem. Not only can computers be used to store and organize sequence information into databases, but they can also be used to analyze sequence data rapidly. The evolution of computing power and storage capacity has, so far, been able to outpace the increase in sequence information being created. Theoretical scientists have derived new and sophisticated algorithms which allow sequences to be readily compared using probability theories. These comparisons become the basis for determining gene function, developing phylogenetic relationships and simulating protein models. The physical linking of a vast array of computers in the 1970's provided a few biologists with ready access to the expanding pool of sequence information. This web of connections, now known as the Internet, has evolved and expanded so that nearly everyone has access to this information and the tools necessary to analyze it.

**Bioinformatic Databases include (but not limited to):**

- Literature
    - PubMed (Medline)
- Nucleic Acid Sequences
    - Genomes
    - Chromosomes
    - Genes and mRNAs (cDNA)
    - Est, GSS, SNP, etc.
        - Est - expressed sequence tag (sequence of an expressed mRNA)
        - SNP - single nucleotide polymorphism
    - RNA
- Proteins
- Structure
    - Proteins (and their interactions)
    - Nucleic Acids
    - Protein-Nucleic acid interactions (protein and DNA or protein and RNA)
- Secondary Databases
    - Protein
        - Protein Families
        - Motifs, consensus sequencs, profiles, etc
        - Protein domains and folds
        - Structural families
        - RefSeq
        - Clusters of Orthologous Groups (COGs)
    - Nucleic Acid
        - Unigene
        - RefSeq
- Expression profiles of mRNAs and proteins
    - Networks of expression patterns
- Interaction Networks (proteins)

These data require storage in a manner that allows the data to be

- Accessible
- Searchable
- Retrievable
- Analyzed using a variety of programs
- Integrated or linked with other database information

Historically individual databases were constructed for protein or nucleic acid sequences.

- In the case of DNA sequences, the main repositories varied in different parts of the world:
    - Within the US in GenBank
    - Japan in DDBJ (DNA Database of Japan),
    - Europe in the EMBL database (European Molecular Biology Laboratory)
- A variety of protein databases were also developed and collated information.
- For years since the information in each database was not exactly the same (some sequences might be present in one database but not in another) it required concerted effort to carry out analyses on each separate database to ensure that one examined all the data.

More recently the databases have shared and integrated information from a variety of databases to produce complex, comprehensive sites with access to information in most databases from a single site, and in some cases an integrated database.

Three of the main integrated databases in use today include

- Entrez (US National Center for Biotechnology = NCBI)
- The Sequence Retrieval System (European Bioinformatics Institute = EBI)
- DBGET/LinkDB (National Institute of Genetics in Japan = NIG)

A great strength of these databases is that they not only allow you to search them but they provide links and handy pointers to additional information in related databases. The three systems differ in their databases and the links they make to other information.

**Entrez at the NCBI** is a collated and interlinked molecular biology database and retrieval system which serves as an entry point for exploring distinct but integrated databases. Entrez provides access to for example nucleotide and protein sequence databases, a molecular modeling 3-D structure database (MMDB), a genomes and maps database, disease information, and the literature. Of the three, Entrez is the simplest to use, although it offers more limited information to search.

The **The Sequence Retrieval System** is available from the European Molecular Biology Organization (EMBO). This integrated database offers greater flexibility in several areas with a homogeneous interface to over 80 biological databases developed at the European Bioinformatics Institute (EBI) at Hinxton. The types of databases included for example are sequence and sequence related, metabolic pathways, transcription factors, applications, protein 3-D structures, genome, mapping, mutations and locus-specific mutations.

**DBGET/LinkDB** is an integrated database retrieval system developed by the Institute for Chemical Research (Kyoto University) and the Human Genome Center of the University of Tokyo. The system provides access to about 20 databases which can be queried one at a time and with the results will provide links to other databases. This includes access to the Enclyclopedia of Genes and Genomes at Kyoto (KEGG). DBGET has simpler but more limited search methods methods than either SRS or Entrez.

# Introduction to NCBI and Entrez

The National Center for Biotechnology Information (**NCBI**) was established in 1988, as a division of the National Library Medicine (NLM) at the National Institutes of Health (NIH). The following information is taken directly from the NCBI web site.

Understanding nature's mute but elegant language of living cells is the quest of modern molecular biology. From an alphabet of only four letters representing the chemical subunits of DNA, emerges a syntax of life processes whose most complex expression is man. The unraveling and use of this "alphabet" to form new "words and phrases" is a central focus of the field of molecular biology. The staggering volume of molecular data and its cryptic and subtle patterns have led to an absolute requirement for computerized databases and analysis tools. The challenge is in finding new approaches to deal with the volume and complexity of data, and in providing researchers with better access to analysis and computing tools in order to advance understanding of our genetic legacy and its role in health and disease.

**NCBI's mission**, as a national resource for molecular biology information, is to develop new information technologies to aid in the understanding of fundamental molecular and genetic processes that control health and disease. More specifically, the NCBI has been charged with creating automated systems for storing and analyzing knowledge about molecular biology, biochemistry, and genetics; facilitating the use of such databases and software by the research and medical community; coordinating efforts to gather biotechnology information both nationally and internationally; and performing research into advanced methods of computer-based information processing for analyzing the structure and function of biologically important molecules.

**To carry out its diverse responsibilities, NCBI:**

- Conducts research on fundamental biomedical problems at the molecular level using mathematical and computational methods,
- Maintains collaborations with several NIH institutes, academia, industry, and other governmental agencies,
- Fosters scientific communication by sponsoring meetings, workshops, and lecture series,
- Supports training on basic and applied research in computational biology for postdoctoral fellows through the NIH Intramural Research Program,
- Engages members of the international scientific community in informatics research and training through the Scientific Visitors Program,
- Develops, distributes, supports, and coordinates access to a variety of databases and software for the scientific and medical communities, and
- Develops and promotes standards for databases, data deposition and exchange and biological nomenclature.

**Databases and Software**

NCBI assumed responsibility for the **GenBank** DNA sequence database in October, 1992. **GenBank is the NIH genetic sequence database and consists of genome, gene, cDNA, Est, and other nucleic acid sequences.** Genbank consists of NCBI staff or contractors with advanced training in molecular biology who build the database from sequences submitted by individual laboratories and by data exchange with the international nucleotide sequence databases, including the European Molecular Biology Laboratory (EMBL) and the DNA Database of Japan (DDBJ). Arrangements with the U.S. Patent and Trademark Office enable the incorporation of patent sequence data.

**Entrez is NCBI's search and retrieval system** that provides users with i**ntegrated access to sequence, mapping, taxonomy, structural data, literature, and other data.** Entrez also provides graphical views of sequences and chromosome maps. A powerful and unique feature of Entrez is the ability to retrieve related sequences, structures, and references.

- What makes Entrez powerful is that most of its records are linked to other records, both within a given database (such as PubMed) and between databases.
    - Links within the entrez database are called "neighbors".
- Protein and Nucleotide neighbors are determined by performing similarity searches using the algorithm BLAST on the amino acid or DNA sequence in the entry and the results saved as above.
    - What this means is that if you find one or a few documents that match what you are looking for, pressing the "Related Articles/Sequences" button will find a great many more documents that are likely to be relevant, in order from most useful to least.
    - This allows you to find what you want with much greater speed and accuracy: instead of having to flip through thousands of documents to assure yourself that nothing germane to your query was missed, you can find just a few, then look at their neighbors.
- In addition, some documents are linked to others for reasons other than computed similarity.
    - For instance, if a protein sequence was published in a PubMed article, the two will be linked to one another.

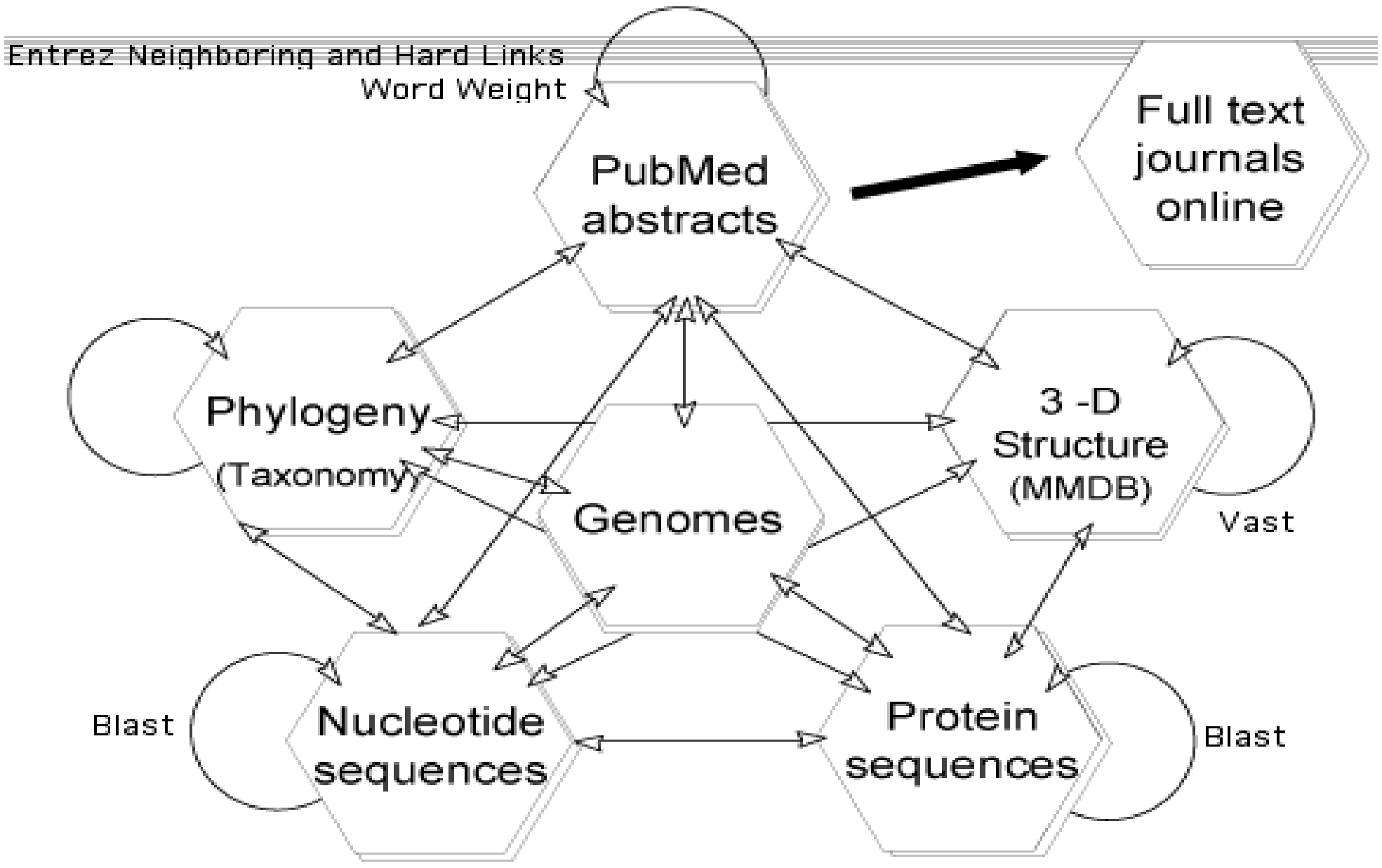
Entrez Neighboring and Hard Links
Word Weight
PubMed abstracts
Full text journals online
Phylogeny (Taxonomy)
Genomes
3 -D Structure (MMDB)
Vast
Blast
Nucleotide sequences
Protein sequences
Blast

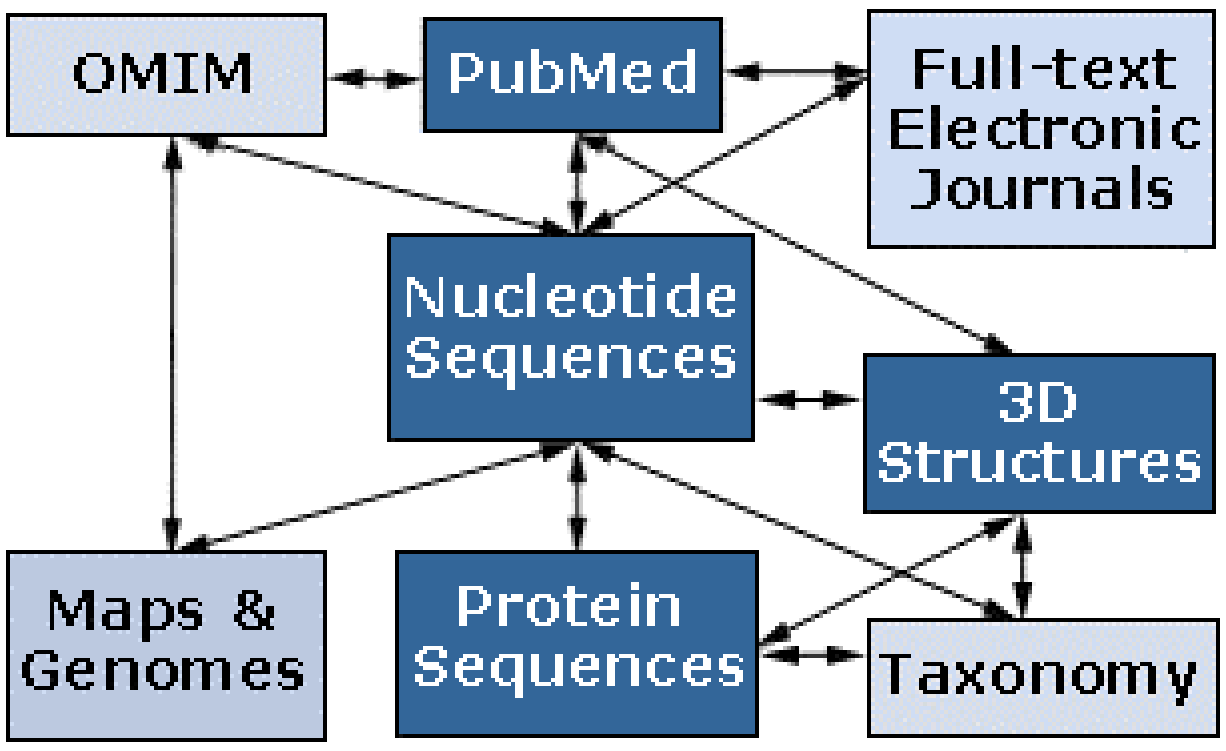

**Entrez Home Page**

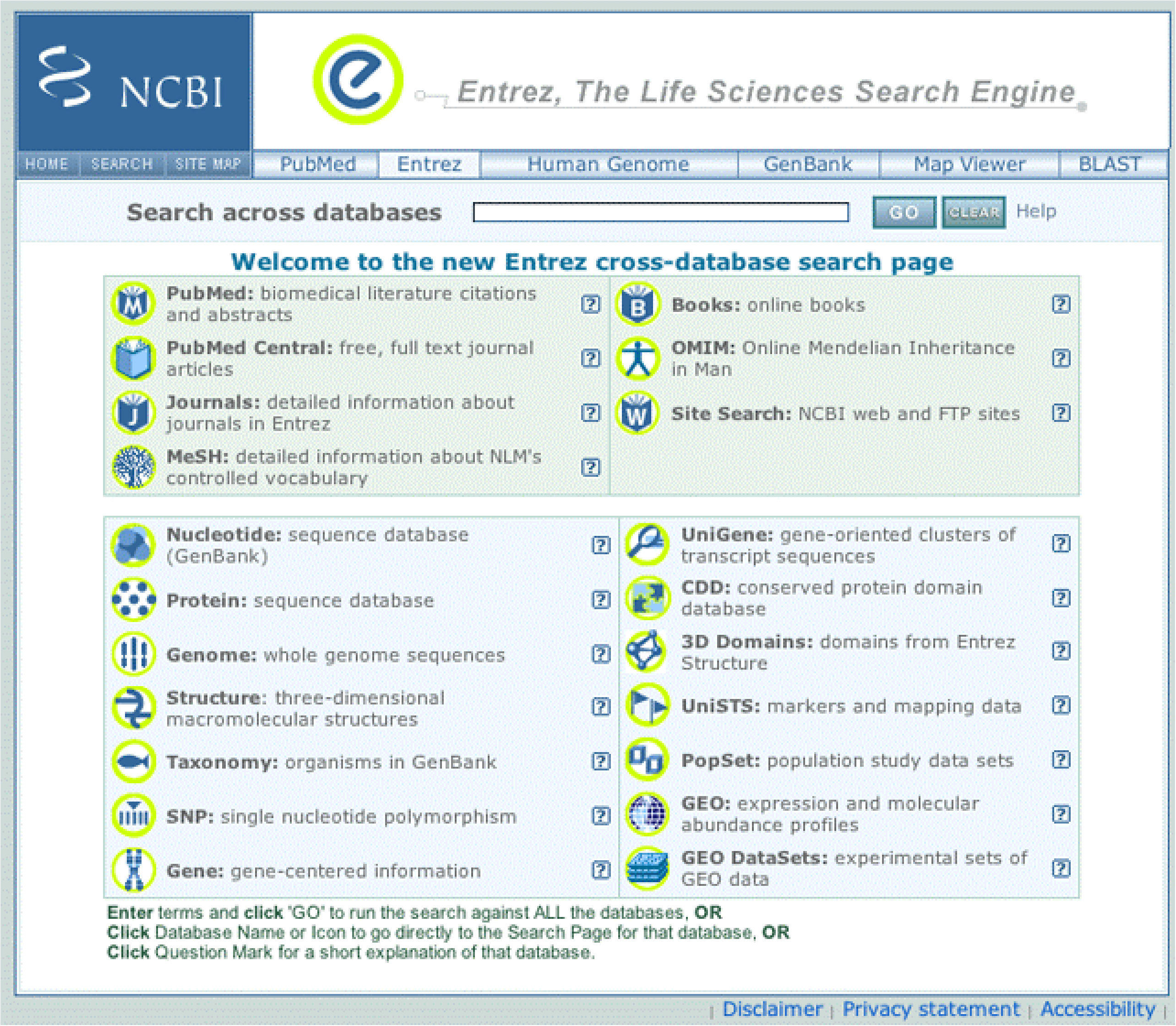

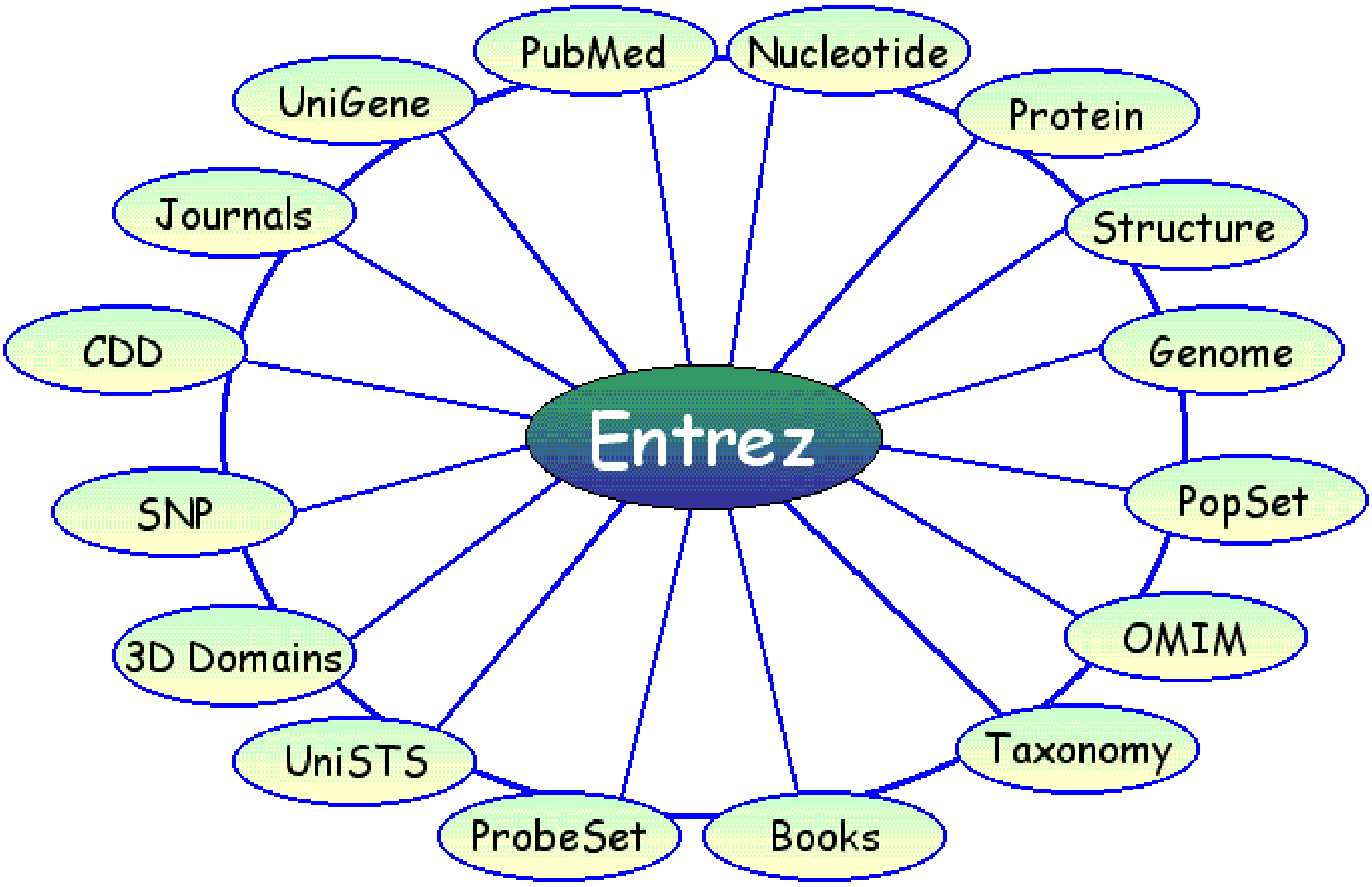

**The (ever) Expanding Entrez System**

**DEMONSTRATION: Example Entrez I**

- **Example**
  - **Author** = *Davis RE*
  - **Organism** = *Schistosoma*
  - **Combined** = *Davis RE AND Schistosoma*
    - *Use of related links options within the*
- **Other examples for literature and sequences related to**
  - a specific protein (telomerase)
  - an RNA (snoRNA or SL RNA)
  - disease (oroticaciduria)
  - obesity
  - the molecular cloning of the human cystic fibrosis gene
- Use MEDLINE with Keyword searching.
- Use neighbor feature to find related articles.
- Switch to Nucleotide database to see sequence.
- Save a copy of sequence to local disk.
- Use MESH terms to find similar articles.
- Search the Nucleotide database by gene name.

**Entrez Databases Support Dual Search Strategies**

- **Iterative searching:** one primary search topic on which you can build constraints upon the initial search.
- ***Phrase Searching*** (forcing PubMed to search for a phrase)

- PubMed consults a phrase index and groups terms into logical phrases. For example, if you enter poison ivy, PubMed recognizes these two words as a phrase and searches it as one search term.
  - However, it is possible that PubMed may fail to find a phrase that is essential to a search.
  - For example, if you enter, single cell, it is not in the Phrase List and PubMed searches for "single" and "cell" separately. To force PubMed to search for a specific phrase enter double quotes (" ") around the phrase, e.g., "single cell".
- **Complex Boolean searching:** Conjunction of searches or terms using Operators
- A search can be performed all at once by specifying the terms to search, their fields, and the boolean operations to perform on them.
  - This is the default (Basic) mode for PubMed
- **Boolean Syntax**
  - search term [tag] BOOLEAN OPERATOR search term [tag]
    - term [field] OPERATOR term [field] ...etc
  - OPERATOR must be upper case
    - AND
    - OR
    - NOT
  - Default is AND between words
  - Use * for truncation
  - [field] identifies the specific axis through which the search will be limited
- Examples of Boolean Search Statements:
  - Find citations on DNA that were authored by Dr. Crick in 1993.
    - dna [mh] AND crick [au] AND 1993 [dp]
  - Find articles that deal with the effects of heat or humidity on multiple sclerosis, where these words appear in all fields in the citation.
    - (heat OR humidity) AND multiple sclerosis
  - Find English language review articles that discuss the treatment of asthma in preschool children.
    - asthma/therapy [mh] AND review [pt] AND child, preschool [mh] AND english [la]

About BLAST

BLAST Services

BLAST Software

Support

NEW 15 Nov 2004 Download the BLAST poster from SC2004!

**Nucleotide**

- Quickly search for highly similar sequences (megablast)
- Quickly search for divergent sequences (discontiguous megablast)
- Nucleotide-nucleotide BLAST (blastn)
- Search for short, nearly exact matches
- Search trace archives with megablast or discontiguous megablast

**Protein**

- Protein-protein BLAST (blastp)
- PHI- and PSI-BLAST
- Search for short, nearly exact matches
- Search the conserved domain database (rpsblast)
- Search by domain architecture (cdart)

**Translated**

- Translated query vs. protein database (blastx)
- Protein query vs. translated database (tblastn)
- Translated query vs. translated database (tblastx)

**Genomes**

- Chicken, cow, pig, dog, sheep, cat
- Environmental samples
- Human, mouse, rat
- Fugu rubripes, zebrafish
- Insects, nematodes, plants, fungi, malaria
- Microbial genomes, other eukaryotic genomes

**Special**

- Search for gene expression data (GEO BLAST)
- Align two sequences (bl2seq)
- Screen for vector contamination (VecScreen)
- Immunoglobin BLAST (IgBlast)
- Human SNP BLAST new!

**Meta**

- Retrieve results by RID
- Get this page with javascript-free links

# Current Research Trends in Genomics, Proteomics and Bioinformatics, and their Relevance to Malting Barley

## Introduction

The "Genomic Era" was ushered in by rapid advances in nucleic acid sequencing, making it possible to sequence whole genomes (all the genetic material in the chromosomes of a particular organism) relatively quickly. The Human Genome Project accelerated development of the various techniques and technologies involved in genomics, and opened up new "-omics" fields, notably Proteomics (the study of the complement of proteins expressed in a given cell, tissue or organism under particular conditions at a particular time). The huge amount of data generated by the systematic analysis and documentation of genomes and proteomes has in turn given rise to the field of Bioinformatics, comprising the informatics capacity and skills needed to organize and annotate the data and to further predict structure, function and inter-relationships among biomolecules of interest. The post-genomics era is characterised by an holistic approach to the study of biological systems, involving many new analytical tools. This short paper will focus on selected technological developments in the areas of genomics and proteomics, and their application in malting barley development.

## Genomic Analysis

The introduction of molecular markers (segments of DNA whose pattern of inheritance can be determined), the polymerase chain reaction (PCR) to amplify genomic DNA fragments, and high throughput DNA sequencers which determine nucleotide sequences of amplified DNA by gel or capillary electrophoresis methods, are all examples of advances in genomic analysis. These advances have led to the sequencing of the genomes of several "model organisms" including human, bacteria (*Escherichia coli*), yeast (*Saccharomyces cerevisiae*), and plants such as *Aradopsis* and rice (*Oryza sativa)*.

Barley researchers and breeders were quick to make use of the molecular markers for barley fingerprinting based on restriction fragment length polymorphisms (RFLP's) and PCR-generated random amplified polymorphic DNAs (RAPDs), sequenced tagged sites (STSs), microsatellites and amplified fragment length polymorphisms (AFLPs). The North American Barley Genome Mapping Project (NABGMP) and other initiatives made great strides in locating quantitative trait loci (QTL) - specific genomic regions that affect quantitative traits (Han et al., 1997; Mather et al., 1997) - and in exploiting molecular markers in breeding for specific traits (Swanston et al., 1999) or for barley malt fingerprinting (Faccioli et al., 1999).

## ESTs and Microarrays

While DNA sequence information defines the genome, it does not of itself provide information on the relative expression levels of genes. This information is more readily obtained by examining cellular RNA content. Reverse transcription of mRNA produces complemcntary DNA (cDNA), a single-stranded form of DNA that can be amplified by PCR for further study. Recent sequencing efforts have been directed towards compiling short sequences of cDNA, known as expressed sequence tags (ESTs), which can be used as genetic probes to relate sequence information to differences in mRNA expression between cell populations. ESTs, or cDNAs of known sequence, can be deposited onto glass slides in known locations to create microarrays, or gene chips. The same reverse transcription process that produces cDNAs for microarrays is used to prepare cDNA fragments from cell populations. These cDNAs can be labelled with different fluorescent tags and allowed to hybridize with the cDNA on the chip so that differences in mRNA expression between the cell populations can be examined. Genome-wide analyses of gene expression patterns are thus enabled, and can be used to study, for example, abiotic stress responses in cereal crops, including wheat and barley. Jones (personal communication) has used the commercially available Rice Chip (Affymetrix Inc., Santa Clara, CA, USA) to investigate signalling pathways and metabolic regulation in barley aleurone layers, and noted an 80% hybridization of barley cDNA to the rice oligochip.

## Proteomic Analysis

The need to go beyond nucleic acid analysis, and reach an understanding of total protein expression and regulation in biological systems is motivating the field of 'Proteomics' (Dutt and Lee, 2000). The field of proteome analysis has largely been driven by technological developments which have improved upon the basic strategy of separating proteins by two-dimensional gel electrophoresis (O'Farrell, 1975). As with genomic information or data derived from microarray analysis, an informatic framework is required to organize proteomic data, as well as to generate structural and predictive models of the molecules and interactions being studied.

The use of 2-D gel elctrophoresis coupled with mass spectrometry (MS) is the most familiar proteomics approach, with advances in the MS ionization techniques and mass analyzers enabling the generation of different types of structural information about proteins of interest. The techniques of nuclear magnetic resonance spectroscopy (NMR) and X-ray crystallography rank alongside mass spectrometry as the major tools of proteomics. The development of protein chips (analogous to microarrays) and protein-protein interaction maps will offer new ways to rapidly characterize expression levels and relationships among proteins in normal or "perturbed" cell populations, helping researchers to bridge the 'genotype-phenotype' gap.

## Perspectives

A combination of functional genomics and proteomics approaches allows researchers to cross interdisciplinary boundaries in "trawling" the genome, transcriptome, proteome and metabolome for new insights into structure and function that can be applied to improving the agronomic and end use traits of malting barley. As Fincher (2001) has pointed out, functional genomics and related technologies are now being used to re-visit the difficult questions of cell wall polysaccharide biosynthesis, involving synthase enzymes that had proved impossible to purify and characterize by classical biochemical methods. In addition to the resolution of their structure, the genes encoding for the more extensively studied hydrolase enzymes, responsible for cell wall breakdown and starch degradation, are being identified, generating knowledge of how characteristics such as thermostability might be enhanced to improve malting and brewing performance (Kristensen et al., 1999; Ziegler, 1999) Structural and functional studies can thus be linked back to protein and mRNA expression patterns, and ultimately to the families of genes that we might wish to conserve or alter in breeding programs.

Ultimately, the extent to which advances in functional genomics and proteomics will be embraced and adopted by the malting and brewing industry depends heavily on the industry's ability to "keep up" with the rapidly moving field, and on the economics (perhaps the most important of the "-omics") of doing so. Brewing Science has a long and distinguished history, closely associated with the need of the industry to profitably apply science in pursuit of product quality and consistency. To this end, generations of chemists, biochemists, microbiologists, botanists and plant breeders have applied their skills to elucidating and manipulating the structural and functional characteristics of barley, yeast and hops. In the case of barley, one of the results of the close linkage between science and industry has been a relatively strong base of research and breeding activities - certainly more than would be expected from barley's position among the world's major crops. It is perhaps ironic that today, when the emerging technologies of genomics and proteomics are allowing us to revisit and build upon the knowledge generated in the past, industry's engagement in the research process has become less active. The danger in moving from a position of active engagement in the generation of new knowledge lies in the potential loss of "absorptive capacity" (Cohen & Levinthal, 1990). According to these authors, once an organization ceases investing in its absorptive capacity in a quickly moving field, it may never assimilate and exploit new information in that field, regardless of the value of that information.

# A Short Primer on Bioinformatics

### Introduction

With its implied double dose of biology and computer science, bioinformatics can be an intimidating topic. The word itself is a somewhat fuzzy recent coinage that covers a whole group of rapidly expanding disciplines, loosely coupled. Many of the people active in bioinformatics prefer the more descriptive ( and older ) term *computational microbiology*, but as the use of computers spreads into other areas of biology, it's likely the more general word will take pride of place.

The goal of this writeup is to provide a quick overview of major topics in bioinformatics, and provide the non-technical reader with a perspective on how bioinformatics relates to more traditional biology and computer science.

## Major Topics in Bioinformatics

Bioinformatics is an umbrella term for a variety of diverse areas of interest. What follows is a survey of some the most active areas of current research.

### Text and Data Archiving

Much of the important work being done in bioinformatics parallels the work taking place in every major discipline - making journal articles, papers, and information accessible over the Internet in electronic form. This challenge has embroiled biologists in the usual battles over standards, metadata, markup and copyright familiar to us from the humanities. While everyone agrees on the laudable goal of providing universal access to print archives, actually making this happen requires time and hard work.

The hard sciences are in a more difficult position than many other disciplines, since they also face the challenge of making available prodigious quantities of raw experimental data. Biologists and computer scientists are actively working on standards to represent gene sequences, protein sequences, molecular structure, medical images, and all the other types of complex data they need to exchange as part of their research, but this work is slow going, made difficult by the existence of mutually incompatible, established formats.

Historically, researchers and research groups have stored their data in ad-hoc fashion, based on the needs of a particular experiment. Everybody recognizes that this variety of incompatible formats has become an obstacle to progress and cooperation, and so a good deal of work is going into setting standards for data compatibility.

One useful tool in this area has been **Perl**, a scripting language first created by Larry Wall in 1987. Originally designed as a language for text processing, Perl has matured into a full-featured computer language that lets programmers interconvert data between many formats with minimal hassle. Recognizing the value of Perl as a tool for processing data, the bioinformatics community has taken steps to standardize on a set of modules for handling sequence data. This initiative, which goes by the name **BioPerl**, is a direct analogue to initiatives like OKI for courseware - an attempt at creating a common set of tools for sharing complex data. Similar efforts are being undertaken in other programming languages, with the goal of eventual interoperability across all environments.

### Ensuring Accessibility

To be useful, it's not enough for data to be in the right format - it also has to be made available on the public Internet, and in such a way that both human beings and computer programs running automated searches can find it. The situation is somewhat delicate given the proprietary nature of much of the research data, particularly in areas of interest to the pharmaceutical industry.

Lincoln Stein has likened the current state of affairs in bioinformatics to that of Italy in the time of the city states - a collection of fragmented, divided, and inward-looking principalities, incapable of sustained cooperation. Much of what is truly revolutionary about computational biology won't get off the ground until there is this seamless framework of data that researchers can combine, sift through, or rearrange without having to worry about its provenance. All of these efforts require time, goodwill, and hard work on the part of people who would probably much rather be doing new research.

### Gene sequencing

The most visible and active branch of bioinformatics is the art of **gene sequencing**, sometimes called ``genomics'', including the much-publicized Human Genome Project and its many offspring. While scientists have long known that DNA molecules carry hereditary information, only in the 1990s did advances in sequencing technology make it feasible to sequence the entire genome of anything more complex than a bacterium.

Understanding the significance of gene sequencing ( as well as its limitations ) requires a little background.

## A Very Short DNA Primer

DNA molecules consist of many hundreds of thousands of **nucleotides**, which form complementary pairs down the length of a DNA strand. There are four of these nucleotides in all (called G, A, C, and T), and the specific sequence of nucleotides in a DNA strand encodes information that tells cells how to build particular proteins.

Machinery within the cell reads DNA in groups of three letters, or **codons**. There are sixty-four possible codons, and together they code for 20 different amino acids (the building blocks of protein molecules), as well as processing instructions to the cell, such as ``cut here'' or ``start reading at this point''. The instructions for building a protein are read sequentially by **ribosomes** that build a protein as they go, adding the appropriate amino acid for each chunk of information they read.

For reasons that are not well understood, much of the DNA in a genome never gets read by the cell at all, and doesn't seem to code for anything useful. The parts that do get read are called **coding regions**, and much of the effort in sequencing goes towards identifying and locating these coding regions on a chromosome.

Very few genes in a cell are actually active in protein production at any given time. Different sections of DNA may be dormant or active over the life of a cell, their expression triggered by other genes and changes in the cell's internal environment. How genes interact, why they express at certain times and not others, and how the mechanisms of gene suppression and activation work are all topics of intense interest in microbiology.

***The Mechanics of Sequencing***

The goal of genome sequencing projects is to record all of the genetic information contained in a given organism - that is, create a sequential list of the base pairs comprising the DNA of a particular plant or animal. Since chromosomes consist of long, unbroken strands of DNA, a very convenient way to sequence a genome would be to unravel each chromosome and read off the base pairs like punch tape.

Unfortunately, there is no machine available that can read a single strand of DNA ( ribosomes are very good at it, but nature has not seen fit to provide them with an output jack ). Instead, scientists have to use a cruder, shotgun technique that first chops the DNA into short pieces ( which we know how to identify ) and then tries to reassemble the original sequence based on how the short fragments overlap.

To illustrate the difficulty of the task, imagine that someone gave you a bin containing ten shredded copies of the United States tax code, and asked you to reconstruct the original. The only way you could do it would be to hunt through the shredded bits, look for other shreds with overlapping regions, and assemble larger and larger pieces at a time until you had at least one copy of the whole thing.

Much of the work involved in DNA sequencing is exactly this kind of painstaking labor, with the added caveat that the short fragments invariably contain errors, and that reliably sequencing a single stretch of DNA might involve combining many dozens of duplicate data sets to arrive at an acceptable level of fidelity.

All of this work is done using computers that sift through sequencing data and apply various alignment algorithms to look for overlaps between short DNA fragments. Since DNA in its computer representation is just a long string of letters, these algorithms are close cousins of text analysis techniques that have been used for years on electronic documents, creating a curious overlap between genetics and natural language processing.

The work of assembling a sequence requires that many data sets be combined together, and common errors ( dropped sequences, duplications, backwards sequences ) detected and eliminated from the final result. Much of the work taking place right now on data from the Human Genome Project is just this kind of careful computerized analysis and correction.

***Finding the Genes***

Once a reliable DNA sequence has been established, there still remains the task of finding the actual genes ( coding regions ) embedded within the DNA strand. Since a large proportion of the DNA in a genome is non-coding ( and presumably unused ), it is important to find and delimit the coding regions as a first step to figuring out what they do.

This search for coding regions is also done with the help of computer algorithms, and once again there is a good amount of borrowing from areas like signal processing, cryptography, and natural language processing - all techniques that are good at distinguishing information from random noise.

Finding the coding regions is an important step in genome analysis, but it is not the end of the road. An ongoing debate in genetics has been the question of how much information is actually encoded in the genome itself, and how much is carried by the complex system surrounding gene expression.

A useful analogy here (borrowed from Douglas Hofstadter) is that of the ant colony. While a single ant has only a small repertoire of behaviors, and responds predictably to a handful of chemical signals, a colony of many thousands of ants will show very subtle and sophisticated behavior in how it forages for food, raises its young, manages resources and deals with external threats. No amount of study of an individual ant can reveal anything about the behavior of a colony, because the properties of the colony don't reside in any single ant - they arise spontaneously out of the interactions between many thousands of ants, all obeying simple rules. This phenomenon of **emergent behavior** is known to play a part in gene expression, but nobody knows to what extent.

Emergent behavior is very hard to simulate, because there is no way to infer the simple rules from the complex behavior - we are forced to proceed by trial and error. Still, computers give us a way to try many different rule sets and test hypotheses about gene interaction that would take decades to work out on pencil and paper.

### *Unexplored Territory*

Surprisingly enough, for all the intense research taking place in genomics, only a very few organisms have had their genome fully sequenced ( we are in the proud company of the puffer fish, the fruit fly, brewer's yeast, and several kinds of bacteria). Many technical and computational challenges remain before sequencing becomes an automatic process - some species are still very difficult for us to sequence, and much remains to be learned about the role and origin of that entire non-coding DNA. Progress will require both advances in the lab and advances in our ability to analyze and process the genome data with computers.

## Molecular Structure and Function

Closely related to gene sequencing is a second major field of interest in bioinformatics - the search for a mapping between the chemical structure of a protein and its function within the cell.

We noted above that proteins are assembled from amino acids based on instructions encoded in an organism's DNA. Even though all proteins are made out of the same building blocks, they come in a bewildering variety of forms and serve many functions within the cell. Some proteins are purely structural, others have special receptors that fit other molecules, and still others serve as signals or messages that can pass information from cell to cell. The role a protein plays depends solely on its shape, and the shape of a protein depends on the sequence of amino acids that compose it.

Amino acid chains start out as linear molecules, but as more amino acids are added, the chain begins to fold up. This spontaneous folding result in a complex, three-dimensional structure unique to the particular protein being generated. The pattern of folding is automatic and reproducible, so that a given amino acid sequence will always create a protein with a certain configuration.

The ability to predict the final shape of a protein from its amino acid composition is the Holy Grail of pharmacology. If we could design protein molecules from scratch, it would become possible to create potent new drugs and therapies, tailored to the individual. Finding treatments for a disease would be as simple as creating a protein to fit around the business ends of an infectious agent and render it harmless.

The **protein folding problem**, as it is called, is computationally very difficult. It may even be an intractable problem. An enormous amount of effort continues to go into finding a mapping between the amino acid sequence of a molecule, its ultimate configuration, and how its structure affects its function.

Distributed computing plays a critical role in studying protein folding. As computing power increases, it will become possible to test more sophisticated models of folding behavior, and more accurately estimate the intramolecular forces within individual proteins to understand why they fold the way they do.

A better understanding of protein folding is also critical to finding therapies for **prion**-based diseases like bovine spongiform encephalopathy ( BSE, or mad cow disease ) which appears to be caused by a 'rogue' configuration of a common protein that in turn reconfigures other proteins it comes into contact with. Prion diseases are poorly understood and present a grave risk, since one protein molecule is presumably enough to infect an entire organism, and common sterilization techniques are not sufficient to destroy infectious particles.

**Molecular Evolution**

A third application of bioinformatics, closely related to genomics and protein analysis, is the study of **molecular evolution**.

Molecular evolution is a statistical science that uses changes in genes and proteins as a kind of evolutionary clock. Over time, any species will accumulate minor mutations in its genetic code, due to inevitable transcription errors and environmental factors like background radiation. Some of these mutations prove fatal; a tiny fraction prove beneficial, but the vast majority have no noticeable effect on the organism.

Because this trickle of small changes is slow, cumulative, and essentially random, it can serve as a useful indicator of how long two species have been out of genetic contact - that is, when they last shared a common ancestor. By comparing molecules across many species, scientists can determine the relative genetic distance between them, and learn more about their evolutionary history.

The study of molecular markers in evolution builds on on the same sequencing and analysis techniques discussed in the section on DNA. Because it relies on tiny changes to the genome, ths kind of statistical analysis requires very precise sequencing data and sophisticated computational models, including advanced **GIS** (geographic information systems) software to correlate a species' genetic history with its historical range.

## Modeling and Simulations

The final field we will cover, computer modeling of biological systems, is the most closely linked to traditional computer science and in many ways the most accessible.

### *Computer modeling*

Ever since Conway's Game of Life in 1970, computer scientists have been creating simulations of complex systems that generate intricate behavior from a simple set of underlying rules. Research in modeling is active in two directions - simulating living systems to gain a better understanding of the rules underpinning their behavior, and applying models observed in nature to help solve abstract problems in unrelated fields.

The first approach has been used to better understand forests, predator-prey relationships, and other complex ecosystems, and has taken its place as an invaluable teaching tool at all levels of biology. Students can use computer simulations to observe firsthand the effects that different constraints and starting conditions can have on a biological system.

Since biological factors are known to have a critical effect on climate, much of the ongoing research on global warming and long-range weather patterns also relies on sophisticated models of plant and ocean ecology.

The second approach to modeling - stealing clever algorithms from nature - has led to some interesting and surprising solutions to thorny problems. For example, researchers in ant behavior have been able to use simulated 'ants' and 'pheromones' running around a shipping map to find more efficient routings for companies like UPS and Federal Express, saving those companies millions of dollars a year in operating expenses.

### *Biophysics*

An interesting and little-mentioned side branch of computer modeling is the study of **biophysics**. Biophysics deals with plants and animals from an engineering perspective, trying to figure out how what holds them up, how they get from place to place, and what ideas we can steal from nature to improve our own building skills. Biophysics also helps scientists make inferences about long-extinct animals based on fossil evidence - figuring out how fast a dinosaur could run, or what kind of environment was necessary for giant insects to flourish.

While not as well-funded or highly publicized as efforts in microbiology, biophysics still represents an innovative application of computer science to biology, and has important applications in an eclectic set of fields: paleontology, structural engineering, hydrodynamics, and sports medicine.

## Bioethics and the Scientific Commons

It's impossible to talk about bioinformatics without discussing the charged legal and ethical context surrounding the field. The legal issues clouding research into microbiology, especially in the area of genetics, are severe and likely to get worse.

### *Biopatents and the research community*

There is a real tension between, on the one hand, the scientific tradition of peer review and open sharing of information, and on the other hand, the very high value of some of this information to business. Private companies have successfully been able to acquire patents on individual genes, and in some cases entire organisms. This unprecedented extension of patent law into the realm of living creatures is not the result of explicit policy decisions by elected representatives, but of administrative decisions made out of the public eye.

Given the enormously lucrative nature of patents and pharmacological research data, many public universities are facing an unpleasant conflict of interest between their need for income and their mission as educational institutions charged with serving the public. More and more often, a university's interest in maintaining a scientific commons gets weighed against the potential monetary value of patents and trade secrets.

### *Genetically modified foods*

Advances in gene sequencing and manipulation techniques have rapidly been put to use by multinational agribusiness concerns. Much of the food supply in this country now comes from genetically modified crops, whose safety has never been demonstrated, and whose potential for causing permanent changes to the ecosystem is poorly understood.

The introduction of bioengineered plants and animals into the wild is a massive, ongoing uncontrolled experiment. Given our marginal understanding of even the most basic aspects of gene expression and cross-species gene transfer, the lack of a public debate on the subject of transgenic crops and animals has been somewhat shocking.

### *Privacy and public health*

Any increase in our understanding of hereditary and congenital traits in the human genome has implications for privacy, individual choice, and public health policy. As screening for potentially fatal diseases becomes available, sometimes years in advance, we will all have to make difficult personal and collective choices about the kind of health care system we want, and the degree to which we are ready to sacrifice our privacy in the interests of public health.

# BIOINFORMATICS AND DRUG DISCOVERY

In recent years, we have seen an explosion in the amount of biological information that is available. Various databases are doubling in size every 15 months and we now have the complete genome sequences of more than 100 organisms. It appears that the ability to generate vast quantities of data has surpassed the ability to use this data meaningfully. The pharmaceutical industry has embraced genomics as a source of drug targets. It also recognises that the field of bioinformatics is crucial for validating these potential drug targets and for determining which ones are the most suitable for entering the drug development pipeline.

Recently, there has been a change in the way that medicines are being developed due to our increased understanding of molecular biology. In the past, new synthetic organic molecules were tested in animals or in whole organ preparations. This has been replaced with a molecular target approach in which in-vitro screening of compounds against purified, recombinant proteins or genetically modified cell lines is carried out with a high throughput. This change has come about as a consequence of better and ever improving knowledge of the molecular basis of disease.

All marketed drugs today target only about 500 gene products. The elucidation of the human genome which has an estimated 30,000 to 40,000 genes, presents immense new opportunities for drug discovery and simultaneously creates a potential bottleneck regarding the choice of targets to support the drug discovery pipeline. The major advances in genomics and sequencing means that finding an attractive target is no longer a problem but finding the targets that are most likely to succeed has become the challenge. The focus of bioinformatics in the drug discovery process has therefore shifted from target identification to target validation.

A lot of factors need to be taken into account concerning a candidate target from a multitude of heterogeneous resources. The types of information that one needs to gather about potential targets include nucleotide and protein sequencing information, homologues, mapping information, function prediction, pathway information, disease associations, variants, structural information, gene and protein expression data and species/taxonomic distribution among others. Different bioinformatics tools can be used to gather this information. The accumulation of this information into databases about potential targets means that the pharmaceutical companies can save themselves much time, effort and expense exerting bench efforts on targets that will ultimately fail. The information that is gathered helps to characterise the different targets into families and subfamilies. It also classifies the behaviour of the different molecules in a biochemical and cellular context. Decisions about which families provide the best potential targets is guided by a number of criteria. It is important that the potential target has a suitable structure for interacting with drug molecules. Structural genomics helps to prioritise the families in terms of their 3D structures.

Sometimes we want to develop broad spectrum drugs that are effective against a wide range of pathogenic species while at other times we want to develop narrow spectrum drugs that are highly specific to a particular organism. Comparative genomics helps to find protein families that are widely taxonomically dispersed and those that are unique to a particular organism.

For example, when we want to develop a broad spectrum antibiotic, we are looking for targets that are present in a large number of bacteria yet have no similar homologues in human. This means that the antibiotic will be effective against many bacteria killing them while causing no harm to the human. In order to determine the role our potential drug target plays in a particular disease mechanism we use DNA and protein chips. These chips can measure the amount of transcript or protein expressed by a cell at different times or in different states (healthy versus diseased).

Clustering algorithms are used to organise this expression data into different biologically relevant clusters. We can then compare the expression profiles from the diseased and healthy cells to help us understand the role our gene or protein plays in a disease process. All of these computational tools can help to compose a detailed picture about a protein family, its involvement in a disease process and its potential as a possible drug target.

Following on from the genomics explosion and the huge increase in the number of potential drug targets, there has been a move from the classical linear approach of drug discovery to a non linear and high throughput approach. The field of bioinformatics has become a major part of the drug discovery pipeline playing a key role for validating drug targets. By integrating data from many inter-related yet heterogeneous resources, bioinformatics can help in our understanding of complex biological processes and help improve drug discovery.

## Drug Design based on Bioinformatics Tools

The processes of designing a new drug using bioinformatics tools have open a new area of research. However, computational techniques assist one in searching drug target and in designing drug in silco, but it takes long time and money. In order to design a new drug one need to follow the following path.

- **Identify Target Disease:** One needs to know all about the disease and existing or traditional remedies. It is also important to look at very similar afflictions and their known treatments.

  Target identification alone is not sufficient in order to achieve a successful treatment of a disease. A real drug needs to be developed. This drug must influence the target protein in such a way that it does not interfere with normal metabolism. One way to achieve this is to block activity of the protein with a small molecule. Bioinformatics methods have been developed to virtually screen the target for compounds that bind and inhibit the protein. Another possibility is to find other proteins that regulate the activity of the target by binding and forming a complex.

- **Study Interesting Compounds:** One needs to identify and study the lead compounds that have some activity against a disease. These may be only marginally useful and may have severe side effects. These compounds provide a starting point for refinement of the chemical structures.
- **Detect the Molecular Bases for Disease:** If it is known that a drug must bind to a particular spot on a particular protein or nucleotide then a drug can be tailor made to bind at that site. This is often modeled computationally using any of several different techniques. Traditionally, the primary way of determining what compounds would be tested computationally was provided by the researchers' understanding of molecular interactions. A second method is the brute force testing of large numbers of compounds from a database of available structures.
- **Rational drug design techniques:** These techniques attempt to reproduce the researchers' understanding of how to choose likely compounds built into a software package that is capable of modeling a very large number of compounds in an automated way. Many different algorithms have been used for this type of testing, many of which were adapted from artificial intelligence applications. The complexity of biological systems makes it very difficult to determine the structures of large biomolecules. Ideally experimentally determined (x-ray or NMR) structure is desired, but biomolecules are very difficult to crystallize.
- **Refinement of compounds:** Once you got a number of lead compounds have been found, computational and laboratory techniques have been very successful in refining the molecular structures to give a greater drug activity and fewer side effects. This is done both in the laboratory and computationally by examining the molecular structures to determine which aspects are responsible for both the drug activity and the side effects.
- **Quantitative Structure Activity Relationships (QSAR):** This computational technique should be used to detect the functional group in your compound in order to refine your drug. This can be done using QSAR that consists of computing every possible number that can describe a molecule then doing an enormous curve fit to find out which aspects of the molecule correlate well with the drug activity or side effect severity. This information can then be used to suggest new chemical modifications for synthesis and testing.

- **Solubility of Molecule:** One need to check whether the target molecule is water soluble or readily soluble in fatty tissue will affect what part of the body it becomes concentrated in. The ability to get a drug to the correct part of the body is an important factor in its potency. Ideally there is a continual exchange of information between the researchers doing QSAR studies, synthesis and testing. These techniques are frequently used and often very successful since they do not rely on knowing the biological basis of the disease which can be very difficult to determine.
- **Drug Testing:** Once a drug has been shown to be effective by an initial assay technique, much more testing must be done before it can be given to human patients. Animal testing is the primary type of testing at this stage. Eventually, the compounds, which are deemed suitable at this stage, are sent on to clinical trials. In the clinical trials, additional side effects may be found and human dosages are determined.

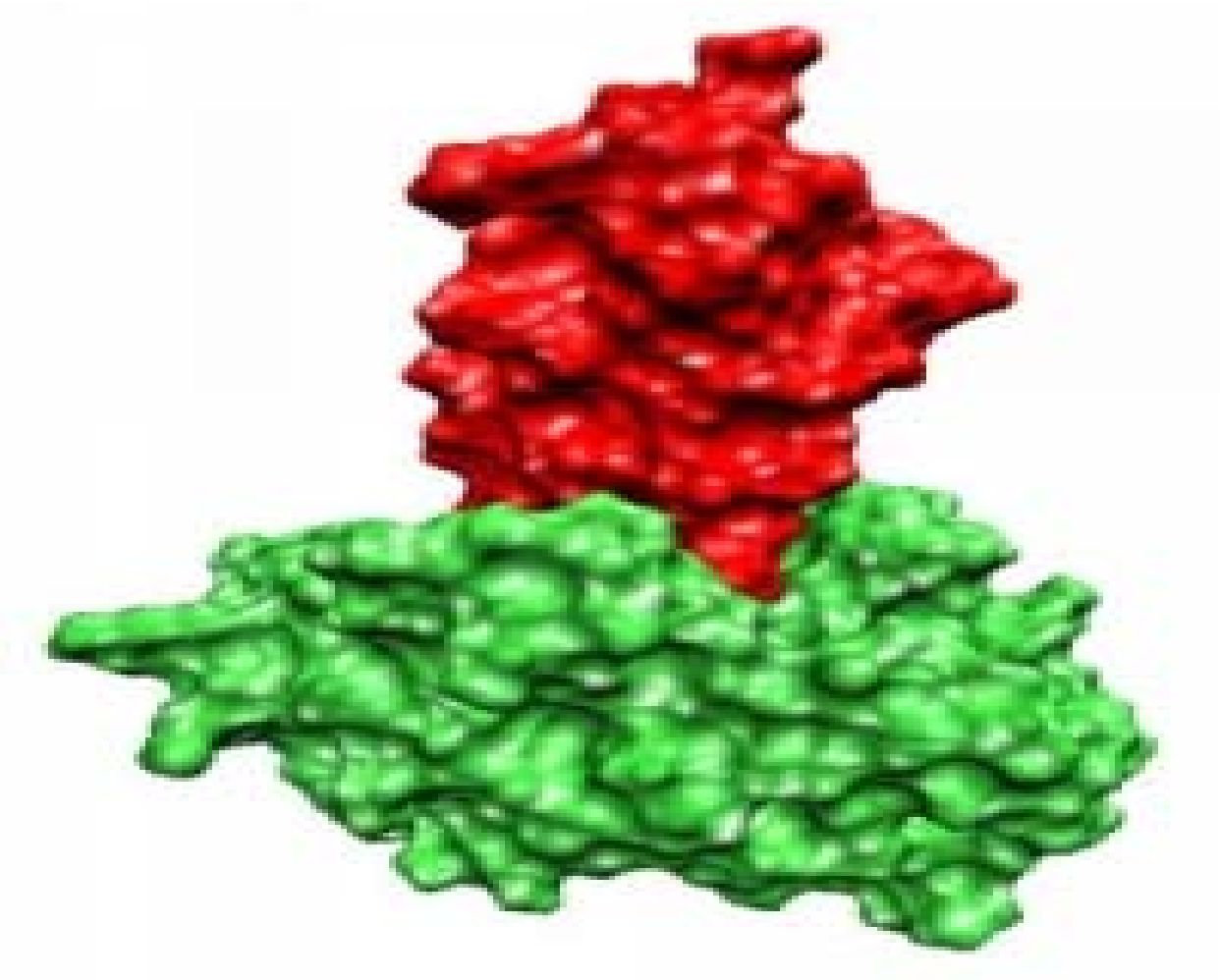

Two proteins whose complex was determined by Protein-Protein-Docking.

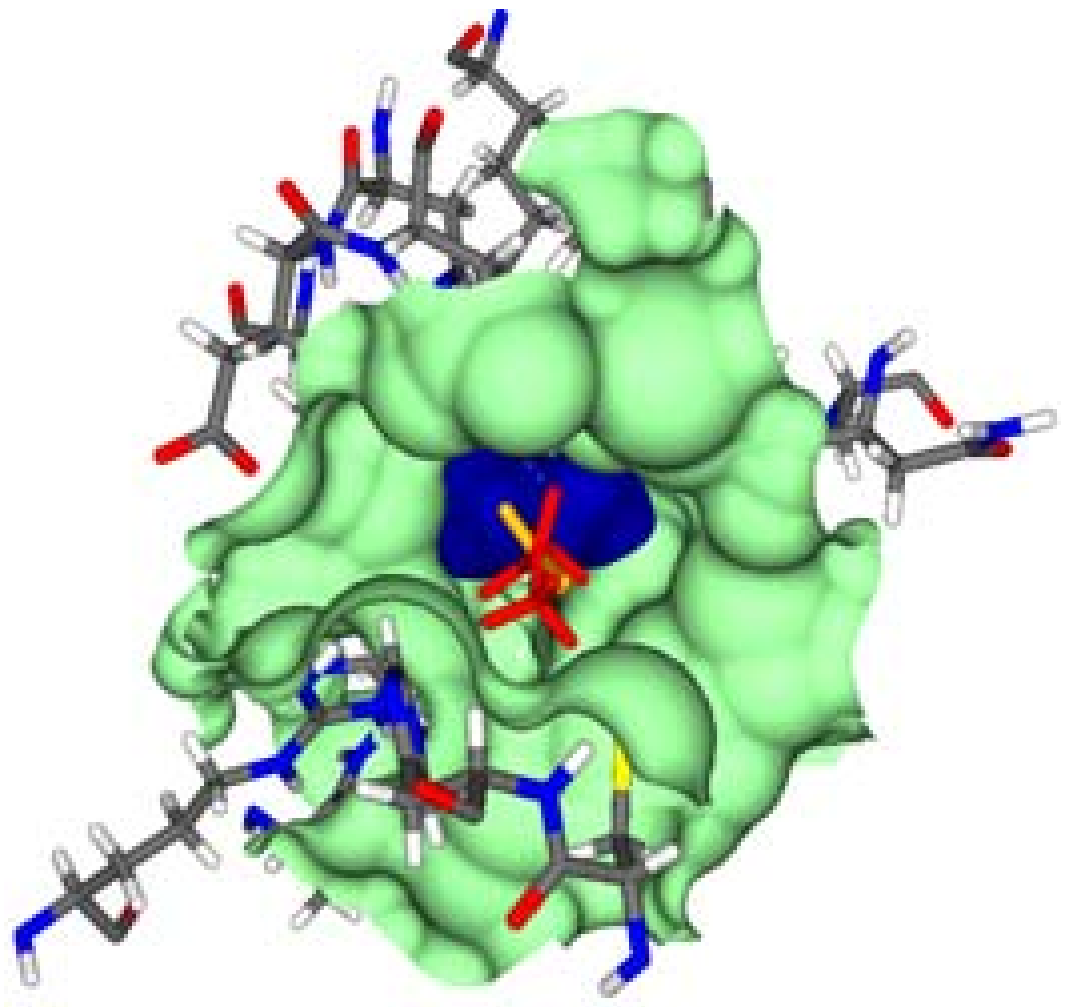

Docking of a small inhibitor to the urease protein.

# DNA - deoxyribonucleic acid

DNA the magic code of life, the form of which was identified in 1953 as a double helix of deoxyribonucleic acid (DNA), is made up of two strands of four bases in varying triplets that repeat over and over in a very long molecule. It is a very complex code, the human DNA is over 3 billion characters long. It can be used to identify specific human beings and has been used to clone animals. In a less dramatic but important application, DNA is being used as an information carrier and as an encryption and computation device. The DNA code could theoretically carry a whole written novel in one single molecule.

DNA is a long fiber, like a hair, only thinner and longer. It is made from two strands that stick together with a slight twist.

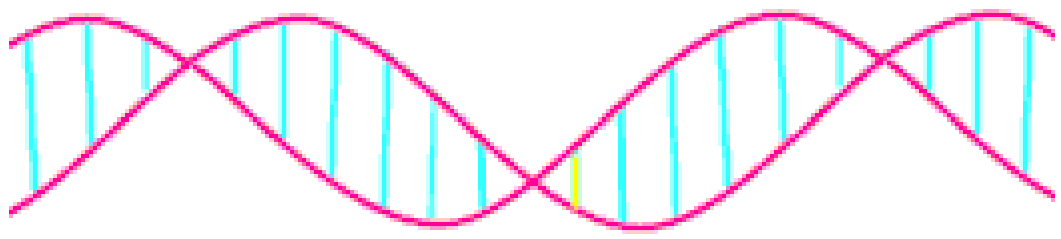

Proteins attach to the DNA and help the strands coil up into a chromosome when the cell gets ready to divide.

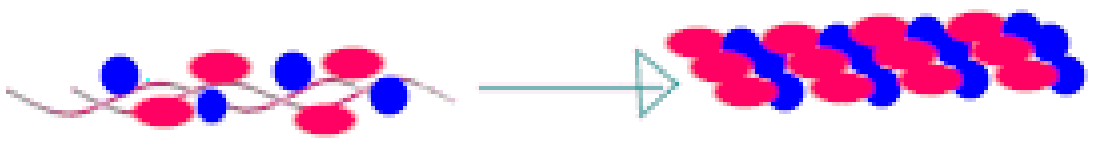

The DNA is organized into stretches of genes, stretches where proteins attach to coil the DNA into chromosomes, stretches that "turn a gene on" and "turn a gene off", and large stretches whose purpose is not yet known to scientists.

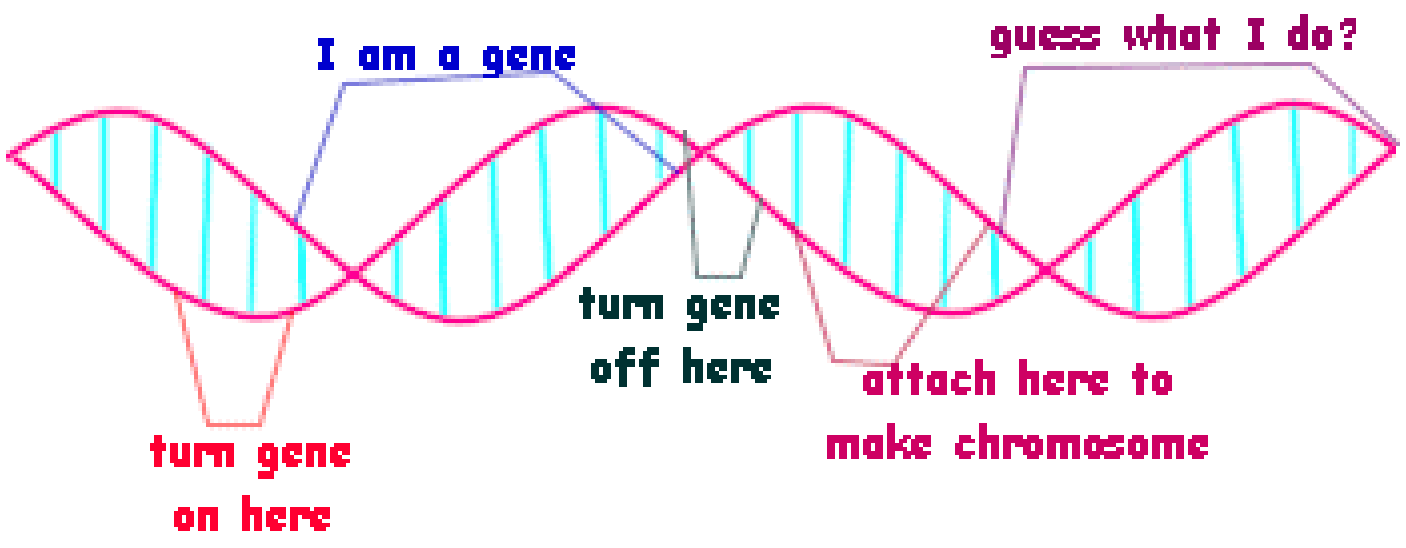

The genes carry the instructions for making all the thousands of proteins that are found in a cell. The proteins in a cell determine what that cell will look like and what jobs that cell will do. The genes also determine how the many different cells of a body will be arranged. In these ways, DNA controls how many fingers you have, where your legs are placed on your body, and the color of your eyes.

A chromosome is made up of DNA and the proteins attached to it. There are 23 pairs of chromosomes in a human cell. One of each pair was inherited from your mother and the other from your father. DNA is a particular bio-molecule. All of the DNA in a cell is found in individual pieces, called chromosomes. This would be like muffins. Muffins are made up of muffin-matter and paper cups. All of the muffin-matter in your kitchen is found in individual pieces, called muffins.

A revolution has occurred in the last few decades that explains how DNA makes us look like our parents and how a faulty gene can cause disease. This revolution opens the door to curing illness, both hereditary and contracted. The door has also been opened to an ethical debate over the full use of our new knowledge.

## Introduction to DNA Structure

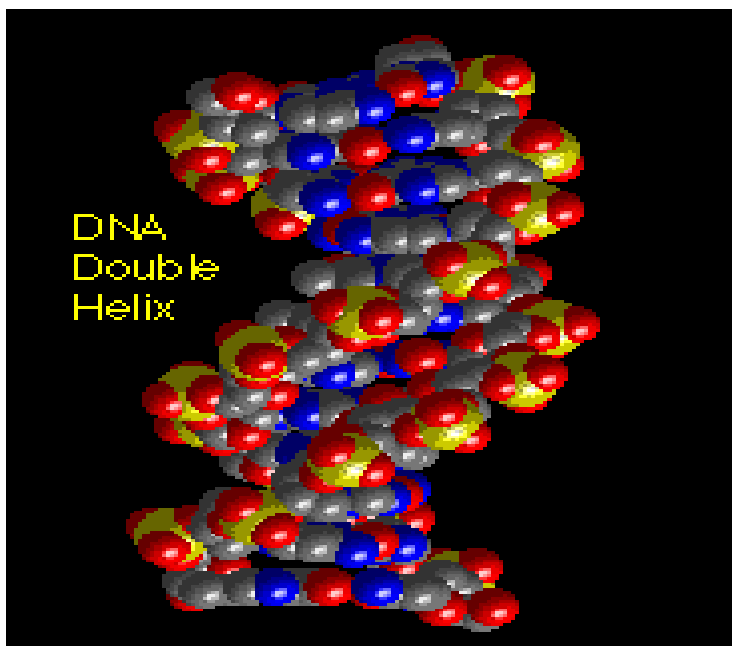

**Components of DNA**
DNA is a polymer. The monomer units of DNA are nucleotides, and the polymer is known as a "polynucleotide." Each nucleotide consists of a 5-carbon sugar (deoxyribose), a nitrogen containing base attached to the sugar, and a phosphate group. There are four different types of nucleotides found in DNA, differing only in the nitrogenous base. The four nucleotides are given one letter abbreviations as shorthand for the four bases.

- A is for adenine
- G is for guanine
- C is for cytosine
- T is for thymine

**Purine Bases**
Adenine and guanine are purines. Purines are the larger of the two types of bases found in DNA. Structures are shown below:

**Structure of A and G**

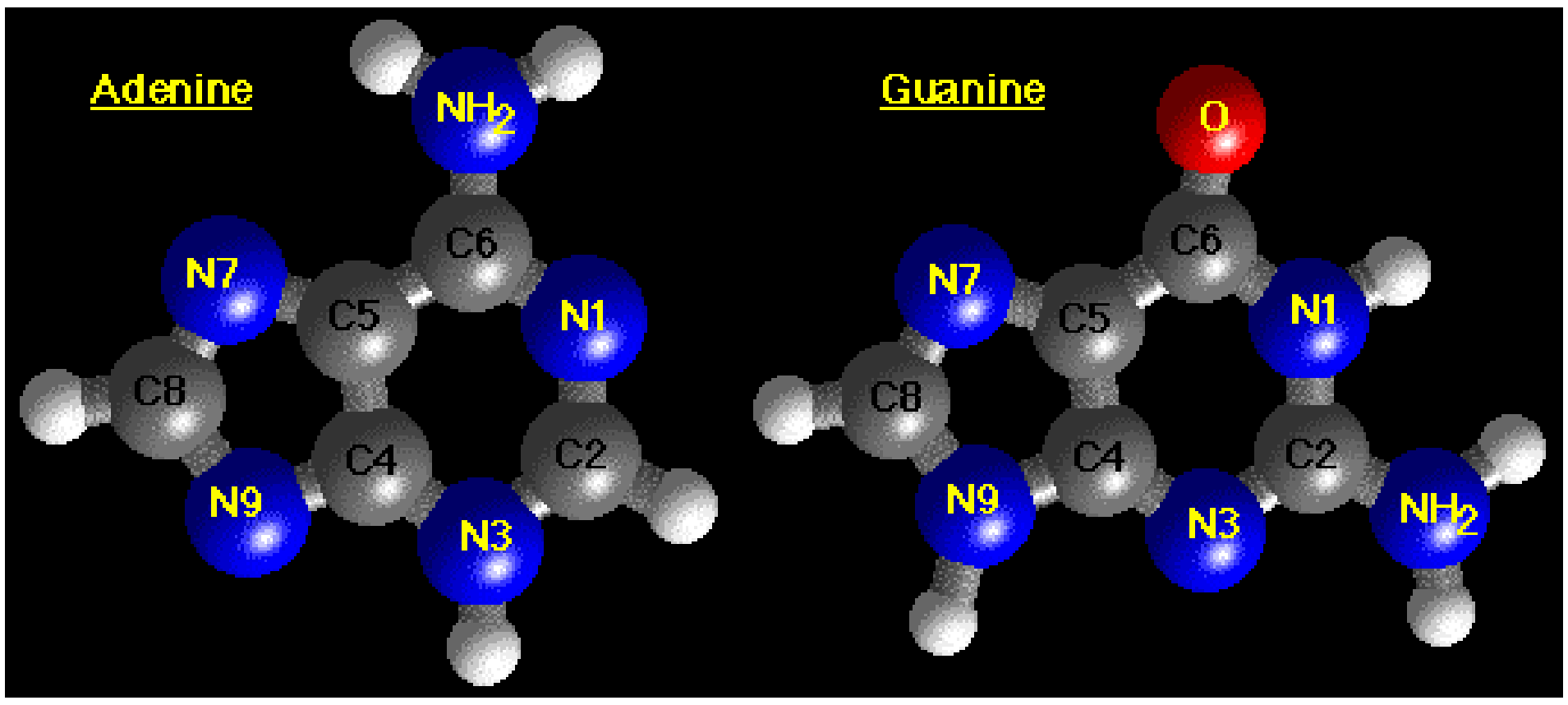

The 9 atoms that make up the fused rings (5 carbon, 4 nitrogen) are numbered 1-9. All ring atoms lie in the same plane.

**Pyrimidine Bases**
Cytosine and thymine are pyrimidines. The 6 stoms (4 carbon, 2 nitrogen) are numbered 1-6. Like purines, all pyrimidine ring atoms lie in the same plane.

**Structure of C and T**

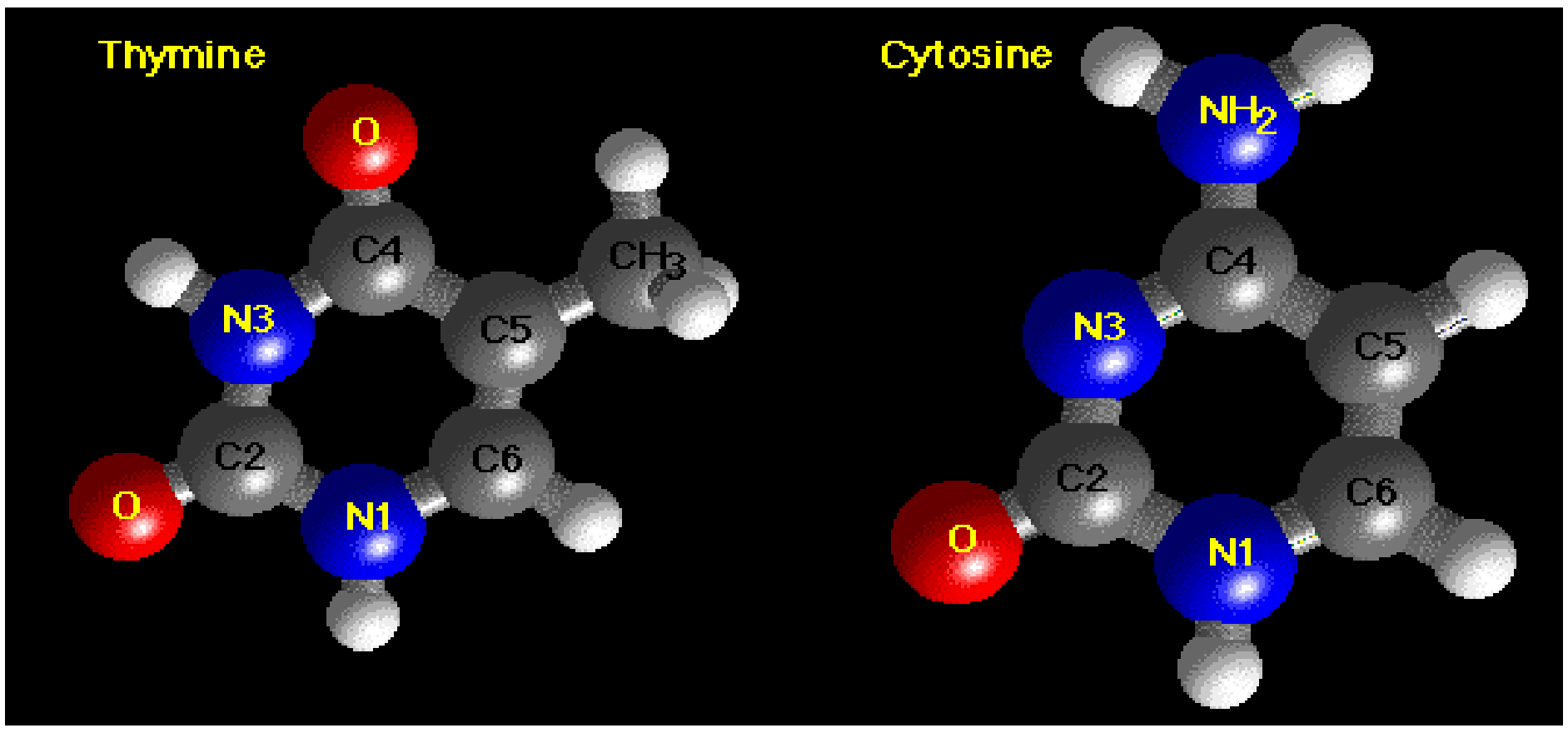

**Deoxyribose Sugar**
The deoxyribose sugar of the DNA backbone has 5 carbons and 3 oxygens. The carbon atoms are numbered 1', 2', 3', 4', and 5' to distinguish from the numbering of the atoms of the purine and pyrmidine rings. The hydroxyl groups on the 5'- and 3'- carbons link to the phosphate groups to form the DNA backbone. Deoxyribose lacks an hydroxyl group at the 2'-position when compared to ribose, the sugar component of RNA.

**Structure of deoxyribose**

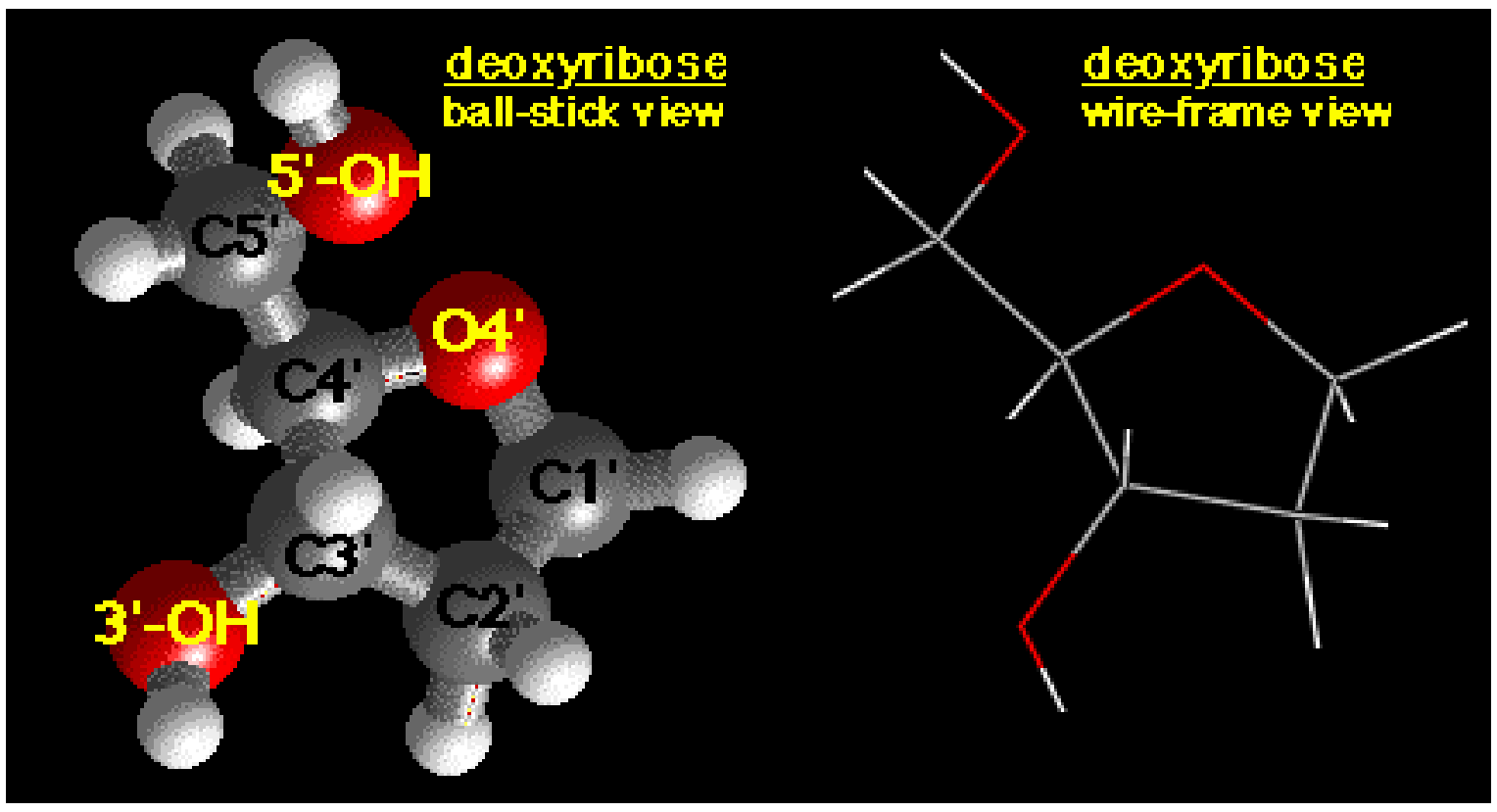

**Nucleosides**
A nucleoside is one of the four DNA bases covalently attached to the C1' position of a sugar. The sugar in deoxynucleosides is 2'-deoxyribose. The sugar in ribonucleosides is ribose. Nucleosides differ from nucleotides in that they lack phosphate groups. The four different nucleosides of DNA are deoxyadenosine (dA), deoxyguanosine (dG), deoxycytosine (dC), and (deoxy)thymidine (dT, or T).

Structure of dA

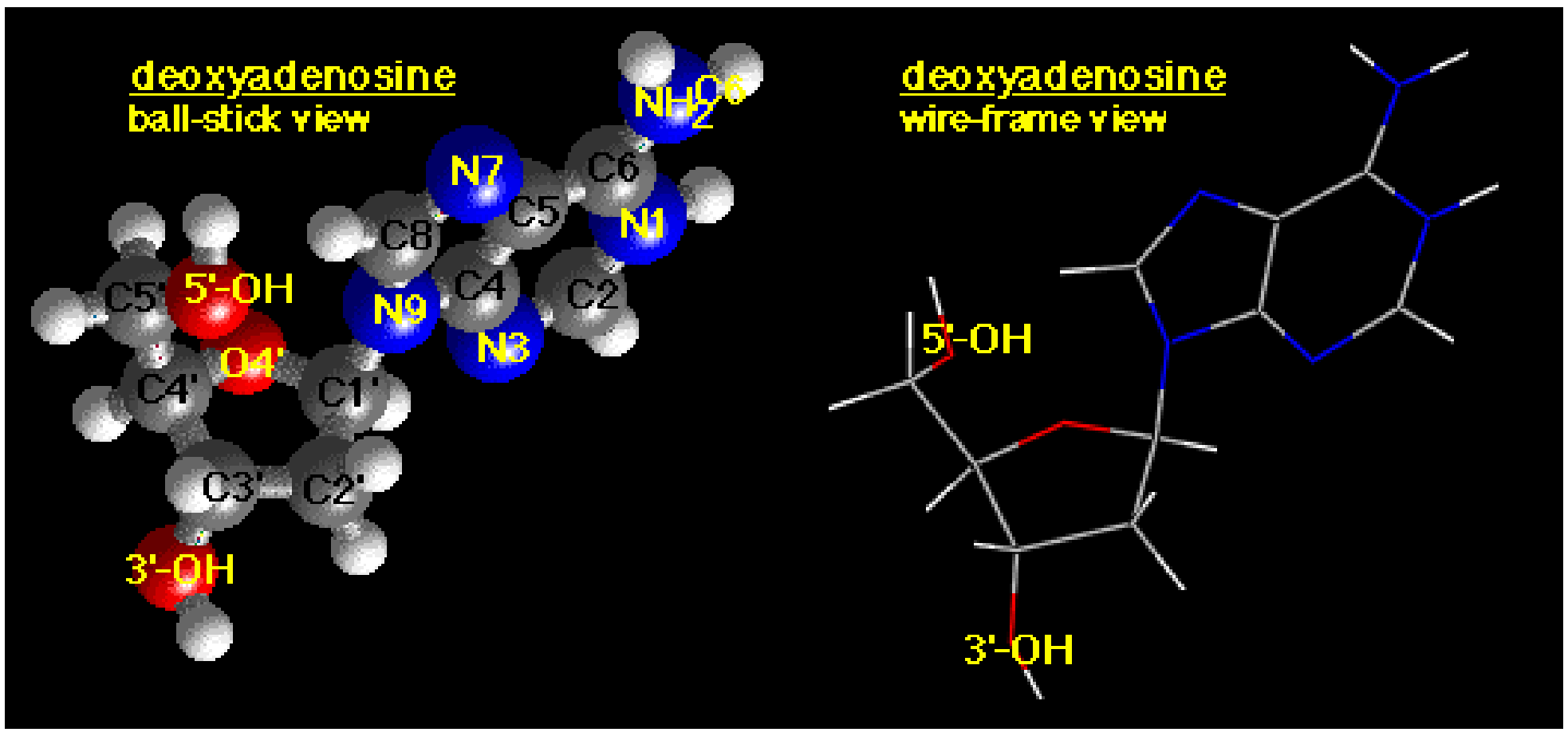

In dA and dG, there is an "N-glycoside" bond between the sugar C1' and N9 of the purine.

**Nucleotides**
A nucleotide is a nucleoside with one or more phosphate groups covalently attached to the 3'- and/or 5'-hydroxyl group(s).

**DNA Backbone**
The DNA backbone is a polymer with an alternating sugar-phosphate sequence. The deoxyribose sugars are joined at both the 3'-hydroxyl and 5'-hydroxyl groups to phosphate groups in ester links, also known as "phosphodiester" bonds.

Example of DNA Backbone: 5'-d(CGAAT):

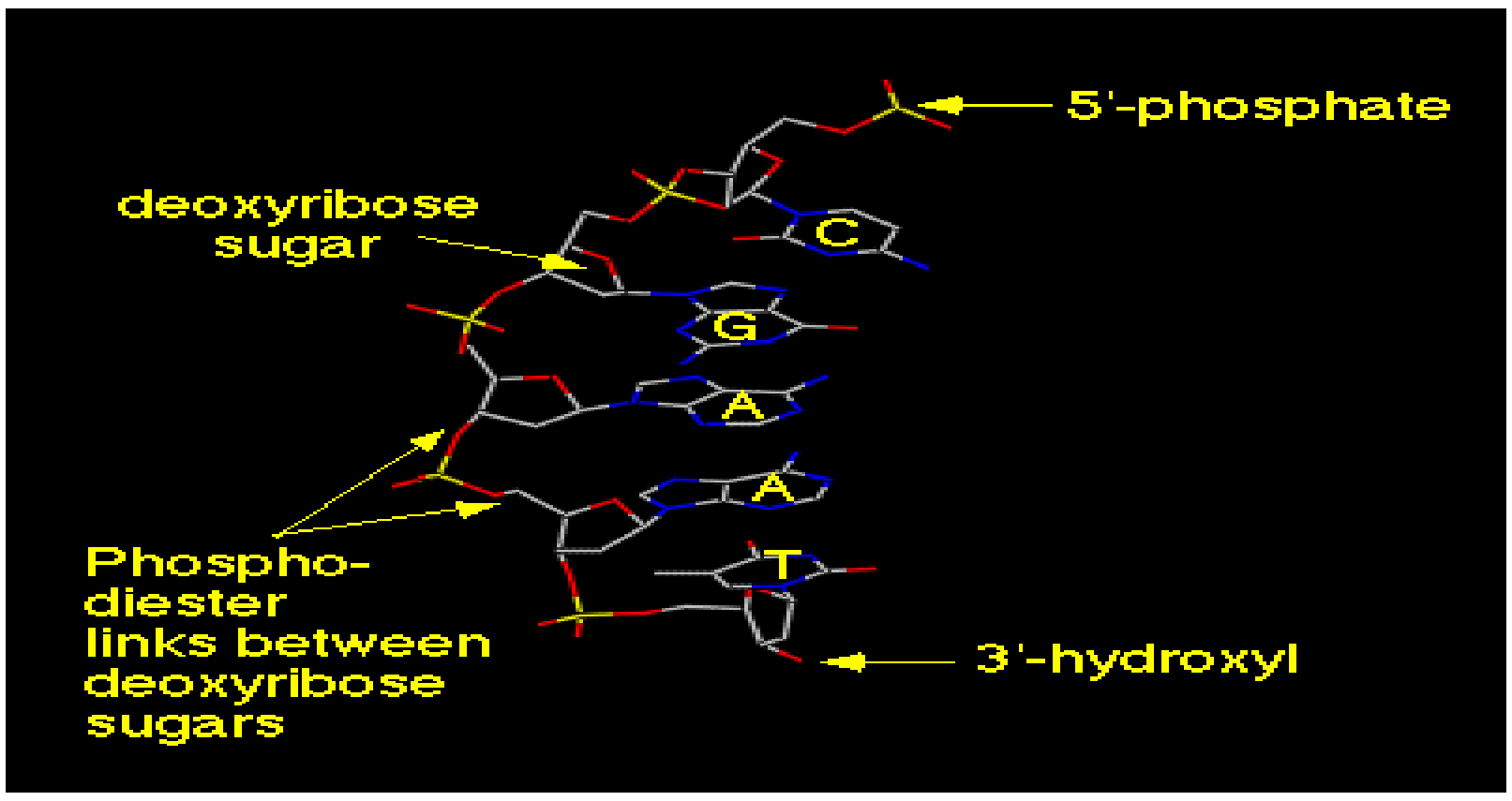

**Features of the 5'-d(CGAAT) structure:**

- Alternating backbone of deoxyribose and phosphodiester groups
- Chain has a direction (known as polarity), 5'- to 3'- from top to bottom
- Oxygens (red atoms) of phosphates are polar and negatively charged
- A, G, C, and T bases can extend away from chain, and stack atop each other
- Bases are hydrophobic

**DNA Double Helix**
DNA is a normally double stranded macromolecule. Two polynucleotide chains, held together by weak thermodynamic forces, form a DNA molecule.

**Structure of DNA Double Helix**

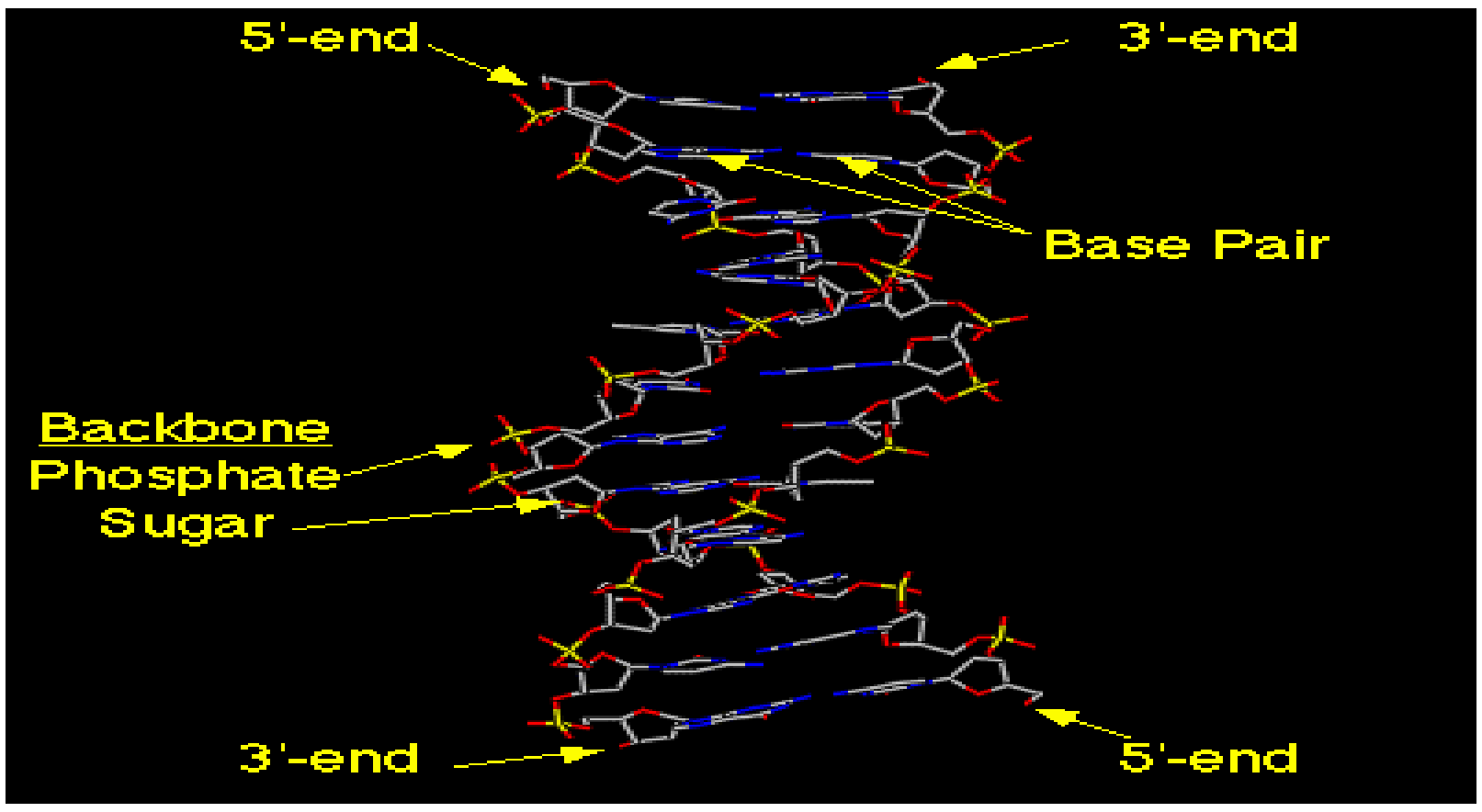

***Features of the DNA Double Helix***

- Two DNA strands form a helical spiral, winding around a helix axis in a right-handed spiral
- The two polynucleotide chains run in opposite directions
- The sugar-phosphate backbones of the two DNA strands wind around the helix axis like the railing of a sprial staircase
- The bases of the individual nucleotides are on the inside of the helix, stacked on top of each other like the steps of a spiral staircase.

**Base Pairs**
Within the DNA double helix, A forms 2 hydrogen bonds with T on the opposite strand, and G forms 3 hyrdorgen bonds with C on the opposite strand.

Example of dA-dT base pair as found within DNA double helix

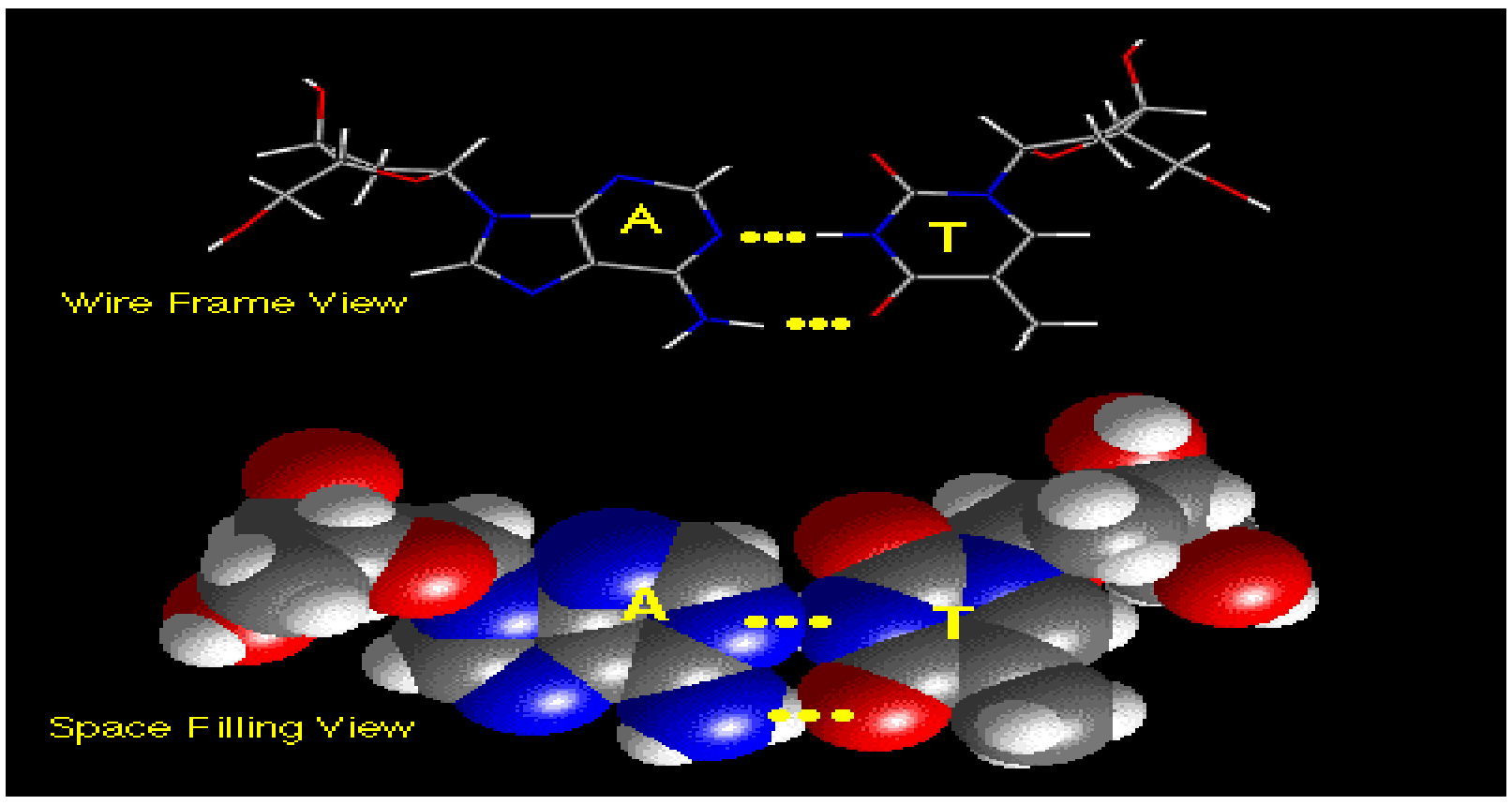

Example of dG-dC base pair as found within DNA double helix

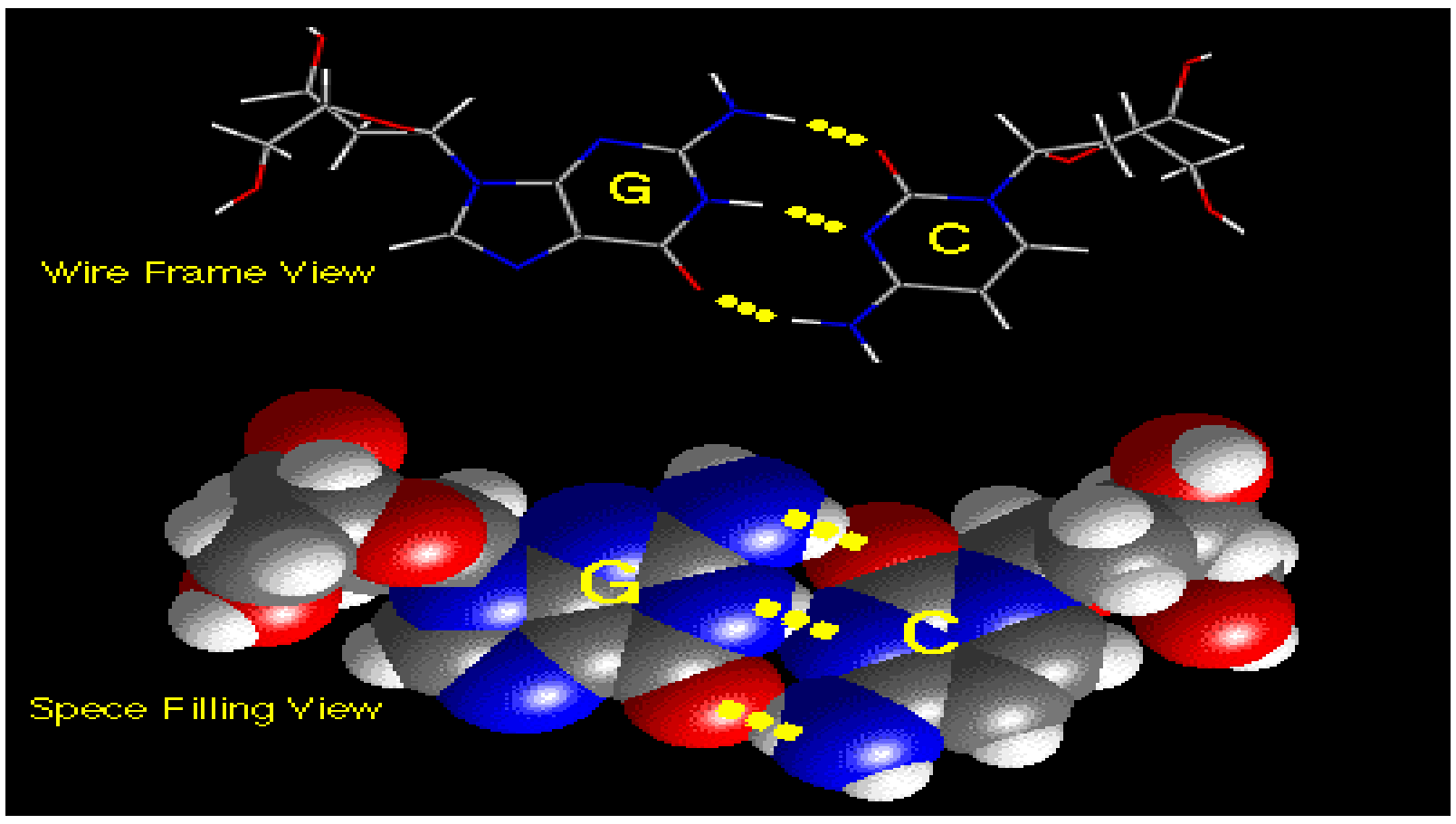

- dA-dT and dG-dC base pairs are the same length, and occupy the same space within a DNA double helix. Therefore the DNA molecule has a uniform diameter.
- dA-dT and dG-dC base pairs can occur in any order within DNA molecules

**DNA Helix Axis**

The helix axis is most apparent from a view directly down the axis. The sugar-phosphate backbone is on the outside of the helix where the polar phosphate groups (red and yellow atoms) can interact with the polar environment. The nitrogen (blue atoms) containing bases are inside, stacking perpendicular to the helix axis.

# SEQUENCE ANALYSIS

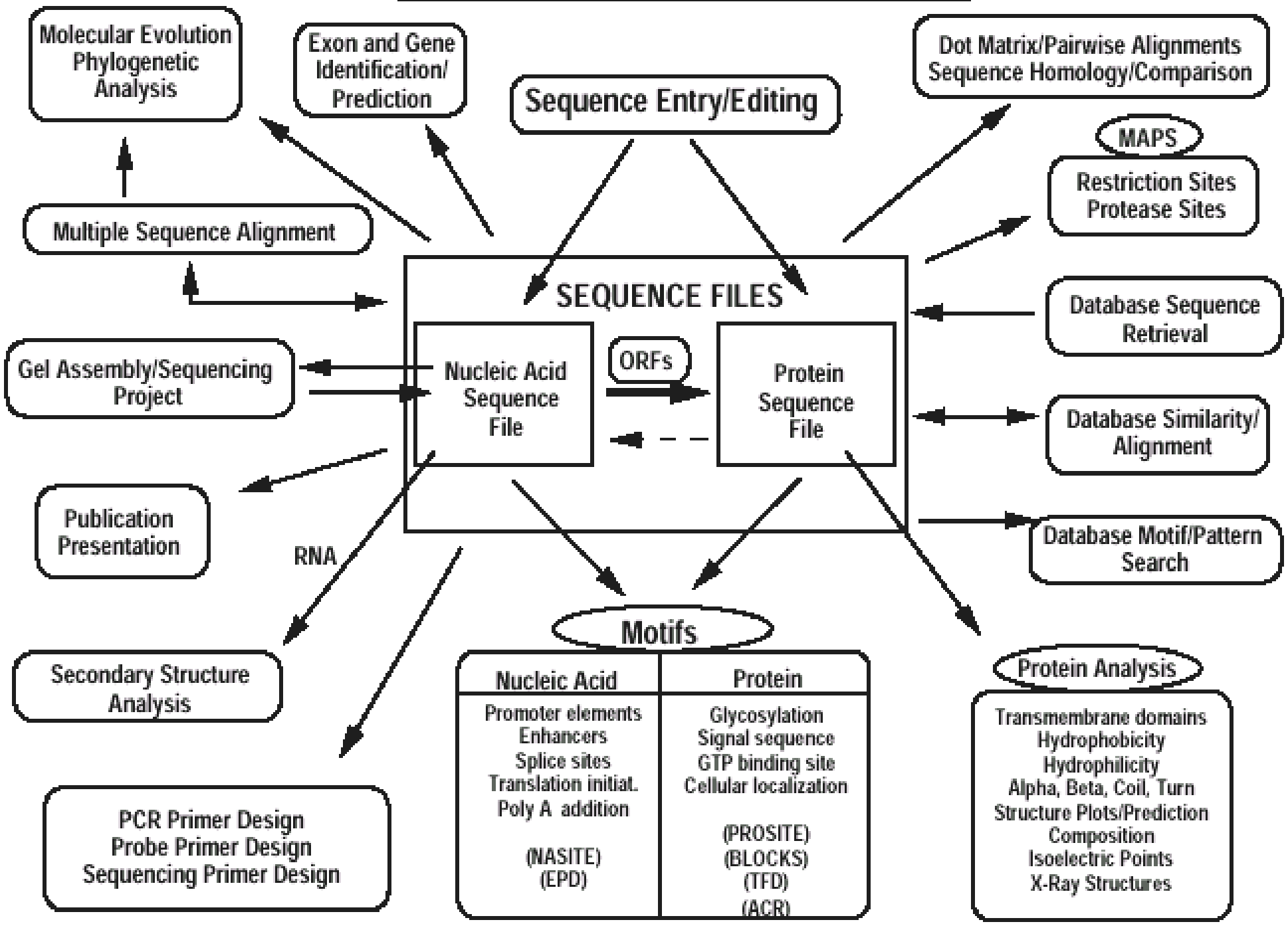

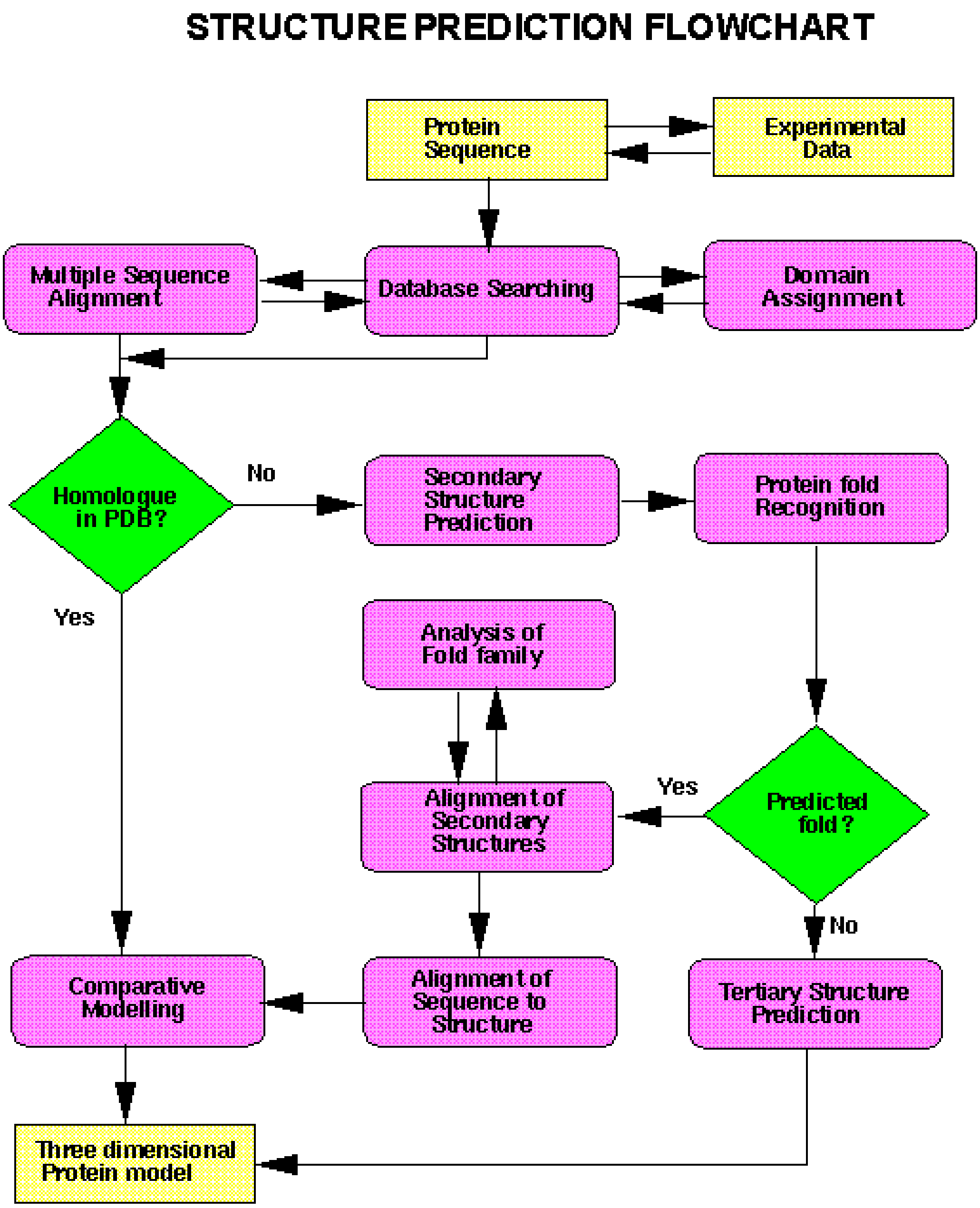

## Experimental Data

Much experimental data can aid the structure prediction process. Some of these are:

- Disulphide bonds, which provide tight restraints on the location of cysteines in space
- Spectroscopic data, which can give you and idea as to the secondary structure content of your protein
- Site directed mutagenesis studies, which can give insights as to residues involved in active or binding sites
- Knowledge of proteolytic cleavage sites, post-translational modifictions, such as phosphorylation or glycosylation can suggest residues that must be accessible
- Etc.

Remember to keep all of the available data in mind when doing predictive work. Always ask yourself whether a prediction agrees with the results of experiments. If not, then it may be necessary to modify what you've done.

## Protein sequence data

There is some value in doing some initial analysis on your protein sequence. If a protein has come (for example) directly from a gene prediction, it may consist of multiple domains. More seriously, it may contain regions that are unlikely to be globular, or soluble. This flowchart assumes that your protein is soluble, likely comprises a single domain, and does not contain non-globular regions.

Things to consider are:

- Is your protein a transmembrane protein, or does it contain transmembrane segments? There are many methods for predicting these segments, including:
  - TMAP (EMBL)
  - PredictProtein (EMBL/Columbia)
  - TMHMM (CBS, Denmark)
  - TMpred (Baylor College)
  - DAS (Stockholm)
- Does your protein contain coiled-coils? You can predict coiled coils at the COILS server .
- Does your protein contain regions of low complexity? Proteins frequently contain runs of poly-glutamine or poly-serine, which do not predict well. To check for this you can use the program SEG .

If the answer to any of the above questions is yes, then it is worthwhile trying to break your sequence into pieces, or ignore particular sections of the sequence, etc.

## Sequence database searching

The most obvious first stage in the analysis of any new sequence is to perform comparisons with sequence databases to find homologues. These searches can now be performed just about anywhere and on just about any computer. In addition, there are numerous web servers for doing searches, where one can post or paste a sequence into the server and receive the results interactively:

There are many methods for sequence searching. By far the most well known are the BLAST suite of programs. One can easily obtain versions to run locally and there are many web pages that permit one to compare a protein or DNA sequence against a multitude of gene and protein sequence databases. To name just a few:

- National Center for Biotechnology Information (USA) Searches
- European Bioinformatics Institute (UK) Searches
- BLAST search through SBASE (domain database; ICGEB, Trieste)
- and others too numerous to mention.

One of the most important advances in sequence comparison recently has been the development of both gapped BLAST and PSI-BLAST (position specific interated BLAST). Both of these have made BLAST much more sensitive, and the latter is able to detect very remote homologues by taking the results of one search, constructing a *profile* and then using this to search the database again to find other homologues (the process can be repeated until no new sequences are found). It is essential that one compare any new protein sequence to the database with PSI-BLAST to see if known structures can be found prior to doing any of the other methods discussed in the next sections.

Other methods for comparing a single sequence to a database include:

- The FASTA suite (William Pearson, University of Virginia, USA)
- SCANPS (Geoff Barton, European Bioinformatics Institute, UK)
- BLITZ (Compugen's fast Smith Waterman search)
- and others.

It is also possible to use multiple sequence information to perform more sensitive searches. Essentially this involves building a *profile* from some kind of multiple sequence alignment. A profile essentially gives a score for each type of amino acid at each position in the sequence, and generally makes searches more sentive. Tools for doing this include:

- PSI-BLAST (NCBI, Washington)
- ProfileScan Server (ISREC, Geneva)
- HMMER Hidden Markov Model searching (Sean Eddy, Washington University)
- Wise package (Ewan Birney, Sanger Centre; this is for protein versus DNA comparisons)
- and several others.

A different approach for incorporating multiple sequence information into a database search is to use a MOTIF. Instead of giving every amino acid some kind of score at every position in an alignment, a motif ignores all but the most invariant positions in an alignment, and just describes the key residues that are conserved and define the family. Sometimes this is called a "signature". For example, "H-[FW]-x-[LIVM]-x-G-x(5)-[LV]-H-x(3)-[DE]" describes a family of DNA binding proteins. It can be translated as "histidine, followed by either a phenylalanine or tryptophan, followed by an amino acid (x), followed by leucine, isoleucine, valine or methionine, followed by any amino acid (x), followed by glycine,... [etc.]".

PROSITE (ExPASy Geneva) contains a huge number of such patterns, and several sites allow you to search these data:

- ExPASy
- EBI

It is best to search a few different databases in order to find as many homologues as possible. A very important thing to do, and one which is sometimes overlooked, is to compare any new sequence to a database of sequences for which 3D structure information is available. Whether or not your sequence is homologous to a protein of known 3D structure is not obvious in the output from many searches of large sequence databases. Moreover, if the homology is weak, the similarity may not be apparent at all during the search through a larger database.

One last thing to remember is that one can save a lot of time by making use of pre-prepared protein alignments. Many of these alignments are hand edited by experts on the particular protein families, and thus represent probably the best alignment one can get given the data they contain (i.e. they are not always as up to date as the most recent sequence databases). These databases include:

- SMART (Oxford/EMBL)
- PFAM (Sanger Centre/Wash-U/Karolinska Intitutet)
- COGS (NCBI)
- PRINTS (UCL/Manchester)
- BLOCKS (Fred Hutchinson Cancer Research Centre, Seatle)
- SBASE (ICGEB, Trieste)

Generally one can compare a protein sequence to these databases via a variety of techniques. These can also be very useful for the domain assignment.

## Locating domains

If you have a sequence of more than about 500 amino acids, you can be nearly certain that it will be divided into discrete functional domains. If possible, it is preferable to split such large proteins up and consider each domain separately. You can predict the location of domains in a few different ways. The methods below are given (approximately) from most to least confident.

- If homology to other sequences occurs only over a portion of the probe sequence and the other sequences are whole (i.e. not partial sequences), then this provides the strongest evidence for domain structure. You can either do database searches yourself or make use of well-curated, pre-defined databases of protein domains. Searches of these databases (see links below) will often assign domains easily.
    - SMART (Oxford/EMBL)
    - PFAM (Sanger Centre/Wash-U/Karolinska Intitutet)
    - COGS (NCBI)
    - PRINTS (UCL/Manchester)
    - BLOCKS (Fred Hutchinson Cancer Research Centre, Seatle)
    - SBASE (ICGEB, Trieste)
- Regions of low-complexity often separate domains in multidomain proteins. Long stretches of repeated residues, particularly Proline, Glutamine, Serine or Threonine often indicate linker sequences and are usually a good place to split proteins into domains.

  Low complexity regions can be defined using the program SEG which is generally available in most BLAST distributions or web servers (a version of SEG is also contained within the GCG suite of programs).

- Transmembrane segments are also very good dividing points, since they can easily separate extracellular from intracellular domains. There are many methods for predicting these segments, including:
    - TMAP (EMBL)
    - PredictProtein (EMBL/Columbia)
    - TMHMM (CBS, Denmark)
    - TMpred (Baylor College)
    - DAS (Stockholm)

- Something else to consider are the presence of *coiled-coils*. These unusual structural features sometimes (but not always) indicate where proteins can be divided into domains.
- Secondary structure prediction methods will often predict regions of proteins to have different protein structural classes. For example one region of sequence may be predicted to contain only alpha helices and another to contain only beta sheets. These can often, though not always, suggest likely domain structure (e.g. an all alpha domain and an all beta domain)

If you have separated a sequence into domains, then it is very important to repeat all the database searches and alignments using the domains separately. Searches with sequences containing several domains may not find all sub-homologies, particularly if the domains are abundant in the database (e.g. kinases, SH2 domains, etc.). There may also be "hidden" domains. For example if there is a stretch of 80 amino acids with few homologues nested in between a kinase and an SH2 domain, then you may miss matches found when searching the *whole* sequence against a database.

## Multiple Sequence Alignment

Regardless of the outcome of your searches, you will want a multiple sequence alignment containing your sequence and all the homologues you have found above.

Some sites for performing multiple alignment:

- EBI (UK) Clustalw Server
- IBCP (France) Multalin Server
- IBCP (France) Clustalw Server
- IBCP (France) Combined Multalin/Clustalw
- MSA (USA) Server
- BCM Multiple Sequence Alignment ClustalW Sever (USA)

Alignments can provide:

- Information as to protein domain structure
- The location of residues likely to be involved in protein function
- Information of residues likely to be buried in the protein core or exposed to solvent
- More information than a single sequence for applications like homology modelling and secondary structure prediction.

## Secondary Structure Prediction methods

With no homologue of known structure from which to make a 3D model, a logical next step is to predict secondary structure. Although they differ in method, the aim of secondary structure prediction is to provide the location of alpha helices, and beta strands within a protein or protein family.

### *Methods for single sequences*

Secondary structure prediction has been around for almost a quarter of a century. The early methods suffered from a lack of data. Predictions were performed on single sequences rather than families of homologous sequences, and there were relatively few known 3D structures from which to derive parameters. Probably the most famous early methods are those of Chou & Fasman, Garnier, Osguthorbe & Robson (GOR) and Lim. Although the authors originally claimed quite high accuracies (70-80 %), under careful examination, the methods were shown to be only between 56 and 60% accurate (see Kabsch & Sander, 1984 given below). An early problem in secondary structure prediction had been the inclusion of structures used to derive parameters in the set of structures used to assess the accuracy of the method.

### *Recent improvments*

The availability of large families of homologous sequences revolutionised secondary structure prediction. Traditional methods, when applied to a family of proteins rather than a single sequence proved much more accurate at identifying core secondary structure elements. The combination of sequence data with sophisticated computing techniques such as neural networks has lead to accuracies well in excess of 70 %. Though this seems a small percentage increase, these predictions are actually much more useful than those for single sequence, since they tend to predict the core accurately. Moreover, the limit of 70-80% may be a function of secondary structure variation within homologous proteins.

There are numerous automated methods for predicting secondary structure from multiply aligned protein sequences. Nearly all of these now run via the world wide web.

### *Manual intervention*

It has long been recognised that patterns of residue conservation are indicative of particular secondary structure types. Alpha helices have a periodicity of 3.6, which means that for helices with one face buried in the protein core, and the other exposed to solvent, will have residues at positions i, i+3, i+4 & i+7 (where i is a residue in an a helix) will lie on one face of the helix. Many alpha helices in proteins are amphipathic, meaning that one face is pointing towards the hydrophobic core and the other towards the solvent. Thus patterns of hydrophobic residue conservation showing the i, i+3, i+4, i+7 pattern are highly indicative of an alpha helix.

For example, this helix in myoglobin has this classic pattern of hydrophobic and polar residue conservation (*i* = *1*):

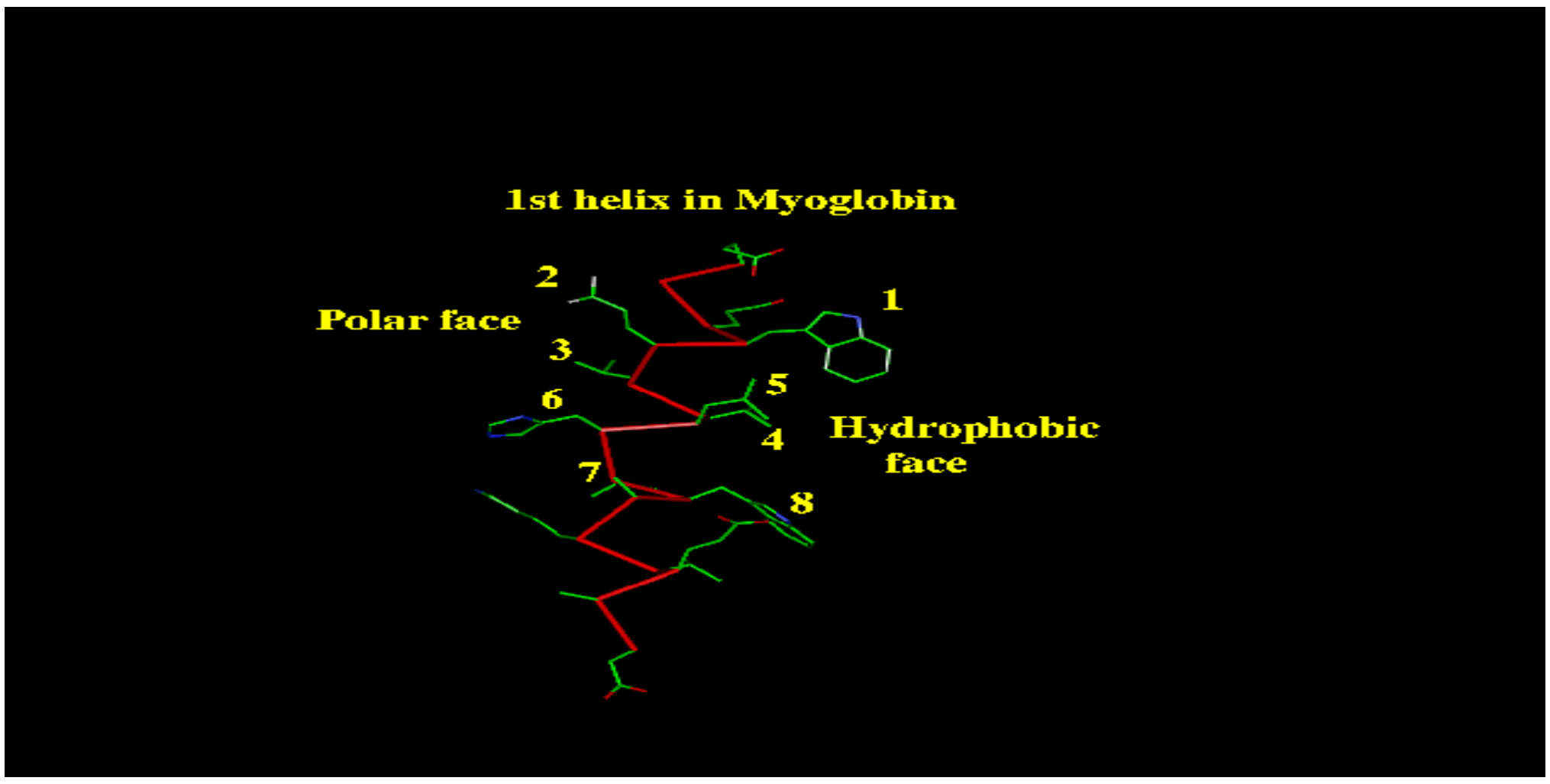

Similarly, the geometry of beta strands means that adjacent residues have their side chains pointing in oppposite directions. Beta strands that are half buried in the protein core will tend to have hydrophobic residues at positions i, i+2, i+4, i+8 etc, and polar residues at positions i+1, i+3, i+5, etc.

For example, this beta strand in CD8 shows this classic pattern:

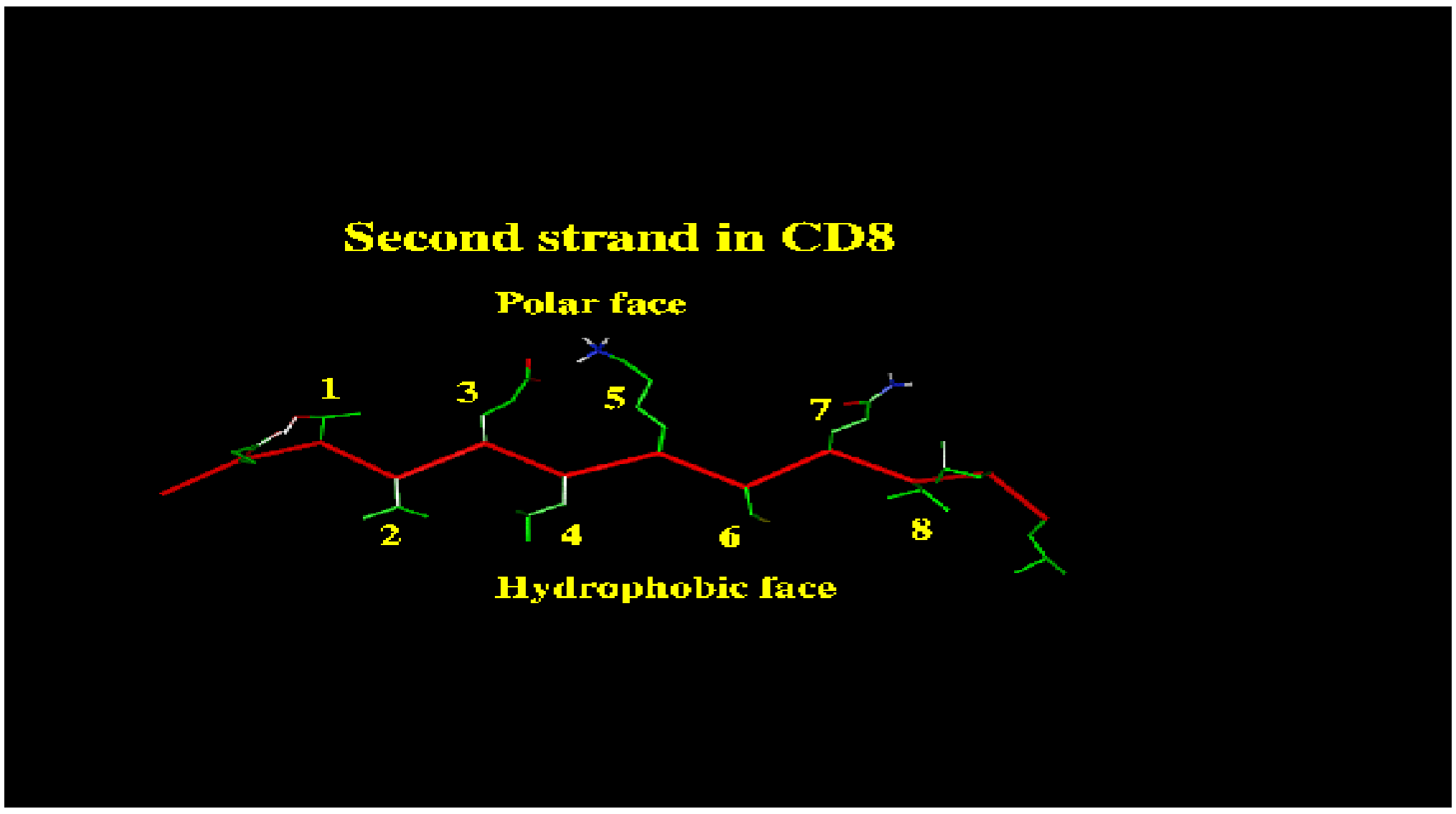

Beta strands that are completely buried (as is often the case in proteins containing both alpha helices and beta strands) usually contain a run of hydrophobic residues, since both faces are buried in the protein core.

This strand from Chemotaxis protein CheY is a good example:

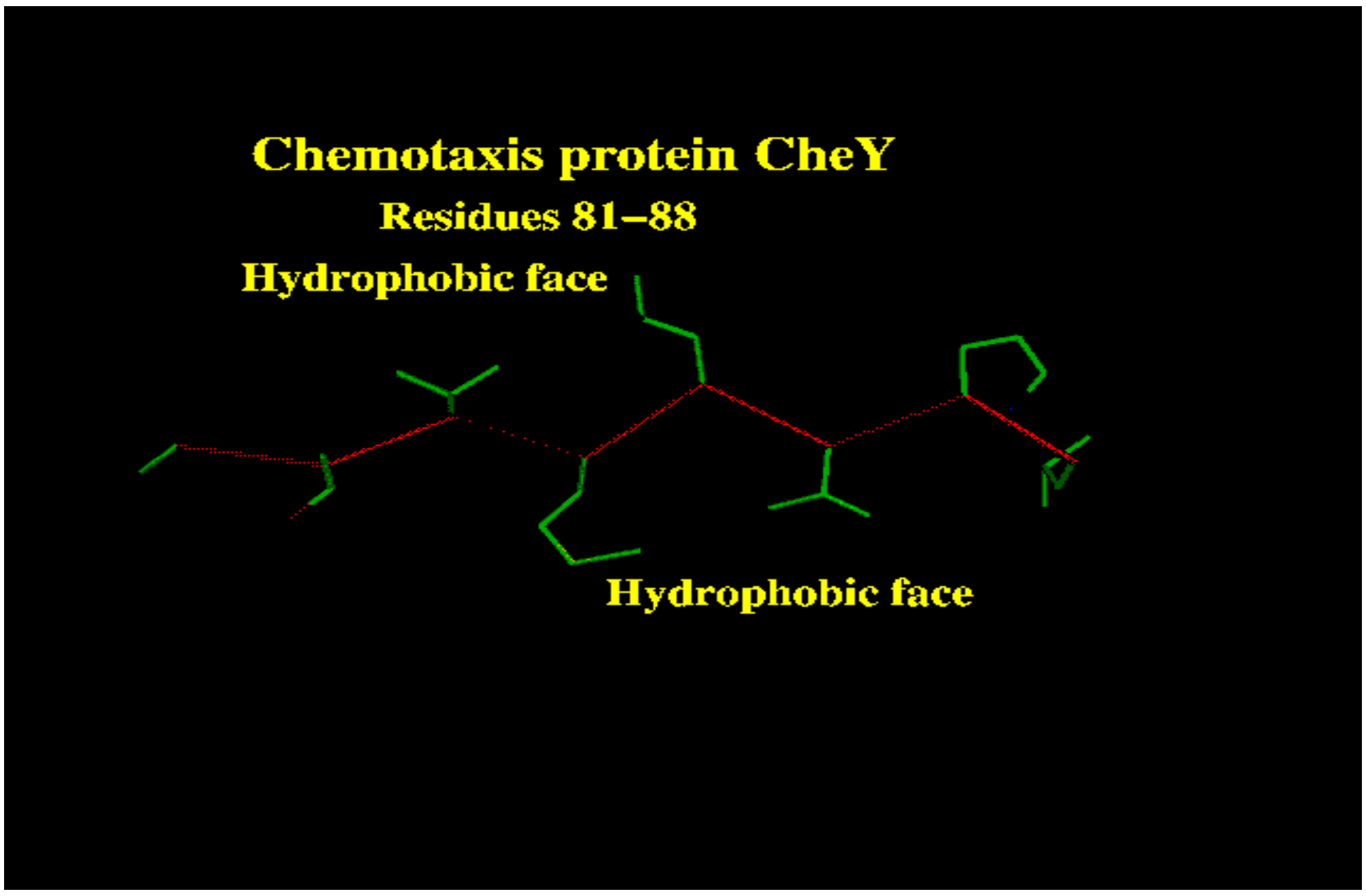

- The principle behind most manual secondary structure predictions is to look for patterns of residue conservation that are indicative of secondary structures like those shown above. It has been shown in numerous successful examples that this strategy often leads to nearly perfect predictions.

## A strategy for secondary structure prediction

For example, here is part of an alignment of a family of proteins I looked at recently:

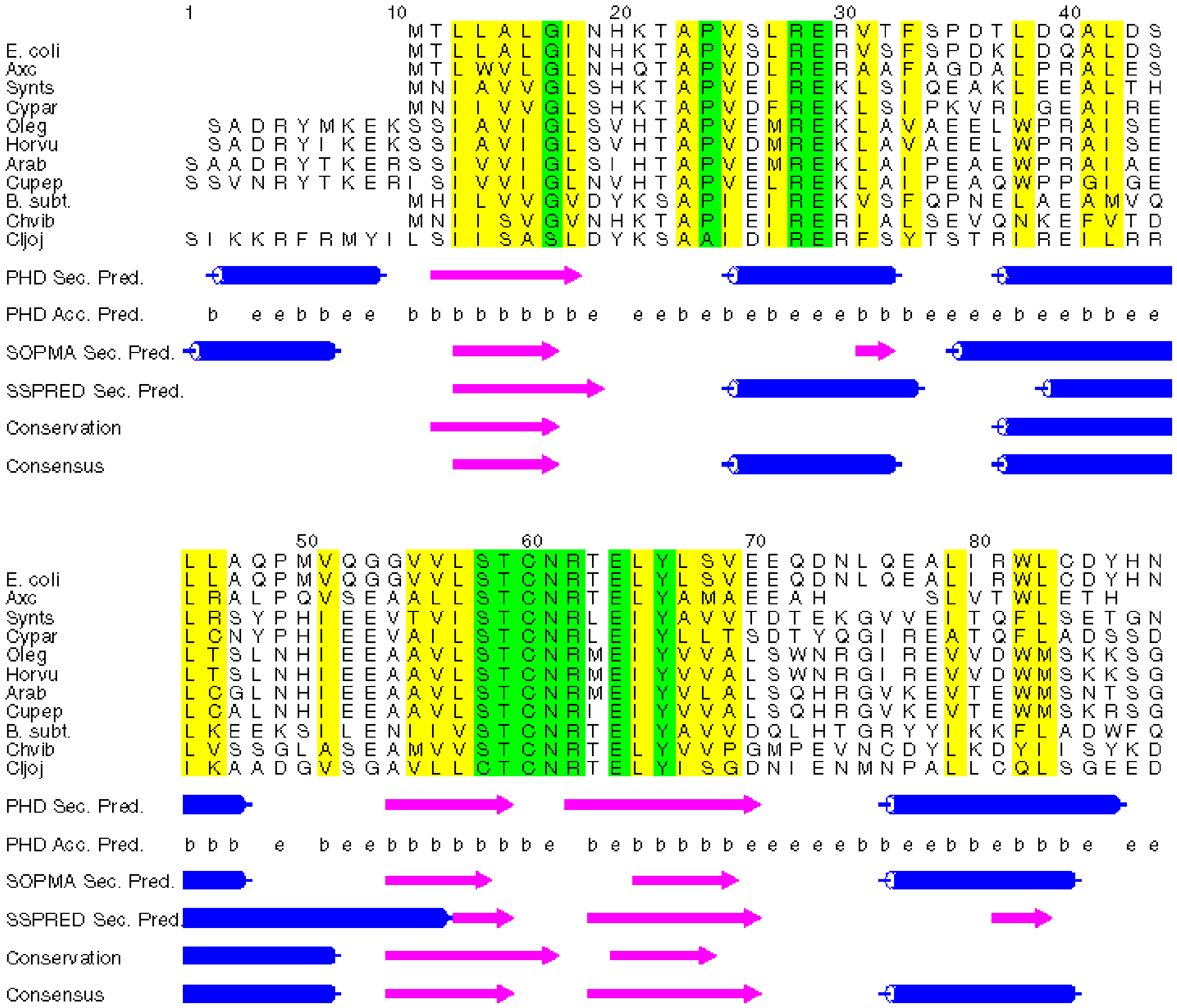

In this figure, three automated secondary structure predictions (PHD, SOPMA and SSPRED) appear below the alignment of 12 glutamyl tRNA reductase sequences. Positions within the alignment showing a conservation of hydrophobic side-chain character are shown in yellow, and those showing near total conservation of non-hydrophobic residues (often indicative of active sites) are coloured green.

Predictions of accessibility performed by PHD (PHD Acc. Pred.) are also shown (b = buried, e = exposed), as is a prediction I performed by looking for patterns indicative of the three secondary structure types shown above. For example, positions (within the alignment) 38-45 exhibit the classical amphipathic helix pattern of hydrophobic residue conservation, with positions i, i+3, i+4 and i+7 showing a conservation of hydrophobicity, with intervening positions being mostly polar. Positions 13-16 comprise a short stretch of conserved hydrophobic residues, indicative of a beta-strand, similar to the example from CheY protein shown above.

By looking for these patterns I built up a prediction of the secondary structure for most regions of the protein. Note that most methods - automated and manual - agree for many regions of the alignment.

Given the results of several methods of predicting secondary structure, one can build up a *consensus* picture of the secondary structure, such as that shown at the bottom of the alignment above.

## Fold recognition methods

Even with no homologue of known 3D structure, it may be possible to find a suitable fold for you protein among known 3D structures by way of *fold recognition methods*

### *3D structural similarities*

*Ab initio* prediction of protein 3D structures is not possible at present, and a general solution to the protein folding problem is not likely to be found in the near future. However, it has long been recognised that proteins often adopt similar folds despite no significant sequence or functional similarity and that nature is apparently restricted to a limited number of protein folds.

There are numerous protein structure classifications now available via the WWW:

- SCOP (MRC Cambridge)
- CATH (University College, London)
- FSSP (EBI, Cambridge)
- 3 Dee (EBI, Cambridge)
- HOMSTRAD (Biochemistry, Cambridge)
- VAST (NCBI, USA)

Thus for many proteins (~ 70%) there will be a suitable structure in the database from which to build a 3D model. Unfortuantely, the lack of sequence similarity will mean that many of these go undetected until after 3D structure determination.

### *The goal of fold recognition*

Methods of protein fold recognition attempt to detect similarities between protein 3D structure that are not accompanied by any significant sequence similarity. There are many approaches, but the unifying theme is to try and find folds that are compatable with a particular sequence. Unlike sequence-only comparison, these methods take advantage of the extra information made available by 3D structure information. In effect, the turn the protein folding problem on it's head: rather than predicting how a sequence will fold, they predict how well a fold will fit a sequence.

### *The realities of fold recognition*

Despite initially promising results, methods of fold recognition are not always accurate. Guides to the accuracy of protein fold recognition can be found in the proceedings of the Critical Assessment of Structure Predictions (CASP) conferences. At the first meeting in 1994 (CASP1) the methods were found to be about 50 % accurate at best with respect to their ability to place a correct fold at the top of a ranked list. Though many methods failed to detect the correct fold at the top of a ranked list, a correct fold was often found in the top 10 scoring folds. Even when the methods were successful, alignments of sequence on to protein 3D structure were usually incorrect, meaning that comparative modelling performed using such models would be inaccurate.

The CASP2 meeting held in December 1996, showed that many of the methods had improved, though it is difficult to compare the results of the two assessments (i.e. CASP1 & CASP2) since very different criteria were used to assess correct answers. It would be foolish and over-ambitious for me to present a detailed assessment of the results here. However, and important thing to note, was that Murzin & Bateman managed to attain near 100% success by the use of careful human insight, a knowledge of known structures, secondary structure predictions and thoughts about the function of the target sequences. Their results strongly support the arguments given below that human insight can be a powerful aid during fold recognition. A summary of the results from this meeting can be found in the *PROTEINS* issue dedicated to the meeting .

The CASP3 meeting was held in December 1998. It showed some progress in the ability of fold recognition methods to detect correct protein folds and in the quality of alignments obtained. A detailed summary of the results will appear towards the end of 1999 in the *PROTEINS* supplement.

For my talk, I did a crude assessment of 5 methods of fold recognition. I took 12 proteins of known structure (3 from each folding class) an ran each of the five methods using default parameters. I then asked how often was a correct fold (not allowing trival sequence detectable folds) found in the first rank, or in the top 10 scoring folds. I also asked how often the method found the correct folding class in the first rank.

Perhaps the worst result from this study is shown below:

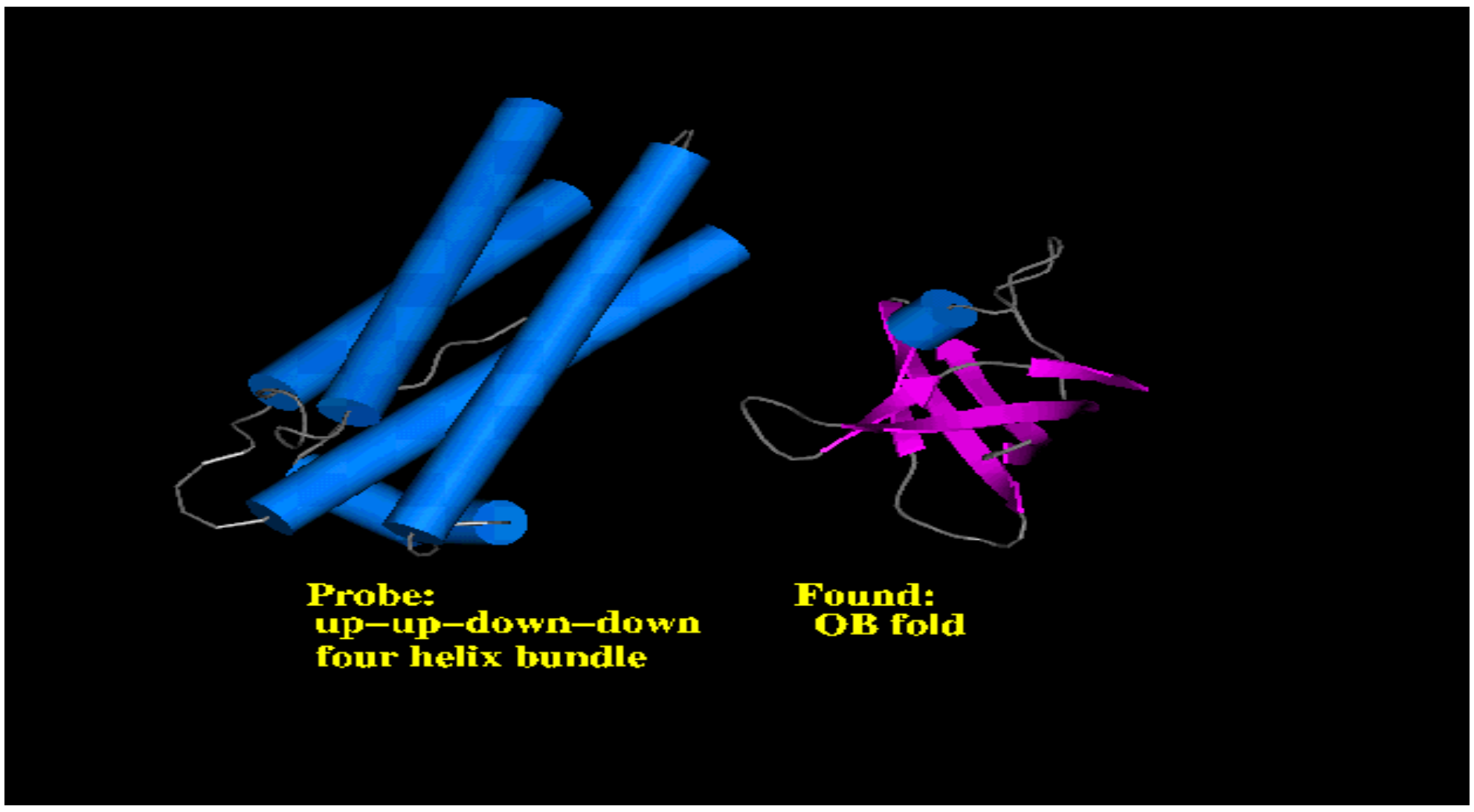

One method suggested that the sequence for the Probe (left) (a four helix bundle) would best fit onto the structure shown on the right (an OB fold, comprising a six stranded barrel).

The results suggest that one should use caution when using these methods. In spite of this, the methods remain very useful.

**A practical approach:**

Although they are not 100 % accurate, the methods are still very useful. To use the methods I would suggest the following:

- Run as many methods as you can, and run each method on as many sequences (from your homologous protein family) as you can. The methods almost always give somewhat different answers with the same sequences. I have also found that a single method will often give different results for sets of homologous sequences, so I would also suggest running each method on as many homologoues as possible. After all of these runs, one can build up a consensus picture of the likely fold in a manner similar to that used for secondary structure prediction above.
- Remember the expected accuracy of the methods, and don't use them as black-boxes. Remember that a correct fold may not be at the top of the list, but that it is likely to be in the top 10 scoring folds.

  Think about the function of your protein, and look into the function of the proteins that have been found by the various methods. If you see a functional similarity, then you may have detected a *weak sequence homologue*, or *remote homologue*. At CASP2, as said above, Murzin & Bateman managed to obtain remarkably accurate predictions by identification of remote homologues.

## Analysis of protein folds and alignment of secondary structure elements

If you have predicted that your protein will adopt a particular fold within the database, then an important thing to consider to which fold your protein belongs, and other proteins that adopt a similar fold. To find out, look at one of the following databases:

- SCOP (MRC Cambridge)
- CATH (University College, London)
- FSSP (EBI, Cambridge)
- 3 Dee (EBI, Cambridge)
- HOMSTRAD (Biochemistry, Cambridge)
- VAST (NCBI, USA)

If your predicted fold has many "relatives", then have a look at what they are. Ask:

- Do any of members show functional similarity to your protein? If there is any functional similarity between your protein and any members of the fold, then you may be able to back up your prediction of fold (possibly by the conservation of active site residues, or the approximate location of active site residues, etc.)
- Is this fold a *superfold*? If so, does this superfold contain a *supersite*? Certain folds show a tendancy to bind ligands in a common location, even in the absense of any functional or clear evolutionary relationships.
- Are there *core* secondary structure elements that should really be present in any member of the fold?

- Are there non-core secondary structure elements that might not be present in all members of the fold?

Core secondary structure elements, such as those comprising a beta-barrel, should really be present in a fold. If your predicted secondary structures can't be made to match up with what you think is the core of the protein fold, then your prediction of fold may be wrong (but be careful, since your secondary structure prediction may contain errors). You can also use your prediction together with the core secondary structure elements to derive an alignment of predicted and observed secondary structures.

## Comparative or Homology Modelling

If your protein sequence shows significant homology to another protein of known three-dimensional structure, then a fairly accurate model of your protein 3D structure can be obtained via homology modelling. It is also possible to build models if you have found a suitable fold via fold recognition and are happy with the alignment of sequence to structure

It is possible now to generate models automatically using the very useful SWISSMODEL server.

Sequence alignments, particularly those involving proteins having low percent sequence identities can be inaccurate. If this is the case, then a model built using the alignment will obvious be wrong in some places.

Once you have a three-dimensional model, it is useful to look at protein 3D structures. There are numerous free programs for doing this, including:

- GRASP Anthony Nicholls, Columbia, USA.
- MolMol Reto Koradi, ETH, Zurrich, C.H.
- Prepi Suhail Islam, ICRF, U.K.
- RasMol Roger Sayle, Glaxo, U.K.

Most places with groups studying structural biology also have commercial packages, such as Quanta, SYBL or Insight, which contain more features than the visualisation packages described above.

### *Have you found a suitable fold?*

If yes then try to get an alignment of secondary structures, if not then look into tertiary structure prediction.

### *Is your protein homologous to a known structure?*

If yes, then you can proceed to comparative or homology modelling otherwise you will probably need to perform a secondary structure prediction .

# PROTEOMICS

Proteomics is a scientific discipline concerned with systematic analyses of proteins present in cells at a given time under given conditions. Proteomics includes the identification, characterization and quantitation of the entire complement of proteins in cells, tissues or whole organisms with a view to understanding their function in relation to the life of the cell.

After the sequencing of the human genome, it is now clear that much of the complexity of the human body resides at the level of proteins rather than the DNA sequences. This view is supported by the unexpectedly low number of human genes (approximately 35,000) and the estimated number of proteins (which is currently about 300,000 - 450,000 and steadily rising), which are generated from these genes. For example, it is estimated that on average human proteins exist as ten to fifteen different post- transitionally modified forms all of which presumably - have different functions. Much of the information processing in healthy and diseased human cells can only be studied at the protein level, and there is increasing evidence to link minor changes in expression of some of these modifications with specific diseases. Together with rapidly improving technology for characterising the proteome, there is now a unique chance to make an impact in this area.

## Proteomics in life sciences

Proteomics has an incredibly wide field of applications and beside the obvious applications in understanding the processes of life, many diverse practical applications exist in the fields of:

- medicine
- biotechnology
- food sciences
- agriculture
- animal genetics and horticulture
- environmental surveillance
- pollution

Within the field of medicine alone in Odense, proteomics has been applied to:

- protein changes during normal processes like differentiation, development and ageing
- abnormal protein expression in disease development (and is especially suited for studies of diseases of multigenic origin)
- diagnosis, prognosis
- identification of novel drug targets
- selection of candidate drugs
- surrogate markers
- targets for gene therapy
- toxicology
- mechanism of drug action

## Challenges in Proteomics

### Proteomics for understanding gene function

Proteomics has been the ideal tool for investigating the function of unknown genes for many years. The deletion (or insertion) of a single gene induces a number of changes in the expression of groups of proteins with related function. These groups are often proteins in the same complex or within the me pathway and so these studies can map the activity of proteins even when their function is unknown or they are expressed at levels too low to be detected. The final step in understanding gene function is the transfection of particular genes into functionally related cells which do not posses the gene (or for example where the endogenous gene is deleted and replaced with a genetically modified version).

### Proteomics for understanding the molecular regulation of the cell

In the organism, cells exist in tissues in a highly specialised state and each cell type within each type of tissue is able to carry out certain particular functions which are often important or essential for the whole organism to perform correctly. When a cell is removed from its normal environment, it rapidly undergoes a process of de- differentiation. During this process the cell looses many of its specialised functions and for this reason alone, much of the early cancer work done until now has very little value as it has been based on established cell lines.

Therefore, one of the key tasks currently in hand is a quest to define growth conditions which press the cell to either retain its differentiated phenotype or even better to induce the cell to develop into a structure that resembles the tissue. Early successes were the derivation of so called "raft cultures" for the development of a multilayered keratinocyte tissue that is indistinguishable from the epidermis (minus the hair and sweat follicles) when examined by the electron microscope. Currently the quest is to induce the differentiation of other cell types, (for example hepatocytes) into a structure that possesses the enzymatic functions of the original tissue. In the case of liver, this would have a huge potential value for studies of for example drug toxicology and environmental chemical surveillance. The use of microgravity devices (developed in collaboration with the North American Space Agency (NASA)) have been shown to induce liver like structures in vitro and thus appear to offer the solution. Current studies now in hand have demonstrated that many more specialised enzyme systems are active in the hepatocytes.

### Proteomics for identification of multiprotein complexes

The classical approach to studying multiprotein complexes has been to isolate them, either by direct purification or by immunoprecipitation and then identify the components of the complex one by one. Their function has been studied by interlink analysis of protein expression, or by site directed mutagenesis of particular proteins in the complex. Novel mass spectrometry based technology now for the first time is making it possible to study a large number of protein interactions in the same experiment. This capability is used commercially to "surround" known disease genes with interaction partners which are potentially more amenable to drug treatment. Furthermore, delineating the role of novel gene products in protein networks and in large multi-protein complexes in the cell has proven to be one of the best ways to learn about their function. This knowledge is indispensable both in basic cellular biology and in biomedicine. Novel proteomics technologies which are now in academic development in Odense and elsewhere promise to further extend the applicability of proteomics to measure changes in the cell. These capabilities include the direct and automated measurement of protein expression level changes (which can be applied to, for example, cancer patient material) as well as the large scale study of signalling processes through the analysis of phosphorylation by mass spectrometry. Together, these developments have the potential to revolutionize the way cellular processes are studied in basic biology and in disease.

**Proteomics for studying cellular dynamics and organization**

Post-translational modifications of proteins (of which there are now known to be over 200 different types) govern key cellular processes, such as timing of the cell cycle, orchestration of cell differentiation and growth, and cell-cell communication processes. Specific labelling techniques and mass spectrometry are exquisitely suited to study post-translationally modified proteins, for example phosphoproteins, glycoproteins and acylated protein, that are present at defined stages of the cells life cycle or targeted to specific organelles within the cell. Specific radioisotopes (e.g. [$^{32}$P]-orthophosphate) or fluorescent dyes (for S-nitrosylation) allow a proteome-wide scanning for all proteins with particular types of modification identifying which proteins are modified, and what percentage of these proteins are modified under which growth conditions. Alternatively, a number of mass spectrometry based methods are in development that will enable systematic studies of whole populations of post-translationally modified proteins isolated from cells or tissues, leading to the concept of "modification-specific proteomics". These developments are pivotal for gaining further insights into the complexity of life. Delineating the dynamics of post-translational modification of proteins is a prerequisite for understanding the organization and dynamics of normal or perturbed cellular systems.

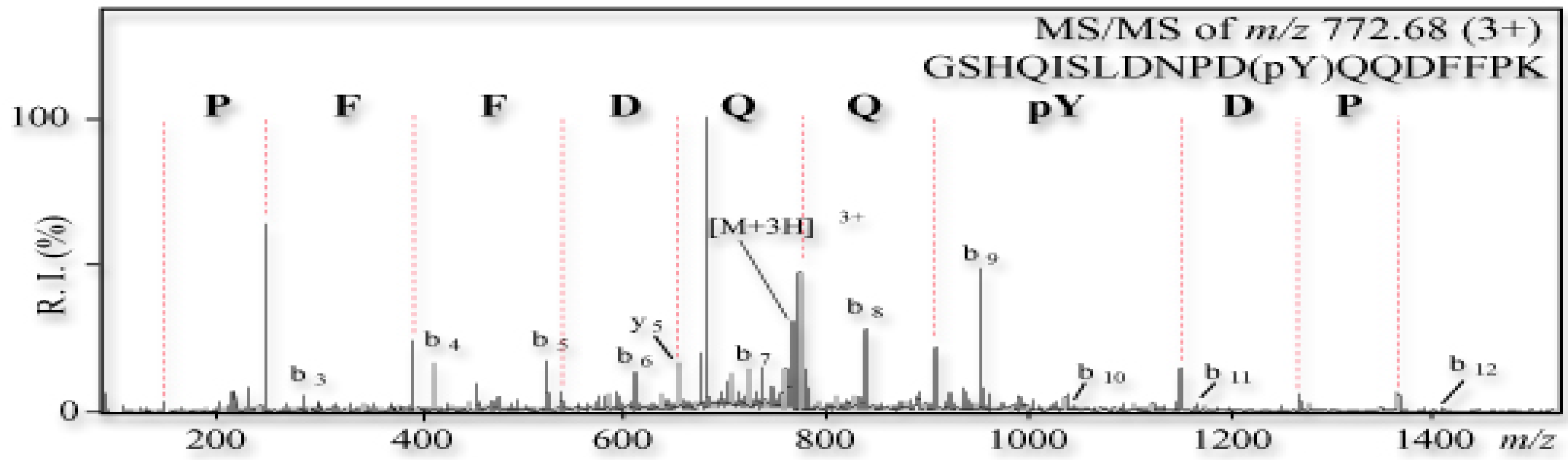

**Proteomics for studying macromolecular interactions**

Structural genomics/proteomics initiatives around the World provides a wealth of data on protein structures by using NMR and X-ray techniques. In parallel, advances in protein biochemistry and biological mass spectrometry and the availability of computational tools for molecular modelling provides novel opportunities to study the dynamics and molecular details of protein folding, protein-ligand and protein-nucleic acid interactions. Current research at SDU is directed towards the development of novel analytical strategies for systematic, structural studies of protein-nucleic acid complexes and other types of protein assemblies. The inclusion of experimentally determined structural constraints in bioinformatics-based simulations of large biomolecular assemblies will provide a firm basis for structure prediction and for identifying novel targets for intervention therapies for diseases.

## *Introduction*

Proteomics is the study of total protein complements, proteomes, e.g. from a given tissue or cell type. Nowadays proteomics can be divided into classical and functional proteomics. Classical proteomics is focused on studying complete proteomes, e.g. from two differentially treated cell lines, whereas functional proteomics studies more limited protein sets. Classical proteome analyses are usually carried out by using two-dimensional gel electrophoresis (2-DE) for protein separation followed by protein identification by mass spectrometry (MS) and database searches. The functional proteomics approach uses a subset of proteins isolated from the starting material, e.g. with an affinity-based method. This protein subset can then be separated by using normal SDS-PAGE or by 2-DE. Proteome analysis is complementary to DNA microarray technology: with the proteomics approach it is possible to study changes in protein expression levels and also protein-protein interactions and post-translational modifications.

## *Methods*

- **Two-dimensional electrophoresis**

Proteins are separated in 2-DE according to their pI and molecular weight. In 2-DE analysis the first step is sample preparation; proteins in cells or tissues to be studied have to be solubilized and DNA and other contaminants must be removed. The proteins are then separated by their charge using isoelectric focusing. This step is usually carried out by using immobilized pH-gradient (IPG) strips, which are commercially available. The second dimension is a normal SDS-PAGE, where the focused IPG strip is used as the sample. After 2-DE separation, proteins can be visualized with normal dyes, like Coomassie or silver staining.

- **Protein identification by mass spectrometry**

Mass spectrometers consist of the ion source, mass analyzer, ion detector, and data acquisition unit. First, molecules are ionized in the ion source. Then they are separated according to their mass-to-charge ratio in the mass analyzer and the separate ions are detected. Mass spectrometry has become a widely used method in protein analysis since the invention of matrix-assisted laser-desorption ionisation/time-of-flight (MALDI-TOF) and electrospray ionisation (ESI) methods. There are several options for the mass analyzer, the most common combinations being time-of-flight (TOF) connected to MALDI and triple quadrupole, quadrupole-TOF, or ion trap mass analyzer coupled to ESI.

In proteome analysis electrophoretically separated proteins can be identified by mass spectrometry with two different approaches. The simplest way is a technique called peptide mass fingerprinting (PMF). In this approach the protein spot of interest is in-gel digested with a specific enzyme, the resulting peptides are extracted from the gel and the molecular weights of these peptides are measured. Database search programs can create theoretical PMFs for all the proteins in the database, and compare them to the obtained one. In the second approach peptides after in-gel digestion are fragmented in the mass spectrometer, yielding partial amino acid sequences from the peptides (sequence tags). Database searches are then performed using both molecular weight and sequence information. PMF is usually carried out with MALDI-TOF, and sequence tags by nano-ESI tandem mass spectrometry (MS/MS). The sensitivity of protein identification by MS is in the femtomole range.

## *Protocols*

➢ **SILVER STAINING**

| | | |
|---|---|---|
| Fixation | 30% EtOH, 10% HAc | 1 h (-overnight) |
| Rinse | 20% EtOH<br>MQ-water | 15 min<br>15 min |
| Sensitization | Sodium thiosulfate, 0.2 g/l | 90 sec |
| Rinse | MQ-water | 2* 20 sec |
| Silver | Silver nitrate, 2.0 g/l | 30 min |
| Rinse | MQ-water | 2* 10 sec |
| Development | 37% Formaldehyde, 0.7 ml/l<br>Potassium carbonate, 30 g/l<br>Sodium thiosulfate, 10 mg/l | to a desired intensity |
| Stop | Tris-base, 50 g/l, 2.5% HAc | 1 min |

Important:
-use plenty of MQ-water for rinsing
-make Sensitizer, Silver and Developer just prior to use
-if you stain the gels in plastic boxes, wipe the boxes with ethanol before use

- **IN-GEL ALKYLATION AND DIGESTION**

1. Excise the stained protein band from the gel with a scalpel. Cut the gel slice into ca. 1x1 mm cubes, and put the pieces in an Eppendorf tube. (Use small tubes that are rinsed with ethanol/methanol.)

2. Wash the gel pieces twice in 200 µl 0.2 M $NH_4HCO_3$/ACN (1:1) for 15 min at 37 $^oC$.

3. Shrink the gel pieces by adding 100 µl ACN (wait until the pieces become white, about 10 min), and then remove ACN. Dry in vacuum centrifuge for 5 min.

4. Rehydrate the gel pieces in 100 µl 20 mM DTT in 0.1 M $NH_4HCO_3$ for 30 min at 56$^oC$. (3.08 mg DTT/1 ml 0.1 M $NH_4HCO_3$)

5. Remove excess liquid, and shrink the pieces with ACN as above (no vacuum is needed). Remove ACN, and add 100 µl 55 mM iodoacetamide in 0.1 M $NH_4HCO_3$. Incubate 15 min in the dark at RT. (10.2 mg IAA/1 ml 0.1 M $NH_4HCO_3$)

6. Remove excess liquid, and wash the pieces twice with 100 µl 0.1 M $NH_4HCO_3$ and shrink again with ACN. Dry in vacuum centrifuge for 5 min.

7. Add 10 µl 0.04 µg/µl mTr solution (in 0.1 M $NH_4HCO_3$ ,10% ACN), and allow to absorb

for 10 min. Then add 0.1 M $NH_4HCO_3$, 10% ACN to completely cover the gel pieces. Incubate overnight at 37 $^oC$ at incubator. (Remember to check the pH of 0.1 M $NH_4HCO_3$, should be ~8)

8. Add 20-50 µl 5% HCOOH (equal volume with the digestion mixture), vortex and incubate 15 min at 37 $^oC$ and collect the supernatant for tipping. HCOOH extraction can be repeated.

OR

8. Add 20-50 µl ACN (equal volume with the digestion mixture), vortex and incubate 5 min at 37 $^oC$ and collect the supernatant. Extract peptides twice at 37 $^oC$ with 150 µl 5 % HCOOH/50 % ACN (pool the supernatant and extracts). Dry down the peptide extracts in a vacum centrifuge.

- **Tipping (MALDI)**

1. Twist the head of a Gel Loader-tip with forcepts

2. Add 20 µl of MeOH into the tip

3. Add 0.5 - 1 µl of Poros R3-slurry (Poros-material in MeOH) into the tip

4. Push MeOH through the tip with syringe

5. Equilibrate with 2 x 20 µl of 0.1% TFA

6. Pipet the sample into the tip and push it through

7. Wash 2 x 10 µl 0.1% TFA

8. Cut the tip just after the twisted part with scissors

9. Add 1.5 µl of matrix material in 0.1% TFA / 60% ACN into the tip

10. Elute the peptides with the matrix straight to the MALDI-plate

11. Let the matrix dry completely

**Q: What are the applications of proteomics?**

**A:** The main applications of proteomics are in drug discovery (identification of drug targets, target validation, toxicology) and in clinical diagnostics (screening of patients for drug responsiveness and side effects). Proteomics makes it possible to obtain and rapidly sort large quantities of biological data from the analysis of biological fluids from healthy and sick individuals. This should allow improved diagnostic methods that may be used to identify molecular abnormalities in diseased states.

**Q: What is proteomics?**

**A:** Proteomics is the study of the full expression of proteins by cells in their lifetime and is the next major step toward understanding how our bodies work and why we fall victim to disease. It aims to study directly the role and function of proteins in tissues and cells and promises to bridge the gap between genome nucleotide sequences and cellular behavior. Proteomics combines know-how in biology, engineering, chemistry and bioinformatics with technologies that include automation, robotics and integrated systems, key strengths of Tecan. Two distinct but complementary strategies have arisen in proteomics – the first monitors the expression of a large number of proteins within a cell or a tissue and follows how this pattern of expression changes under different circumstances, for example in the presence of a drug or in a diseased tissue; the second identifies the structure of proteins and, in particular, characterises proteins that interact with each other. The second approach is suited to the detailed study of particular pathways working in cells; the first offers a broader picture of the proteome. Both will be indispensable tools for drug development.

**Q: What are the key bottlenecks in proteomics?**

**A:** Key bottlenecks exist in the current methods used for proteomics research. The sensitivity (for example difficulties in detecting unknown, low-abundant proteins that could be interesting drug targets), reproducibility (lack of standardized methods for two-dimensional electrophoresis) and the high throughput capacity (lack of automation) of all the current methods employed will have to be significantly improved to fulfil the potential of the technologies. Tecan is ideally positioned in these areas to provide workstations and solutions that will systematically overcome the key bottlenecks facing the industry.

**Q: What is the total available market in proteomics?**

**A:** The proteomics market had an estimated value of CHF 1.7 billion in 2000. The total available market for Tecan is estimated at CHF 760 million. The proteomics market segment is expected to grow by 55% annually to CHF 2.8 billion by 2003.

**Q: What is the role of Tecan Proteomics GmbH?**

**A:** Tecan Proteomics GmbH was established in January 2001 at the European Biotech Center in Munich, Germany. Tecan will commit more than 20% of its total R&D budget for 2001 to this enterprise. The role of Tecan Proteomics is to address the key technological bottlenecks that exist in this field, ranging from the sensitivity of analytical tools and reproducibility of results from one laboratory to the next, to the lack of high throughput techniques. Leading the team of top scientists is Dr Christoph Eckerskorn, who pioneered key analytical approaches used today in proteomics while at the Max Planck Institute in Germany. When Tecan Proteomics was established, a majority share in Dr Weber GmbH was also secured. Dr Weber GmbH develops and commercializes proprietary free-flow electrophoresis, a key enabling technology in protein fractionation. This technology significantly reduces the complexity of a proteome, a major hurdle in proteomics research today.

***Q. Can you tell us more about proteomics, which I understand is a relatively new term?***

A. The more recent developments in recombinant DNA technologies and other biological techniques have endowed scientists with the unprecedented power of expressing proteins in large quantities, in a variety of conditions, and in manipulating their structures. While scientists were usually accustomed to studying proteins one at a time, proteomics represents a comprehensive approach to studying the total proteomes of different organisms. Thus, proteomics is not just about identification of proteins in complex biological systems, a huge task in itself, but also about their quantitative proportions, biological activities, localization in the living cells and their small compartments, interactions of proteins with each other and with other biomolecules. And ultimately, their functions. Because even the lower organisms can feature many thousands of proteins, proteomics-related activities are likely to keep us busy for two or three decades.

***Q. You are a biochemist. What is the connection in proteomics between biology and chemistry?***

A. With the boundaries between traditional scientific disciplines blurring all the time, this is somewhat difficult to answer. While proteomics has some clear connections to modern biological research, its current research tools and methodologies are chemical, some might even say, physio-chemical. You prominently see there are large, sophisticated machines like mass spectrometers, which in some parts of the world are still considered within the 'physics domain.' But things are continuously changing in this dynamic field. Genomic research, a distinct area of modern biology, has significantly changed the way in which we view the task of various proteomes.

***Q. What has changed in the past five or so years?***

A. The field of genomics, with its major emphasis on sequencing the basic building blocks of the 'central molecule,' DNA, already has yielded highly significant information on the previously unknown secrets of living cells. The stories of newly discovered genotype-phenotype relationships and strategies for understanding genetic traits and genetic diseases now regularly flood top scientific journals and popular literature alike. In parallel with providing the blueprint of the human genome, the genomes of many bacterial species, yeast and fruit flies for instance, have been sequenced, providing a valuable resource for modern biomedical research. The mouse genome also has been completed recently. Likewise, in the area of plant sciences, some important genomic advances have been reported. Yet, only a part of genetic information is of a direct use to proteomics.

***Q. How is that?***

A. While the entire protein sequences are encoded in the respective genomes, numerous variations occur as well. Some proteins may remain intact as the direct products of genes, but most will undergo structural alterations due to a variety of epigenetic factors. Thus, a given proteome is not merely a protein complement dictated by a genome, but rather a reflection of the dynamic situations in which different cells express themselves under various conditions of their environment. Here lie some of the most challenging problems of contemporary proteomics—to understand the dynamics, altered structural and quantitative attributes of proteins and their effects on metabolic pathways.

***Q. What other areas are you working in?***

A. Many proteins must become post-translationally modified to fulfill their biological roles. To assess such modifications, in a qualitative and quantitative sense, is thus at the heart of this field. Fortunately, we have been active for a number of years in studying one of the most prevalent and perhaps the most important posttranslational modification, called "glycosylation," which is the attachment of complex carbohydrate molecules to selected sites on a protein. Glycosylation is now being increasingly implicated in a number of disease-related conditions, including cancer, cardiovascular disease, neurological disorders and a great number of other medically interesting conditions.

Also, in collaboration with members of the medical faculty in Indianapolis, we will investigate the molecular attributes of human diseases. Studying the proteomics of highly complex multicellular systems of mammals provides some unusually exciting opportunities and challenges.

Proteomics is bound to become a more 'quantitative science' in the future. Unlike with the negatively charged nucleic acids that are relatively easy to handle in solutions and on surfaces, many proteins suffer from surface interactions and consequently, poor recoveries. We certainly wish to improve acquisition of the quantitative profiles of proteins for different protein classes.

***Q. What are some practical applications of proteomics research?***

Due to the multilateral importance of proteins in living systems, the scope of biomedical application is apparently wide. In fine-tuned functioning of the human body, various proteins act as the catalysts of chemical reactions, cell growth mediators, molecular transporters and receptors, immune agents against microorganisms and more. Consequently, various human diseases manifest themselves in the altered concentrations or structures of proteins, so that finding protein markers of a disease can, for example, result in devising better means of diagnosis or follow-up therapy. Numerous proteins have now been used therapeutically, so that there have to be perfect ways of manufacturing quality control for such therapeutics, or even to trace their action in the human body. Pharmaceutical companies, thus, have considerable interest in proteomics. Various activities in this area also provide considerable stimulus to the instrument industry. Consequently, coming up with new ways and better means to analyze complex protein mixtures is a high priority. These are just a few examples of how proteomics can impact our future.

# Proteomics and cancer diagnosis

**Large-scale protein analysis can identify cancer-related proteins that can be used as diagnostic markers and perhaps as targets for drug development.**

One of the major application of proteomics is the diagnosis and treatment of human diseases. The main proteomic technology platform, two-dimensional gel electrophoresis, is used to separate complex protein mixtures allowing individual protein spots on the gel to be identified by mass spectrometry.

If two related samples are compared, for example some normal healthy tissue and a disease sample, differences in the abundances of particular proteins may be evident. A protein present in the disease sample but not in the healthy sample may be a useful drug target. It may represent an aberrant form of a protein or a normal protein expressed at an unusually high level or in the wrong place.

Even if the protein is unsuitable as a drug target, it may still be useful as a diagnostic marker.

Although proteomics is a relatively new research area, there have already been some promising results in the diagnosis of cancer. For example, proteins have been identified that can be used to diagnose breast cancer, colon cancer and bladder cancer. A protein called stathmin has been identified that is expressed at unusually high levels in cases of childhood leukaemia.

Interestingly, the stathmin protein is phosphorylated in cancer patients – it has an additional phosphate group added to it. Many proteins are modified by phosphorylation after they are synthesized. Such modifications change the chemical properties of the protein allowing the phosphorylated and non-phosphorylated forms to be separated by two-dimensional gel electrophoresis. In the case of stathmin, only the phosphorylated form is associated with childhood leukaemia. This emphasises the importance of proteomics in disease diagnosis, because a change in protein modification associated with cancer cannot be detected using DNA arrays.

In the case of bladder cancer, proteomics analysis has identified several keratin proteins that are expressed in different amounts as the disease progresses from the early transitional epithelium stage to full blown squamous cell carcinoma.

The measurement of keratin levels in bladder cancer biopsies can therefore be used to monitor the progression of the disease. Another protein, psoriasin, is found in the urine of bladder cancer patients and can be used as an early diagnostic marker for the disease. This provides another example of how proteomics, but not DNA arrays, can be used in cancer diagnosis. Urine, in common with most bodily fluids, contains proteins but no RNA.

## CS905– BIOINFORMATICS 4

Introduction to Bioinformatics, Biological Databanks, Sequence Analysis, Structure Prediction, Protein Folding, Emerging Areas in Bioinformatics.